\documentclass{aa}
\usepackage[T1]{fontenc}
\usepackage{ae,aecompl}
\usepackage{graphicx}   
\usepackage{amsmath}    
\usepackage{amssymb}    
\usepackage{subcaption}
\usepackage{color}
\usepackage{nccmath}
\usepackage{float}
\usepackage{placeins}
\usepackage{arydshln}
\usepackage{pdflscape}
\usepackage{hyperref}

\hypersetup{colorlinks=true,linkcolor=blue,citecolor=blue,filecolor=blue,urlcolor=blue}
\begin{document}
\setcitestyle{citesep={,}}

\title{LOFAR observations of the XMM-LSS field}

\author{C. L. Hale \inst{\ref{inst1}}
\and W. Williams \inst{\ref{inst2}}
\and M. J. Jarvis \inst{\ref{inst1}, \ref{inst3}}
\and M. J. Hardcastle \inst{\ref{inst2}}
\and L. K. Morabito \inst{\ref{inst1}}
\and T. W. Shimwell \inst{\ref{inst4}}
\and C. Tasse \inst{\ref{inst5}, \ref{inst5a}}
\and P. N. Best \inst{\ref{inst6}}
\and J. J. Harwood \inst{\ref{inst2}}
\and I. Heywood \inst{\ref{inst1}}
\and I. Prandoni \inst{\ref{inst7}}
\and H. J. A. R\"ottgering \inst{\ref{inst3}}
\and J. Sabater \inst{\ref{inst6}}
\and D. J. B. Smith \inst{\ref{inst2}}
\and R. J. van Weeren \inst{\ref{inst4}}
}
\institute{
University of Oxford, Denys Wilkinson Building, Keble Road, Oxford, OX1 3RH, UK \label{inst1}
\and Centre for Astrophysics Research, School of Physics, Astronomy and Mathematics, University of Hertfordshire, College Lane, Hatfield AL10 9AB, UK \label{inst2}
\and Department for Physics, University of the Western Cape, Bellville 7535, South Africa \label{inst3}
\and Leiden Observatory, Leiden University, PO Box 9513, NL-2300 RA Leiden, the Netherlands \label{inst4}
\and GEPI and USN, Observatoire de Paris, Universit\'e PSL, CNRS, 5 Place Jules Janssen, 92190 Meudon, France \label{inst5}
\and Department of Physics and Electronics, Rhodes University, PO Box 94, Grahamstown, 6140, South Africa \label{inst5a}
\and SUPA, Institute for Astronomy, Royal Observatory, Blackford Hill, Edinburgh, EH9 3HJ, UK \label{inst6}
\and INAF-Istituto di Radioastronomia, Via P. Gobetti 101, 40129 Bologna, Italy \label{inst7}
}

\date{Received  / Accepted  }

\abstract{We present observations of the XMM Large-Scale Structure (XMM-LSS) field observed with the LOw Frequency ARray (LOFAR) at 120-168 MHz. Centred at a J2000 declination of $-4.5^{\circ}$, this is a challenging field to observe with LOFAR because of its low elevation with respect to the array. The low elevation of this field reduces the effective collecting area of the telescope, thereby reducing sensitivity. This low elevation also causes the primary beam to be elongated in the north-south direction, which can introduce side lobes in the synthesised beam in this direction. However the XMM-LSS field is a key field to study because of the wealth of {ancillary information, encompassing most of the electromagnetic spectrum}. The field was observed for a total of 12 hours from three four-hour LOFAR tracks using the Dutch array. {The final image presented encompasses $\sim 27$ deg$^2$, which is the region of the observations with a $>$50\% primary beam response}. Once combined, the observations reach a central rms of {280 $\mu$Jy beam$^{-1}$ at 144 MHz} and have an angular resolution of {$ 7.5 \times \ 8.5 \arcsec$}. {We present our catalogue of detected sources and investigate how our observations compare to previous radio observations}. This includes investigating the flux {scale} calibration of these observations compared to previous measurements, the implied spectral indices of the sources, the observed source counts and corrections to {obtain} the true source counts, and finally the clustering of the observed radio sources. }

\maketitle

\titlerunning{LOFAR observations of the XMM-LSS field}
\authorrunning{C. L. Hale et al.}

\section{Introduction}
\label{sec:introduction}
Radio observations of galaxies provide a view of the Universe that can be very different from other regions of the electromagnetic spectrum. At low frequencies ($\lesssim 5$ GHz), radio surveys typically trace the synchrotron emission from galaxies \citep[][]{Condon1992}. This can be due to the jets of active galactic nuclei (AGN) or star formation in galaxies, where particles are accelerated by either AGN or supernovae, respectively, and spiral in the magnetic field of a galaxy. {This supernova emission can subsequently} be used as a proxy for star formation in galaxies \citep[see e.g.][]{Condon1992, Bell2003}. The brightness of the emission from AGN jets and star formation means the objects can often be observed to high redshifts, {especially because dust attenuation, which may affect observations at other wavelengths, does not affect radio observations at these frequencies. }

{Our} ability to detect many star-forming galaxies (SFGs) and {AGN through their radio emission} over a large redshift range has many uses in astrophysical studies. These include tracing large-scale structure throughout the universe \citep[e.g.][]{Blake2002,Lindsay2014,Hale2018} and {measuring} the evolution of different radio galaxy populations \citep[][]{Best2012, Ineson2015, Williams2018}. With the {coming availability} of future large, deep radio continuum surveys \citep[e.g.][]{Shimwell2017, Jarvis2017, Norris2017}, radio sources will be even more widely used and important for studying galaxy evolution \citep[see][for a review]{Simpson2017,Norris2017}. 

{At $<10$ GHz, synchrotron emission dominates the emission, which is usually quantified as a power law ($S_{\nu} \propto \nu^{-\alpha}$; where $\alpha$ is the spectral index); the spectrum is} typically steep ($\alpha \gtrsim 0.5$) and thus higher flux densities are generally observed compared to higher frequency observations. {This is especially important for very steep spectrum sources ($\alpha \gtrsim 1.0$), which allow the study of the older cosmic-ray electron population and are preferentially observed at lower frequencies.}

The LOw Frequency ARray \citep[LOFAR;][]{LOFAR} is one of the radio telescopes that is revolutionising observations of the low-frequency universe. {Operating in two low-frequency regimes, 110 -- 240 MHz (using the High Band Antenna; HBA) and 10 -- 90 MHz (using the Low Band Antenna; LBA)}, this array provides a unique window into the low-frequency emission of radio galaxies. The combination of stations in LOFAR, some in a dense core (the superterp) and others distributed across the Netherlands and further afield across Europe, has allowed for high-resolution images to be obtained, while also providing sensitivity to extended structure. Low-frequency observations allow investigations of the following: SFGs and AGN \citep[][]{Hardcastle2016, CalistroRivera2017, Williams2018} in regimes in which their emission is dominated by synchrotron radiation, the detection of carbon radio recombination {lines \citep[][]{Morabito2014, Salas2017}, clusters \citep[][]{Savini2018, Wilber2018},} low-frequency observations of {pulsars \citep[][]{Bassa2017,Polzin2018}. Measurements of the epoch of reionisation (EoR) is also expected with LOFAR, by observing the red-shifted 21 cm hydrogen line \citep[][]{Yatawatta2013,Patil2017}. 

At these low frequencies, however, difficulties arise in the process of data analysis. Changes in the properties of the ionosphere (e.g. resulting from differences in the distribution of electrons) give large phase changes in the incoming radio waves, leading to, for example, variations in the apparent position of sources. These phase changes are dependent on the position within the field owing to differing variations in the ionosphere across the field of view. The ionospheric distortions and artefacts induced from the time varying beam, means that direction-dependent calibration needs to be performed to correct for these effects and obtain sensitive, high angular resolution images \citep[][]{Cotton2004,Intema2009,vanWeeren2016}. 

Previous observations with LOFAR have shown that using the Dutch stations with no direction-dependent calibration, $\sim 20-25"$ resolution high quality images are achievable \citep[e.g.][]{Shimwell2017, Mahony2016}. When direction-dependent calibration is performed, images of $\sim$5" angular resolution can be produced at {$\sim 100 \ \mu$Jy beam$^{-1}$ rms} sensitivity for a typical eight-hour observation \citep[][]{Williams2016, vanWeeren2016, Hardcastle2016}. This not only fulfils the requirements for the wide area LOFAR Tier 1 survey {\citep[see][]{Shimwell2017}}, but with deeper observations of specific fields that have a wealth of ancillary information, the flux densities that will be reached by LOFAR will allow the detection of much fainter galaxies {($\sim 20 \ \mu$Jy beam$^{-1}$ rms  in Tier 2 and $\sim 6 \ \mu$Jy beam$^{-1}$ in Tier 3)}. This will allow the detection of many more SFGs and radio quiet quasars (RQQs), which dominate the low flux density source counts \citep[see e.g.][]{Wilman2008, White2015}.
{Additionally}, with the inclusion of the international stations across Europe, LOFAR can produce sub-arcsecond resolution images of sources \citep[see e.g.][]{Varenius2015, Varenius2016, Morabito2016}.

For studies of galaxies, knowledge of their redshifts are important. Radio continuum observations on their own, however, are unable to provide constraints on the redshift of a radio source due to its {predominantly} featureless power-law synchrotron spectrum. With a wealth of information at various wavelengths, including in the radio \cite{Tasse2007}, near-IR \citep[][]{Jarvis2013}, mid-IR \citep[][]{Mauduit2012,Lonsdale2003}, far-IR \citep[][]{Oliver2012}, optical \citep[][]{Erben2013, DES2005, HSC2018}, and X-ray \citep[][]{Pierre2004, Chen2018}, the  XMM Large-Scale Structure (XMM-LSS) field has excellent multiwavelength coverage. This makes it an excellent field to study galaxies and their evolution by providing information on, for example photometric redshifts, stellar masses, and star formation rates. This field presents challenges for observations with LOFAR, however, because the low elevation of the field reduces the resolution and sensitivity of the observations. This arises from the reduced effective collecting area of the telescope and {the effect of projection on baselines}. However for future large sky surveys with LOFAR it is important to show that we are able to observe equatorial regions. A further challenge when combining radio and multiwavelength observations is that the typically poorer resolution of radio surveys compared to those in the optical/IR can make identifying the source of the radio emission challenging, and some radio AGN may appear as large objects with jets, whose core (especially at low frequencies) may not be observed. Because of the wealth of ancillary data in the XMM-LSS region, however, it should be possible to find host galaxy identifications and redshifts for the majority of sources. 

In this paper we present LOFAR observations of $\sim27$ deg$^2$ of the sky that encompasses the XMM-LSS field at $\sim$150 MHz with 12 hours of observational data. The details of the observations are discussed in Section~\ref{sec:observations} before an overview of the data reduction process is given in Section~\ref{sec:reduction}. In Section~\ref{sec:images_catalogues} we present the final reduced image of the field and the associated galaxy catalogue; comparisons to previous radio observations of this field at both $\sim$150~MHz and at other frequencies are presented in Section~\ref{sec:comparisons}. {The source counts derived from our final catalogue and the procedures we use to derive them are discussed in Section~\ref{sec:sourcecounts}. We then investigate the clustering of the radio sources in the field in Section~\ref{sec:clustering} and finally we draw conclusions in Section~\ref{sec:conclusions}.}

\section{Observations}
\label{sec:observations}

We carried out 12~hours of observations of the XMM-LSS field in three separate  four-hour tracks using the HBA in dual inner mode. In dual inner mode, each core station is split into two substations \citep[see e.g.][]{LOFAR}. Each observation covered a frequency range between 120  MHz and 168 MHz and contained $\sim230$ sub-bands. The first of these observations was taken on  22 December 2015 (L424611) and two {further} observations were made 14 months later on 27 February and 1 March 2017 (L569673 and L570753 respectively). All three observations also had an associated calibrator source that was observed for 10 minutes. For the first observation (L424611), 3C48 was used as the calibrator, whilst the {later} observations (L569673 and L570753) used 3C147. The parameters for these observations can be seen in Table \ref{tab:observations}. 

The uv coverage of these data sets, using the Dutch array, can be seen in {Figure \ref{fig:uvcoverage}, plotted} using five frequency bands centred at 121, 132, 144, 156 ,and 168 MHz. {In reality,} the broad frequency coverage fills the UV plane between these bands. All three observations have similar uv coverage, although there are gaps in some regions of the uv plane as a consequence of the lack of baselines and limited individual track time. These inevitably limit the quality of the final images. In this paper, we only use the data from the Dutch stations; we do not include the international baselines, {although data for these were recorded}. Processing of the international-baseline data will be used to obtain sub-arcsecond resolution imaging of the field (Morabito et al. in prep). 

\begin{table}
\begin{center}
\begin{tabular}{l l}

\centering
& \\ \hline \hline \\
Band & HBA (High Band Antenna) \\
& 110-190 MHz \\ \\ \hline \\
Observation IDs & L424607 (3C48) \\
& L424611 (XMM-LSS)\\
Observation Date & 22 December 2015 \\
Observation Times &4 hours - XMM-LSS \\ 
& 10 minutes - 3C48 \\
$N_{\textrm{sub-bands}}$ & 232 \\ \\ \hdashline \\
Observation IDs &  L569673 (XMM-LSS)\\
& L569679 (3C147)\\ 
Observation Date & 27 February 2017 \\
Observation Times &4 hours - XMM-LSS \\
& 10 minutes - 3C147 \\
$N_{\textrm{sub-bands}}$ & 231 \\ \\ \hdashline \\
Observation IDs &  L570753 (XMM-LSS)\\
& L570759 (3C147)\\ 
Observation Date & 1 March 2017 \\
Observation Times &4 hours - XMM-LSS \\ 
& 10 minutes - 3C147 \\
$N_{\textrm{sub-bands}}$ & 231 \\ \\ \hdashline \\
Positions & 02:20:00.00 -04:30:00.0 (XMM-LSS) \\
& 01:37:41.30 +33:09:35.1 (3C48) \\
& 05:42:36.155 +49 51 07.28 (3C147) \\ \\ \hline \hline
\end{tabular}
\caption{Observation parameters for the target field (XMM-LSS) and the associated calibrator sources for the three observations{, including} information about the observation period, times, dates, and number of sub-bands and antennas.}
\label{tab:observations}
\end{center}
\end{table}

\begin{figure}
\begin{center}
\centering
\begin{minipage}[b]{0.5\textwidth}
\centering
\includegraphics[width=7cm]{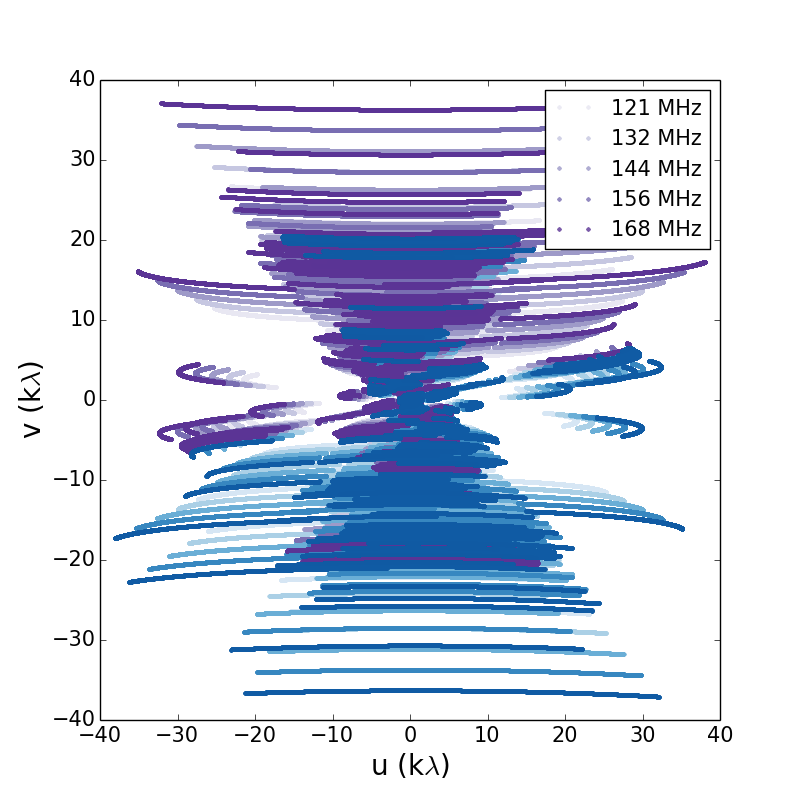}
\subcaption{L424611}
\end{minipage}% 
\newline
\begin{minipage}[b]{0.5\textwidth}
\centering
\includegraphics[width=7cm]{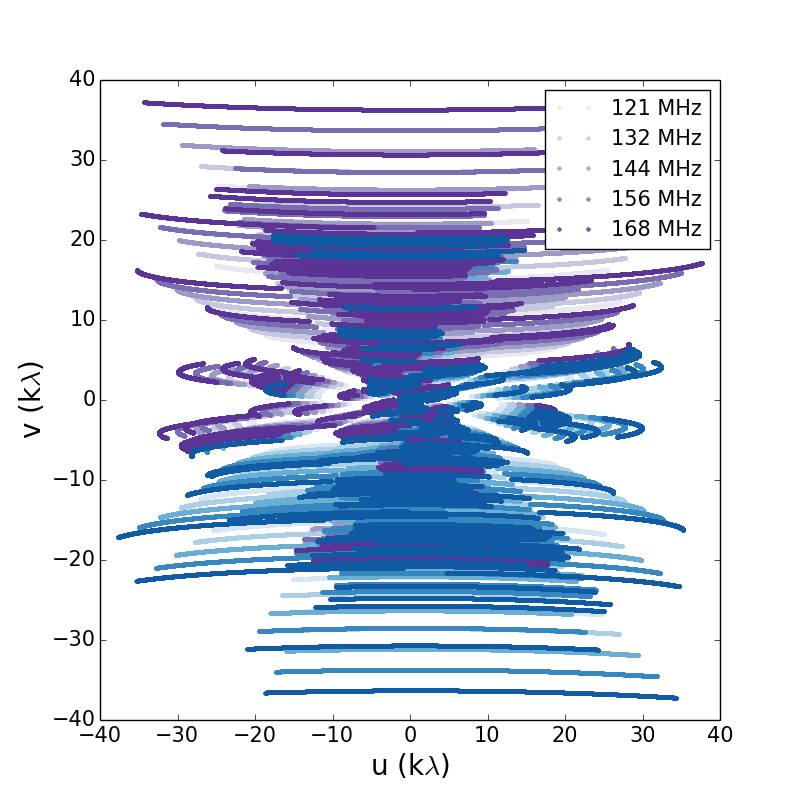}
\subcaption{L569673}
\end{minipage}% 
\newline
\begin{minipage}[b]{0.5\textwidth}
\centering
\includegraphics[width=7cm]{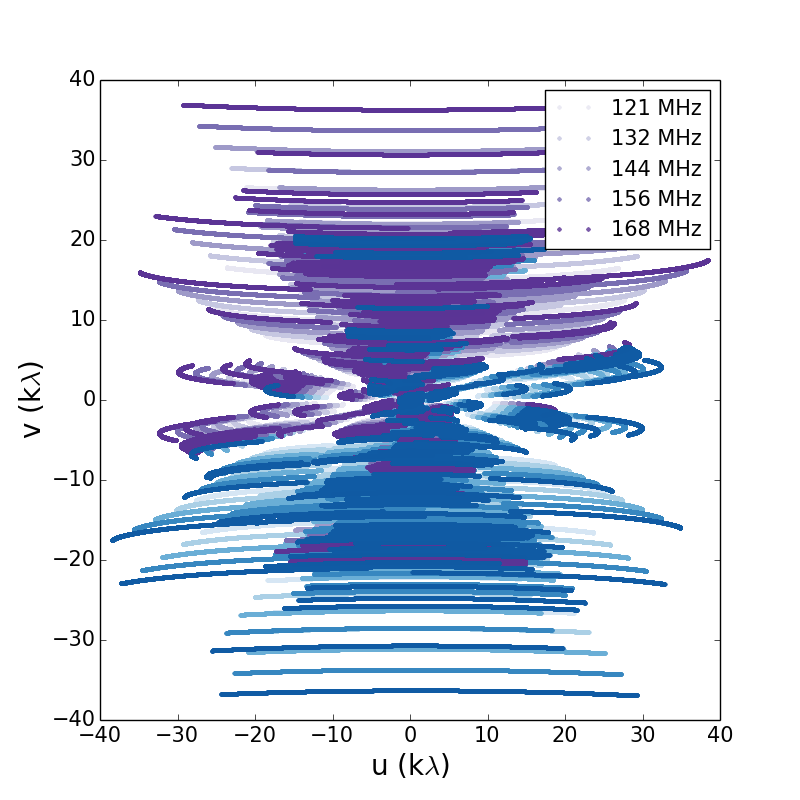}
\subcaption{L570753}
\end{minipage}%
\end{center}
\caption{uv coverage of the data in k$\lambda$ for the three observations. Figure (a) is for observation L424611, (b) for L569673, and (c) for L570753. The shades of purple represent five different frequency bands corresponding to those at 121 MHz (lightest), 132 MHz, 144 MHz, 156 MHz, and 168 MHz (darkest). The conjugate points are plotted in blue.}
\label{fig:uvcoverage}
\end{figure}

\begin{figure}
\begin{center}
\centering
\includegraphics[width=6cm]{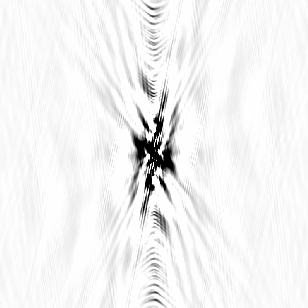}
\end{center}
\caption{ {$20 \times 20 \arcmin$} image of the direction-independent PSF. This highlights the north-south artefacts that are likely to be {visible} around bright sources in the field. These arise because of to the low elevation of the XMM-LSS field with respect to the LOFAR telescope.}
\label{fig:psf}
\end{figure}

\section{Data reduction}
\label{sec:reduction}
The data were first processed using \textsc{Prefactor}\footnote{\url{https://github.com/lofar-astron/prefactor}} \citep[see][]{deGasperin2018}. This is the part of a pipeline used by LOFAR that uses the observations of the calibrator source to calculate the response of the antennas in each of the different LOFAR stations. {This calculates amplitude and phase solutions for each sub-band of the calibrator observations, which are then used for initial calibration of the target observations}. These observations were processed on \textsc{SURFsara}\footnote{\url{https://www.surf.nl/en/about-surf/subsidiaries/surfsara/}} and averaged to 8 s integration time and two channels/sub-band \citep[see][]{Mechev2017}. The sub-bands were also grouped together into 25 bands. Each band typically contained 10 sub-bands, however at the edge of the frequency bands fewer sub-bands were grouped on account of the larger frequency bandwidths. Direction-independent calibration {was also performed before direction}-dependent calibration was carried out. 

Many of the LOFAR studies that include direction-dependent calibration have used the method of facet calibration \citep[][]{vanWeeren2016}, which splits the sky into smaller regions (facets) around bright sources and calibrates each facet separately. The brightest source (or sources) in these facets are used as calibrators for the facets, and four rounds of self-calibration (two phase only and two amplitude and phase) are applied. The solutions from this calibrator source are then applied to the whole facet. The facet calibration method has also been combined into a single pipeline, \textsc{Factor}\footnote{\url{https://github.com/lofar-astron/factor}}.

Because of the increased speed, reduced user input, and ability to produce more focussed sources, in this paper we used two packages, \textsc{DDFacet}\footnote{\url{https://github.com/saopicc/DDFacet}} \citep[][]{Tasse2018} and \textsc{KillMS} \citep[hereafter \textsc{kMS};][]{Tasse2014b,Tasse2014a,Smirnov2015}, to perform direction-dependent calibration and imaging. The package \textsc{kMS} is the calibration gain solver, whereas \textsc{DDFacet} is the corresponding imager.  \cite{Tasse2014a, Tasse2014b, Smirnov2015} and \cite{Tasse2018} provide detailed descriptions of the packages and we present only an overview in this work. These packages have been combined together into the DDF pipeline \citep[DDF; see][]{Shimwell2018} \footnote{\url{https://github.com/mhardcastle/ddf-pipeline}}.

By combining \textsc{DDFacet} and \textsc{kMS} into the \textsc{DDF} pipeline, this allows for the entire LOFAR field to be imaged at once but allows for different calibration solutions in different directions (facets). By splitting the field into facets, the radio interferometry measurement equation (RIME) can be solved for smaller areas within which directional effects are approximately the same. The Jones matrices (gains) are solved using \textsc{kMS}, where directions are solved simultaneously and the interactions between the different directions are taken into account. In \textsc{DDFacet} the direction-dependent gain solutions are applied to the visibilities during imaging. These facets do not have strict boundaries and instead taper towards the edges. This means that the facets have continuity in flux and rms at the facet boundaries. This is not the case in \textsc{Factor}. Finally, {\textsc{DDFacet}} is able to take into account point spread function (PSF) variations across the field. This is important as we are performing the observations at low elevation to the LOFAR stations and so there may be PSF smearing across the field. However, unlike \textsc{Factor}, DDF imaging parameters cannot be tailored on a facet by facet basis, which may present challenges if there are complex sources in regions of the field.

The DDF pipeline \citep[][]{Shimwell2018} takes \textsc{DDFacet} and \textsc{kMS} and {combines} them into a pipeline capable of {reducing LOFAR} data using cycles of calibration and imaging including phase-only calibrations as well as amplitude and phase calibrations. A bootstrapping step is also applied in which the fluxes of sources were bootstrapped \citep[see e.g.][]{Hardcastle2016, Shimwell2018} to those at 325~MHz and 74~MHz from the Very Large Array \citep[VLA;][]{Cohen2003, Tasse2006}. The TIFR GMRT Sky Survey Alternate Data Release (TGSS-ADR) 7$\sigma$ Catalogue \citep[][]{Intema2017} was also used within the DDF pipeline to provide an initial \textsc{clean} mask that encompasses known sources. Once the final image has been generated by DDF, a primary beam correction is also applied.

Running DDF requires little human intervention, which allows it to be easily run for many observations and, as well as this, DDF allows multiple observations to be combined. Therefore all three observations presented in this paper were processed simultaneously to produce a single deep image. {For the reduction we used Briggs weighting with a final robustness of -0.5.} \\ \\
\begin{figure*}
\begin{center}
\centering
\includegraphics[height=20cm]{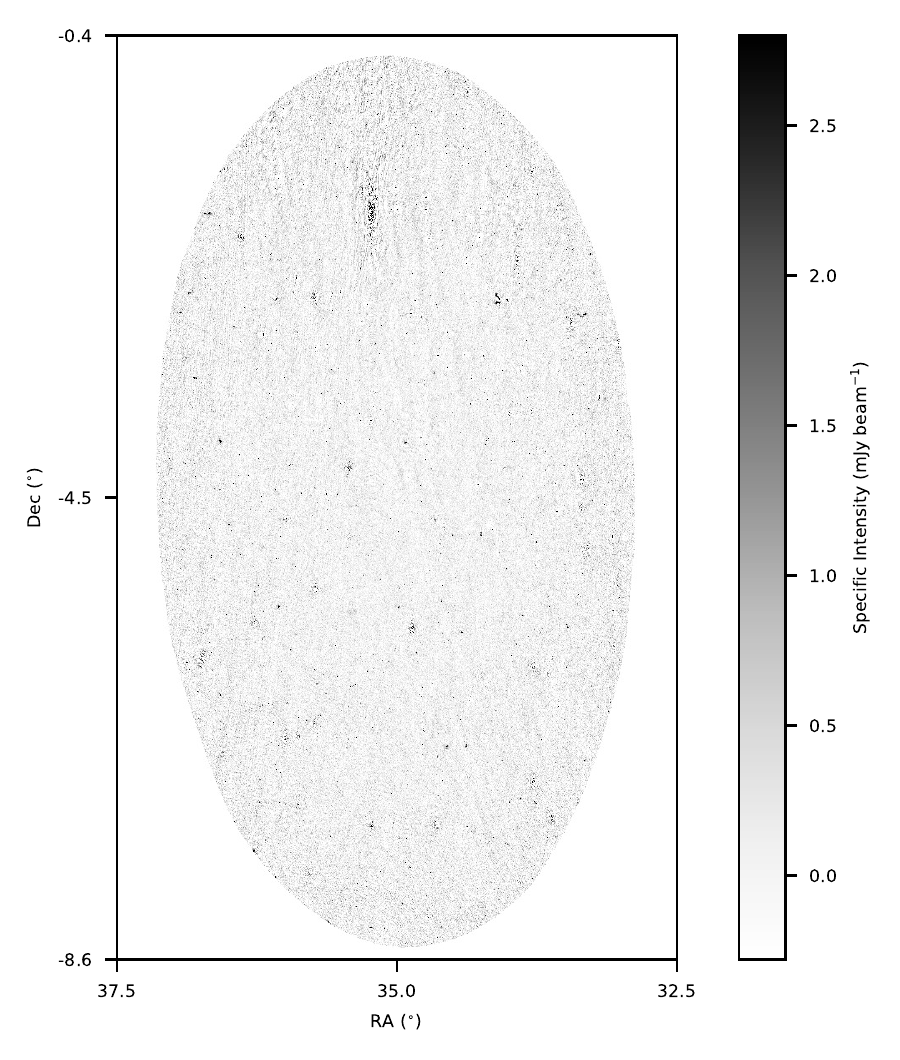}
\end{center}
\caption{{Image of the XMM-LSS} field observations imaged with LOFAR at $\sim$150MHz covering $\sim$27 deg$^2$. The associated flux scale is shown on the right-hand side in mJy beam$^{-1}$. The flux scale was chosen to range between -$\sigma_{\textrm{cen}}$ and 10$\sigma_{\textrm{cen}}$, where $\sigma_{\textrm{cen}}$ is the central rms of the image, {which was determined to be} 0.28 mJy beam$^{-1}$. }
\label{fig:fullfield}
\end{figure*}

\begin{figure*}
\begin{center}
\centering
\includegraphics[height=13.5cm]{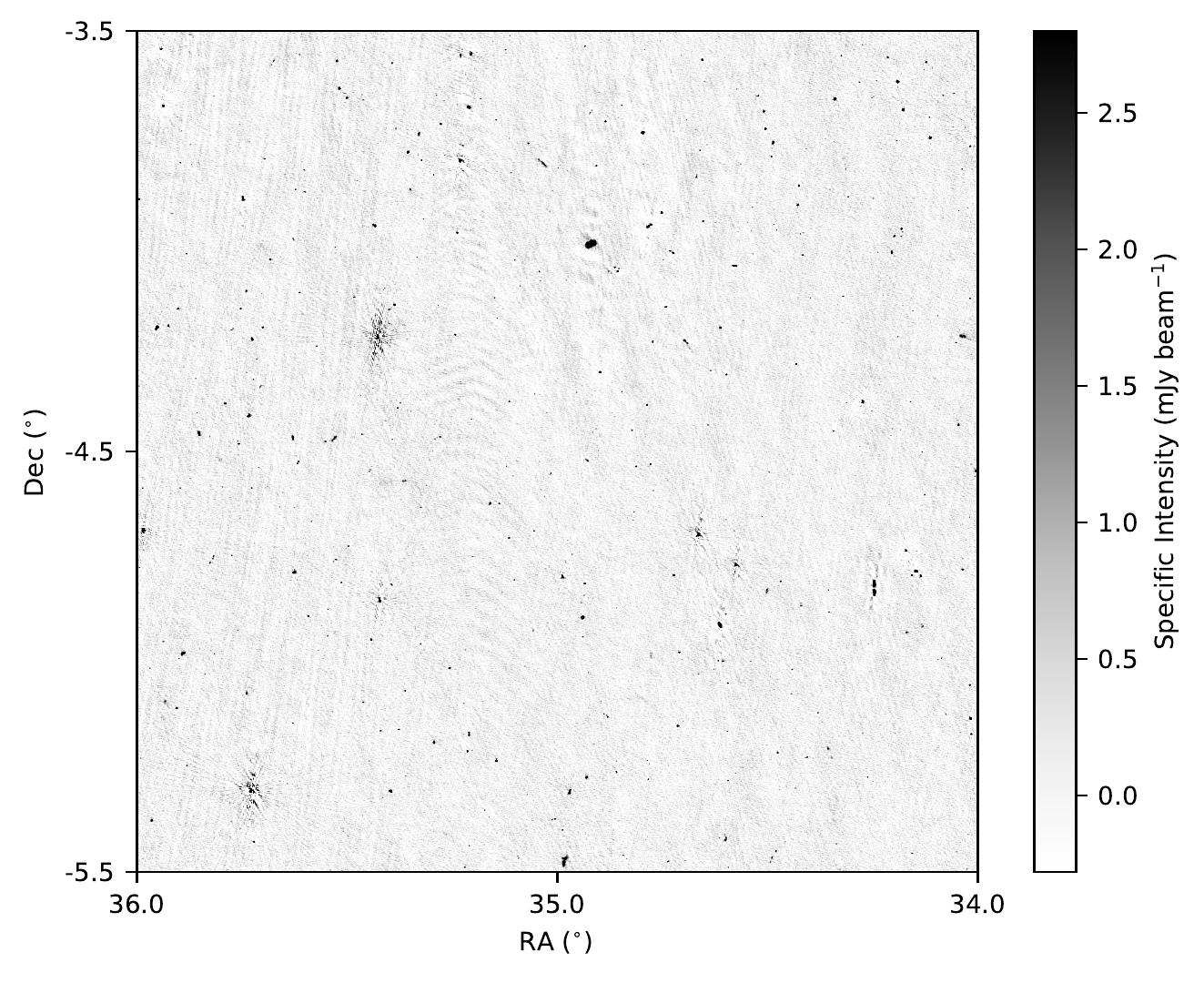}
\end{center}
\caption{Image of the central 4 deg$^2$ of the field, containing $\sim$ 600 sources. This has the same flux scale as Figure \ref{fig:fullfield}, ranging between -$\sigma_{\textrm{cen}}$ and 10$\sigma_{\textrm{cen}}$, where $\sigma_{\textrm{cen}}$ is the central rms of the image, which is 0.28 mJy beam$^{-1}$. This zoomed-in image allows the presence of multi-component sources in this field to be seen.}
\label{fig:fieldcentre}
\end{figure*}
\section{Final images and catalogues}
\label{sec:images_catalogues}

The final image was generated {over} a reduced region of the primary beam corrected image. This was {done} to remove areas with high primary beam corrections, where the telescope is less sensitive. Regions with a primary beam correction of 2 or less (i.e. the primary beam response $\geq$ 50\% or more of the response in the centre of the image) were used in the final image. This produced a $\sim$27 deg$^2$ elliptical image of the sky around the XMM-LSS field at 7.5$\times$8.5" resolution. The ellipse is aligned in the north-south direction. This {image} has a central frequency of $\nu = 144$~MHz and a central rms of 0.28 mJy beam$^{-1}$. Previous observations with LOFAR at higher elevations have found rms values  {of $\sim 0.1$ mJy beam$^{-1}$} for a typical eight-hour observation. Taking into account differences in integration times as well as declination, assuming a $\cos(\delta_{best}-\delta)^2$ sensitivity relation centred on the optimal declination of LOFAR $\sim 53^{\circ}$,  a central rms of $\sim$0.3 mJy beam$^{-1}$ is similar to the expected value. This is therefore promising for future LOFAR Two-metre Sky Survey \citep[LoTSS;][]{Shimwell2018} equatorial observations.

As the field is close to equatorial, most uv tracks are in the east-west direction, which also leads to elongation of the synthesised beam in the north-south direction. The final image of the full field can be seen in Figure~\ref{fig:fullfield} and the central 4~deg$^2$ {is shown} in Figure~\ref{fig:fieldcentre}, which shows the multi-component nature of some radio sources. 

We {extracted sources} and their parameters using \textsc{PyBDSF} \citep[Python Blob Detector and Source Finder;][]{PyBDSF}\footnote{\url{http://www.astron.nl/citt/pybdsf/}}. This package not only allows sources to be detected and output to a catalogue file with parameters relating to their fluxes, sizes, and positions but also generates associated rms maps and residual maps. The rms map is an image of the noise across the field, whilst the residual map is an image of the field with all modelled sources removed. The apparent image (pre-primary beam correction) was {used as the detection image, as this has more uniform noise}, but source parameters are calculated from the true sky image. The PyBDSF parameters chosen for source extraction, as {used in} \cite{Williams2016}, are shown in Table \ref{tab:PyBDSF}.

\begin{table}
\begin{center}
\centering
\begin{tabular}{l l}
\hline \hline
 thresh\_isl=3.0 & thresh\_pix=5.0 \\
 rms\_box=(160,50) & rms\_map=True \\
 mean\_map=`zero' & ini\_method=`intensity' \\
 adaptive\_rms\_box=True & adaptive\_thresh=150 \\
 rms\_box\_bright=(60,15) & group\_by\_isl=False \\
 group\_tol=10.0 & atrous\_do=True \\
 flagging\_opts=True & flag\_maxsize\_fwhm=0.5 \\
 advanced\_opts=True & blank\_limit=None \\
 atrous\_jmax=3 \\ \hline \hline
 \end{tabular}
 \caption{Specified input properties of \textsc{PyBDSF} that were {used to} generate the source catalogue and associated rms and residual images. For more information on what the different inputs mean see \url{http://www.astron.nl/citt/pybdsf/}.}
 \label{tab:PyBDSF}
 \end{center}
\end{table}

The result of the source detection was a final catalogue of 3,169 radio sources. This includes some sources that consist of multiple components whose Gaussian components overlapped and were grouped together by \textsc{PyBDSF}. The number of sources in the \textsc{PyBDSF} output catalogue and the central and average rms properties of the image are given in Table \ref{tab:rmsproperties}. {The rms is known} to vary across the map because of, for example, reduced sensitivity away from the pointing centre, residual noise around sources, and leftover artefacts. The variation in the rms coverage across the field is shown in Figure \ref{fig:rmscoverage} where the left panel shows the rms image of the field as output from \textsc{PyBDSF}, in which the higher noise regions {are} apparent. In the top of the field in particular there is a large region of high noise around 3C63. This was the brightest and most challenging source to image in the field and has the most artefacts surrounding it (as can be seen in Figure \ref{fig:fullfield}). The right panel of Figure \ref{fig:rmscoverage} shows the area, and the corresponding percentage, of the image that has an rms value less than a given value. Roughly 85\% of the area of the image has an rms of $<0.5$ mJy beam$^{-1}$.

\pagebreak
\begin{table*}\ContinuedFloat
\caption{First 10 sources and one later source from the final catalogue. This was generated from the \textsc{PyBDSF} catalogue but where multiple component sources have been combined together and artefacts have been removed. The columns include the IAU source ID, position, flux densities, sizes (deconvolved sizes are not shown), and information on whether the source had been matched or flagged. All fluxes have been corrected to TGSS-ADR as described in Section \ref{sec:fluxes}.}
\label{tab:catalogue}
\begin{tabular}{ccccccccc}
\hline \\
IAU\_source\_ID &Source\_ID & Pre\_matched & RA & E\_RA & DEC & E\_DEC & Total\_flux & E\_Total\_flux  \\
 && \_source\_ID &  &   &  & &  &    \\
&&& ($^{\circ}$) &($^{\circ}$)& ($^{\circ}$) & ($^{\circ}$) & (mJy) & (mJy)  \\ \hline \hline
J022831.74-044607.2 & 0 & 0 & 37.13225 & 0.00003 & -4.76867 & 0.00004 & 48.38 & 1.05 \\
J022831.48-041625.0 & 1 & 1 & 37.13118 & 0.00008 & -4.27361 & 0.00006 & 7.42 & 0.89 \\
J022827.74-042647.2 & 2 & 2 & 37.11558 & 0.00024 & -4.44643 & 0.00021 & 5.34 & 0.80 \\
J022826.77-041752.8 & 3 & 3 & 37.11155 & 0.00005 & -4.29800 & 0.00004 & 13.20 & 0.89 \\
J022826.44-050442.1 & 4 & 4 & 37.11017 & 0.00025 & -5.07836 & 0.00021 & 8.66 & 0.79 \\
J022825.23-032639.9 & 5 & 5 & 37.10513 & 0.00008 & -3.44442 & 0.00006 & 10.56 & 0.81 \\
J022825.36-043441.3 & 6 & 6 & 37.10567 & 0.00004 & -4.57814 & 0.00004 & 22.05 & 0.83 \\
J022824.21-042709.9 & 7 & 7 & 37.10086 & 0.00007 & -4.45275 & 0.00005 & 30.66 & 0.74\\
J022823.25-043022.8 & 8 & 8 & 37.09689 & 0.00005 & -4.50634 & 0.00003 & 103.46 & 1.15 \\
J022821.60-032343.5 & 9 & 9 & 37.08998 & 0.00004 & -3.39541 & 0.00005 & 20.96 & 0.81\\ 
...& ... & ... & ... & ... & ... & ... & ... & ...  \\
J022814.35-050242.1 & 17 & 17 & 37.05885 & 1.10995 & -5.04834 & -0.15114 & 85.58 & 1.33 \\ \hline \\ \hdashline \hdashline \\ 

\end{tabular}
\end{table*}
\begin{table*}\ContinuedFloat
\begin{tabular}{ccccccccccc}
\hline \\
IAU\_source\_ID & ... & Peak\_flux & E\_peak\_flux &  Maj & E\_Maj & Min & E\_Min & PA & E\_PA & ... \\
&&(mJy/beam)&(mJy/beam) & (") & (") & (") & (")&   ($^{\circ}$) & ($^{\circ}$)&  \\ \hline \hline 
J022831.74-044607.2 & ... & 22.87 & 0.49 & 13.12 & 0.37 & 8.39& 0.17& 123.56& 2.77& ... \\
J022831.48-041625.0 & ... & 7.09 & 0.52 & 9.01 & 0.73 & 7.42& 0.50& 77.26& 16.65&  ... \\
J022827.74-042647.2 & ... & 3.08 & 0.55 & 10.95 & 2.01 & 10.09& 1.74& 90.29& 93.35&  ... \\
J022826.77-041752.8 & ... & 11.92 & 0.53 & 9.00 & 0.43 & 7.85& 0.33& 114.95& 14.14&  ... \\
J022826.44-050442.1 & ... & 3.64 & 0.58 & 13.23 & 2.20 & 11.48& 1.75& 109.38& 49.89& ... \\
J022825.23-032639.9 & ... & 7.96 & 0.51 & 10.10 & 0.70 & 8.38& 0.49& 111.96& 15.21&  ... \\
J022825.36-043441.3 & ... & 15.55 & 0.53 & 9.97 & 0.35 & 9.07& 0.30& 85.41& 15.32&  ... \\
J022824.21-042709.9 & ... & 13.04 & 0.54 & 14.87 & 0.69 & 10.09& 0.37& 117.80& 5.36& ... \\
J022823.25-043022.8 & ... & 37.52 & 0.52 & 20.21 & 0.46 & 10.23& 0.17& 158.41& 1.80& ... \\ 
J022821.60-032343.5 & ... & 13.83 & 0.53 & 10.63 & 0.43 & 9.10& 0.33& 9.01& 10.78&  ... \\
... & ... & ... & ... & ... & ... & ... & ... & ... & ... & ... \\
J022814.35-050242.1 & ... & 21.29 & 0.49 & nan & nan & nan& nan& nan& nan&  ... \\ \hline \\ \hdashline  \hdashline \\
\end{tabular}
\end{table*}
\begin{table*}\ContinuedFloat
\begin{tabular}{cccccccccccc}
\hline \\

IAU\_source\_ID & ... & Composite & N & Matched\  & ... & Bright & Edge & rms & RA & DEC\\
&&\_Size& \_sources& \_ID\_1& &&&  \_central&   \_FIRST& \_FIRST   \\
&&(") & & && &&  (mJy/beam) &  ($^{\circ}$) & ($^{\circ}$)  \\ \hline \hline 
J022831.74-044607.2 & ...& nan & 1  & 0  &...& 1 & 0 & 0.49 & 37.13234 & -4.76854 \\
J022831.48-041625.0 & ... & nan & 1 & 0&... & 0 & 0 & 0.52 & 37.13127 & -4.27348 \\
J022827.74-042647.2 & ... & nan & 1 & 0 &...& 0 & 0 & 0.51 & 37.11568 & -4.44630 \\
J022826.77-041752.8 & ... & nan & 1 & 0 &...& 0 & 0 & 0.52 & 37.11164 & -4.29787 \\
J022826.44-050442.1 & ... & nan & 1 & 0 &...& 1 & 0 & 0.54 & 37.11026 & -5.07823 \\
J022825.23-032639.9 & ... & nan & 1 & 0 &...& 1 & 0 & 0.48 & 37.10522 & -3.44429 \\
J022825.36-043441.3 & ... & nan & 1 & 0  &...& 0 & 0 & 0.50 & 37.10576 & -4.57801 \\
J022824.21-042709.9 & ... & nan & 1 & 0  &...& 0 & 0 & 0.51 & 37.10095 & -4.45262 \\
J022823.25-043022.8 & ... & nan & 1& 0  &...& 0 & 0 & 0.52 & 37.09698 & -4.50621 \\
J022821.60-032343.5 & ... & nan & 1& 0  &... & 0 & 0 & 0.49 & 37.09008 & -3.39528 \\
... & ... & ... & ... & ... & ... & ... & ... & ... & ... & ... \\
J022814.35-050242.1 &  ... & 36.59 & 2  & 20 & 0 & 0 & 0 &0.53 & 37.05894 & -5.04821 \\ \hline
\end{tabular}
\end{table*}
\FloatBarrier
\pagebreak

 \begin{table}
\begin{center}
\begin{tabular}{l l l}
\centering
& \\ \hline \hline
Sources (\textsc{PyBDSF} catalogue) & : & 3,169 \\
Sources (post inspection) & : & 3,044 \\
Median rms ($\mu$Jy beam$^{-1}$) & :& 394 \\ 
Central rms ($\mu$Jy beam$^{-1}$) & :& 277 \\
Area (deg$^2$) & : &26.9 \\ 
Central frequency (MHz)& : & 144 \\ \hline \hline
\end{tabular}
\caption{Properties of the final image of the XMM-LSS field and its associated rms map as generated by \textsc{PyBDSF}.}
\label{tab:rmsproperties}
\end{center}
\end{table}

Each source was visually inspected to ensure that the object appeared to be a real source and not an artefact and to check whether multiple components of the same source (that {were not associated} {by} \textsc{PyBDSF}) could be combined together into a single source. {For $\sim 50$ sources, there existed two \textsc{PyBDSF} sources that were approximately co-located; one of these was a much larger ellipse that did not seem to contain any sources other than the co-located source. As a check of these sources, they were cross-matched with TGSS-ADR using a 5" search radius. The ratio of the LOFAR flux with and without the larger source was compared to the flux from TGSS-ADR and compared to the ratio that was observed for the full population}. This suggested that these large ellipses should be {combined with the central source to form one object}. In regions for which recent VLA observations over the XMM-LSS field \citep[][]{HeywoodVLA} were available, these were used to help with the eyeballing process. However, these covered only $\sim 7$ deg$^2$ of the field; see Figure \ref{fig:crossmatch_coverage}. 
{The sources that needed to be amalgamated as determined by this visual inspection were combined together.} We followed \cite{Williams2016} to assign properties to the combined sources (e.g. their position and {flux densities}) and used the following methods:

\begin{itemize}
\item Positions: {Flux density} weighted averaged RA and Dec (and errors) \\
\item {Total flux density}: The sum of the individual {flux densities} of the components that need combining ({and errors summed in quadrature}) \\
\item {Peak flux density per beam}: {The} maximum {peak flux density per beam} from the components that need combining (max($S_{{\textrm{peak}}, i}$) {and its associated error}) 
\end{itemize}

Source IDs were reassigned for the newly combined catalogue and given an ID based on the IAU naming convention. The concatenated output table however records the pre-matched Source ID output from the \textsc{PyBDSF} catalogue and the source ID of any sources that were merged together to form that source. For those sources that had not been combined with other sources we included information from PyBDSF on their measured and deconvolved sizes (major/minor axis and position angle). As these are not easy to describe for amalgamated sources, they were not included for these sources. Instead a column (`Composite\_size') was included for the sources that had been combined together. This was the largest size between {the components for sources} that were combined together \citep[in arcsec; as in][]{Hardcastle2016}\footnote{{We note that for the sources described earlier, which were combined with a large elliptical component, this size may underestimate the true source size.}}.

To highlight whether a source had been combined together with other components, an additional column was added labelled `N\_sources'. This column lists a value of 1 if it was a single source, or a value $>$1 if multiple components were combined together from the \textsc{PyBDSF} catalogue. In the cases where sources had been combined together, we  recorded which sources were combined together. These were recorded in four columns (`Matched\_ID\_X'; where X is an integer [1,4]) to list any of the source IDs from the original PyBDSF catalogue. For radio sources for which no PyBDSF sources were combined together (i.e. `N\_sources'=1) then all Matched\_ID\_X columns were assigned the value 0 to indicate this. For radio sources for which multiple PyBDSF sources were combined together (i.e. `N\_sources'>1), the `Matched\_ID' columns indicate the source IDs from the original \textsc{PyBDSF} catalogue that were combined together. If the number of sources combined together were less than four any unnecessary Matched\_ID\_X columns were designated 0. {We also made note of any bright sources that had artefacts around using the column `Bright' (and a value of 1) and those near or cut off by the edge of the field with a column `Edge' (using a value of 1). }

Combining multiple components together into single radio sources led to {3,092} individual sources. Once those sources that visually appeared to be {artefacts were removed} this led to {3,044} sources in the final catalogue. An example of the first ten lines from this catalogue can be found in Table \ref{tab:catalogue}. We also include in this a later line from the table as an example of where sources had been combined together; this is to highlight the differences in the catalogue for these sources. The table also includes a value of the rms noise at the central (RA, Dec) of the source and a correction to the positions to correct these to {FIRST \citep[][]{Becker1995,Helfand2015}}, which is discussed in Section \ref{sec:positions}. Example cut-outs of the largest sources from this final catalogue can be seen in Appendix \ref{sec:large_sources}. The final catalogue is published on-line.

\begin{figure*}
\begin{center}
\begin{minipage}[b]{0.55\textwidth}
\includegraphics[width=9.4cm]{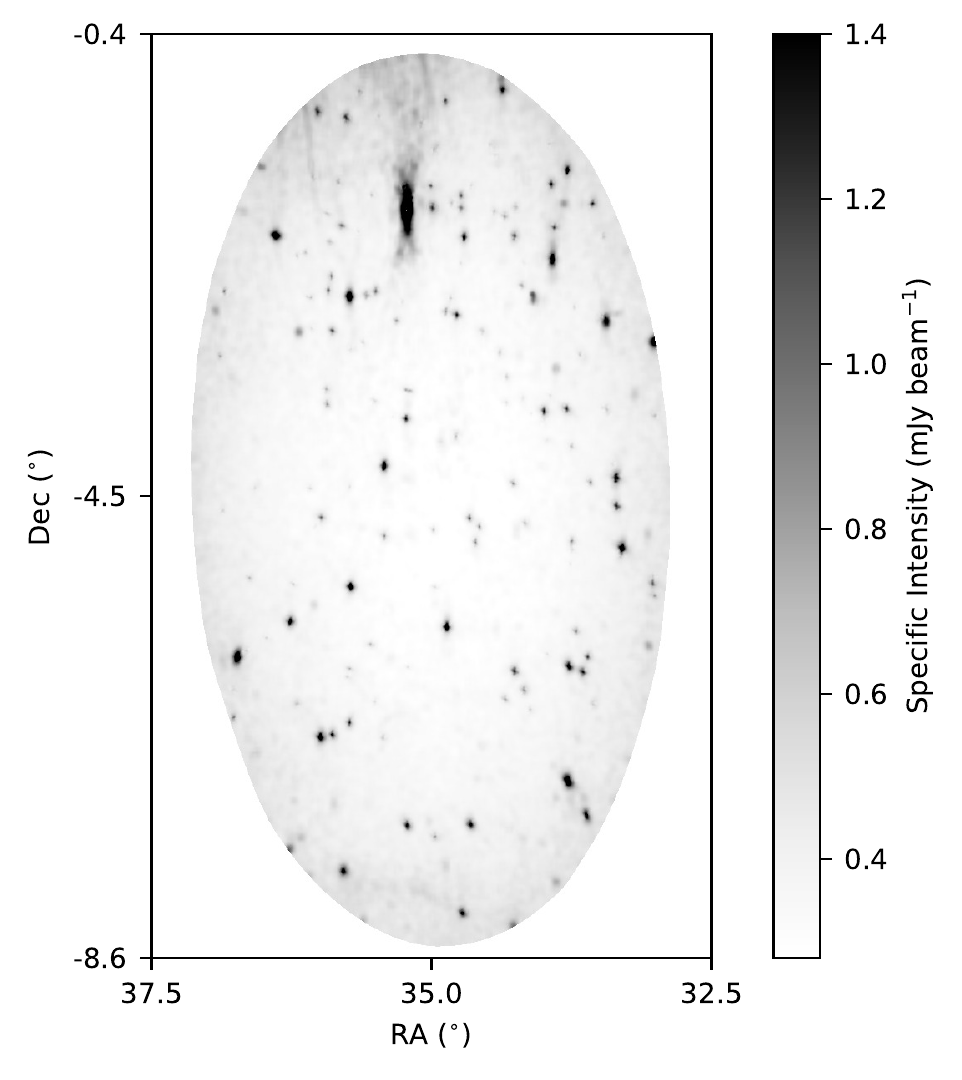}
\subcaption{}
\end{minipage}% 
\begin{minipage}[b]{0.45\textwidth}
\includegraphics[width=8cm]{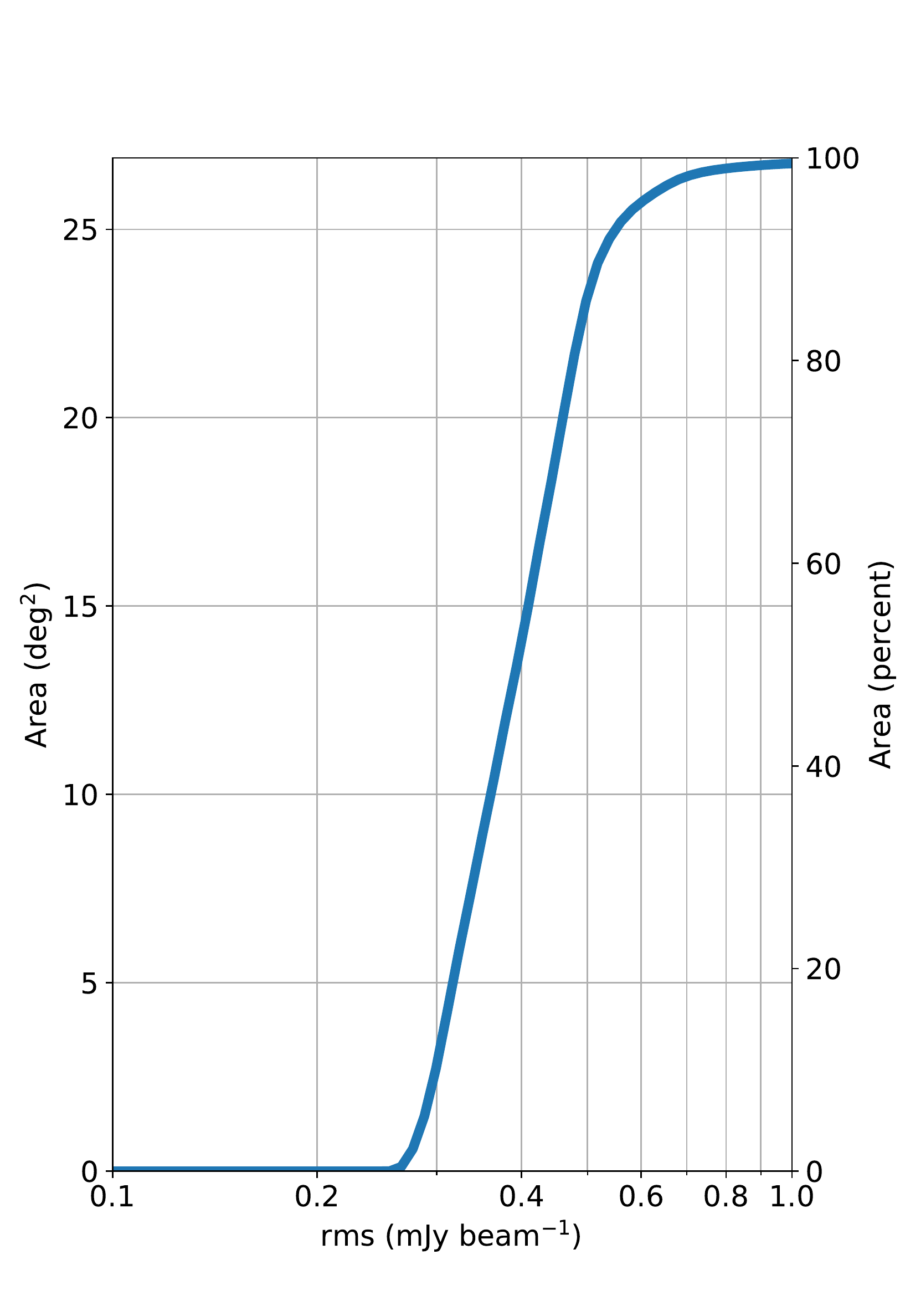}
\subcaption{}
\end{minipage}% 
\end{center}
\caption{rms coverage of the LOFAR XMM-LSS images. Panel (a) shows an image of the rms coverage across the field in which the flux scale varies between $\sigma_{\textrm{cen}}$ and 5$\sigma_{\textrm{cen}}$, where $\sigma_{\textrm{cen}}$ is the central rms of the image that was found to be 0.28 mJy beam$^{-1}$. Panel (b) shows a plot of the area of the field that has an rms less than the given value. On the left-hand y-axis this is as a function of area in deg$^2$ and on the right-hand y-axis it is given as a percentage of the total area. The noise can be relatively high around {bright} sources in the field.}
\label{fig:rmscoverage}
\end{figure*}

\section{Comparisons to other radio observations}
\label{sec:comparisons}

In this section we compare our catalogue to previous radio catalogues. Five other radio catalogues were used for these comparisons: {FIRST \citep[1.4 GHz;][]{Becker1995, Helfand2015}}; {GMRT observations \citep[240 and 610 MHz;][]{Tasse2007}}; TGSS-ADR \citep[150 MHz;][]{Intema2017}; {GLEAM \citep[$\sim$ 70-230 MHz;][we use the $\sim$ 150 MHz observations for this work]{HurleyWalker2017},} and finally recent VLA observations over the XMM-LSS field \citep[1.5 GHz;][]{HeywoodVLA}. Whereas FIRST and TGSS-ADR are large sky surveys, the observations in \cite{HeywoodVLA} and \cite{Tasse2007} were targeted observations of the XMM-LSS field, covering smaller areas than the LOFAR observations presented in this work. \cite{HeywoodVLA} covered $\sim 7$ deg$^2$ \citep[specifically targeting the VIDEO field;][]{Jarvis2013}, whereas \cite{Tasse2007} observed areas of $\sim$18 and $\sim$13 deg$^2$ at 240~MHz and 610~MHz, respectively. The LOFAR observations presented in this paper are slightly off centre compared to these previous observations, centred at a slightly lower right ascension. The locations of these other surveys compared to the LOFAR observations presented in this paper are shown in Figure \ref{fig:crossmatch_coverage}. 

The multi-frequency observations allow us to learn about whether there are systematic flux and positional offsets. Comparing positional information with higher frequency data is especially useful as ionospheric distortions are significantly less problematic at higher frequencies. All the previous radio catalogues were cross-matched with our catalogue with a 5" positional search radius. {Because of differences in flux density limits} of the different observations and the intrinsic spectral indices of sources, not all sources within the same imaged regions have counterparts in the different surveys. Therefore we provide information on the different flux limits of the surveys and the corresponding 150~MHz flux limit, assuming a spectral index of $\alpha = 0.7$, in Table \ref{tab:matches}. 

\begin{table*}
\begin{center}
\begin{tabular}{c c c c c c c c}
\centering
Catalogue& Citation &Frequency   & Area overlap & rms depth & Scaled 150 MHz & Resolution \\ 
&&& & & rms depth  & \\ 
&&(MHz)& {(deg$^2$)} & {(mJy beam$^{-1}$)} &  {(mJy beam$^{-1}$)} & (") \\ \hline \hline
LOFAR & This work & 144 &   26.9 & \  0.28 (central) & \  0.28 (central) & 7.5 $\times$ 8.5 \\
 &  &  &  & $\sim$ 0.4 (median) &$\sim$ 0.4 (median) & \\  \hdashline
TGSS-ADR & \cite{Intema2017}& 150 &  26.9 & 3.5 & 3.5 & 25\\
{GLEAM} & {\cite{HurleyWalker2017}} & {$\sim$70--230} & {26.9} & {6--10} & {$\sim 8$} & {120} \\
GMRT & \cite{Tasse2007} &240 &   $\sim$14.4 & 2.5 &  3.5 & 14.7 \\ 
GMRT & \cite{Tasse2007} &610 &  $\sim$10.8 & 0.3 & 0.8 & 6.5 \\
FIRST & {\cite{Helfand2015}} &1400 &     26.9 & 0.15 &  0.7 & 5 \\
VLA & \cite{HeywoodVLA} &1500 & $\sim$6.8 & 0.02 &  0.1 & 4.5 \\
\end{tabular}
\caption{{Information on the other radio catalogues that our catalogue is compared to}. We list the  name of the catalogue  (as used to describe in this work), frequency of the catalogue, overlap area, and rms depth. The depth is given at the frequency of the survey and scaled to 150~MHz using $\alpha=0.7$. }
\label{tab:matches}
\end{center}
\end{table*}

\subsection{Flux comparison}
\label{sec:fluxes}
{First}, we {compared} our observed fluxes to those previously measured at 150~MHz. Although our observations are centred at 144~MHz, this is a minimal correction to a flux at 150~MHz; however, we  corrected the {TGSS-ADR \citep[][]{Intema2017}} flux density to 144 MHz assuming $\alpha = 0.7$. TGSS-ADR is a 3.6$\pi$ steradian survey of the complete northern sky north of Dec {$\sim -53^{\circ}$}. This survey therefore covers the entirety of the field observed in this work with LOFAR. It has 25" resolution with an rms of 5~mJy\,beam$^{-1}$.

As TGSS-ADR has a poorer angular resolution than our observations, we made further cuts on the sources to be used in our comparison. This was to ensure that we compared fluxes like-for-like between surveys. Following the criteria of \cite{Williams2016}, we compared only isolated, small, and high signal-to-noise sources. We used the same criteria as \cite{Williams2016} for these, defining isolated sources as those that do not have a neighbouring LOFAR source within 25" {(the TGSS-ADR beam size)}. This ensures that multiple sources that may have been blended together in TGSS-ADR are not included. We also only considered sources that are small, $< 50$" in angular size (or {double the TGSS-ADR beam size}); this prevented differences in fluxes arising from extended emission that may be detected with LOFAR but not with TGSS-ADR (or detected in TGSS-ADR but not in LOFAR) as a consequence of differing baselines. Finally we used only the sources in the LOFAR catalogue presented, and the comparison catalogue where possible, with high signal to noise ($S_{peak}/\sigma_{rms, central} \ge 10$). This left a total of {319} sources to use for comparison of LOFAR and TGSS-ADR flux densities. The offsets between the measured flux densities are shown in Figure~\ref{fig:fluxoffsets1}.

{For further comparison, we also investigated how our flux density ratio compares to the Murchison Widefield Array (MWA) GLEAM Survey \citep{HurleyWalker2017}.  This is a 2' resolution survey that covers the majority of the southern sky south of declinations of 30$^{\circ}$, but to much higher flux densities. We only considered sources that fulfilled the same high signal-to-noise thresholds as before, but with the respective beam size for MWA used for the small and isolated conditions. This left {45} sources for comparison. Again the flux density ratios can be seen in  Figure~\ref{fig:fluxoffsets1}.}

\begin{figure}
\begin{center}
\centering
\includegraphics[width=8.5cm]{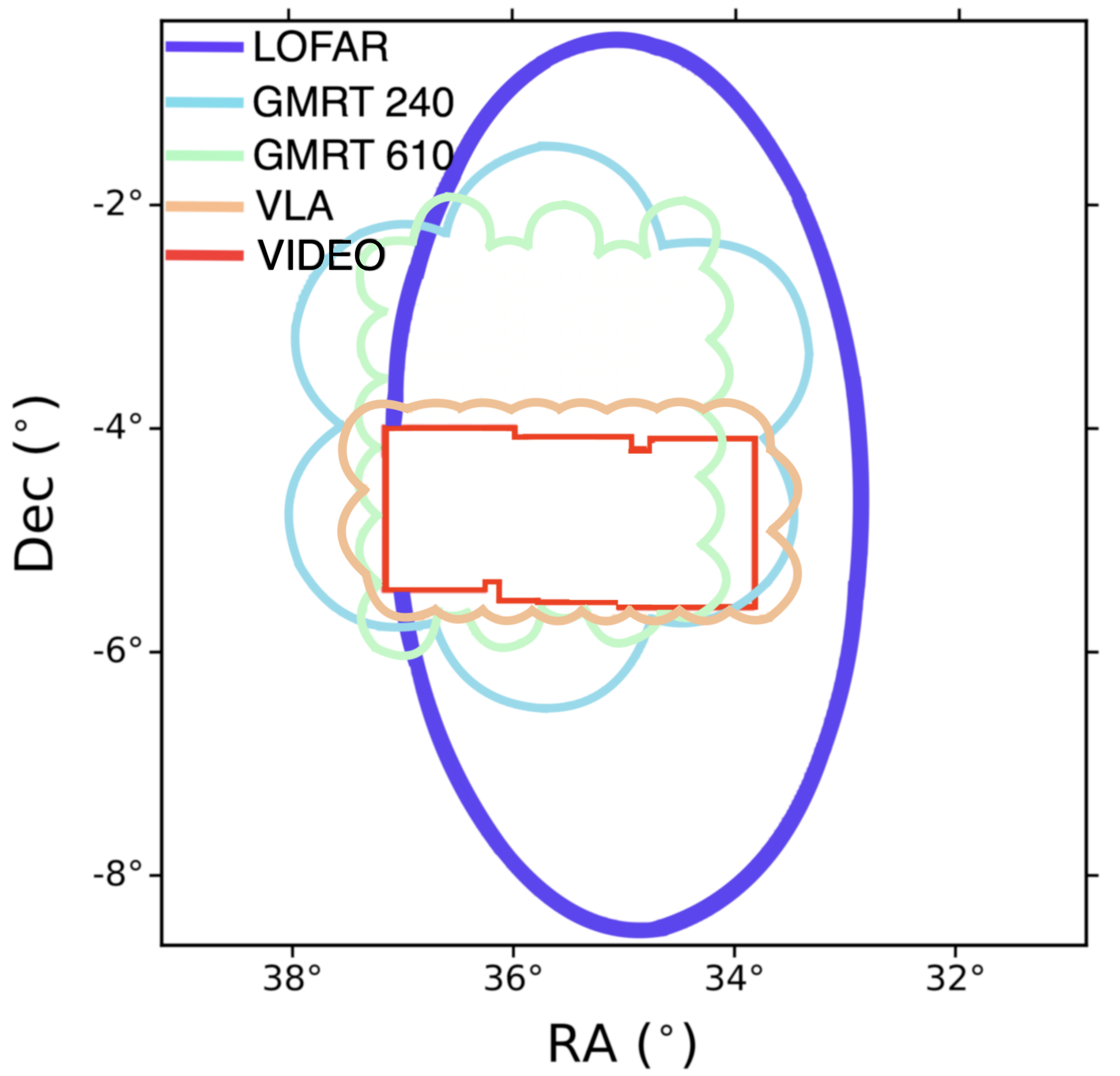}
\end{center}
\caption{{Locations of targeted radio observations that have covered the XMM-LSS field at various radio frequencies. Shown are the observations presented in this paper (purple), observations from \protect \cite{Tasse2007} at 240 MHz (blue) and 610 MHz (green), recent JVLA observations of the VIDEO field of \protect \cite{HeywoodVLA} (orange), and the near-IR VIDEO observations from \cite{Jarvis2013} (red).}}
\label{fig:crossmatch_coverage}
\end{figure}

{Comparing these fluxes we find a median offset for $S_{\textrm{LOFAR}}/S_{\textrm{TGSS}}$ to be  $1.23_{-0.19}^{+ 0.28}$ with errors from the 16th and 84th percentiles; this is seen in black in Figure \ref{fig:fluxoffsets1}(a). For our comparison with the MWA (cyan in Figure \ref{fig:fluxoffsets1}a), we instead find a ratio of $1.05_{-0.19}^{+ 0.11}$, much closer to a ratio of 1. However this is for a much smaller number of sources,} all of which are at high fluxes. The results from TGSS-ADR suggest that we are measuring higher fluxes in LOFAR than we expect}. This is higher than has been found in previous LOFAR observations \citep[e.g.][]{Hardcastle2016, Shimwell2017}, were ratios more similar to $S_{\textrm{LOFAR}}/S_{\textrm{TGSS}} \sim 1.1$ were found. This could be a result of our low-declination observations. 

{On account of the large differences in the flux ratios observed from TGSS-ADR and MWA, we also looked at the comparison with the corrected TGSS-ADR catalogue that was generated by \cite{Hurley-Walker2017-TGSS}. In their analysis, \cite{Hurley-Walker2017-TGSS} suggested that there may be systematic offsets to the TGSS-ADR catalogue. Over the XMM-LSS field, the correction factors that need to be applied to TGSS can range from $\sim 1.05-1.25$. The flux correction factors that these then suggest to LOFAR can be seen in Figure \ref{fig:fluxoffsets1}(b). This now suggests a flux ratio of $1.10_{-0.18}^{+ 0.28}$, which is smaller than suggested previously and now similar to the ratios seen in \cite{Hardcastle2016, Shimwell2017}. }

\begin{figure*}
\begin{center}
\centering
\begin{minipage}[b]{0.5\textwidth}
\includegraphics[width=8.5cm]{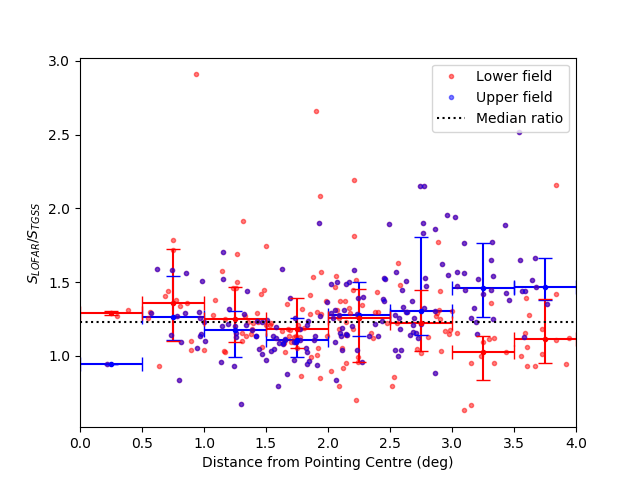}
\subcaption{Using the TGSS-ADR catalogue from \cite{Intema2017}}
\end{minipage}%
\begin{minipage}[b]{0.5\textwidth}
\includegraphics[width=8.5cm]{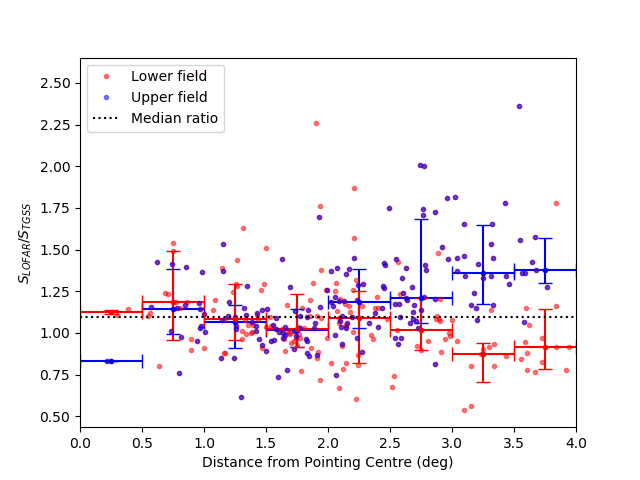}
\subcaption{Using the rescaled TGSS-ADR catalogue from \cite{Hurley-Walker2017-TGSS}}
\end{minipage}%
\end{center}
\caption{{Comparisons of the integrated flux recorded by LOFAR and TGSS for sources matched within a 5" radius as a function of distance from the pointing centre. Comparisons are shown, in the left-hand panel, from the TGSS-ADR catalogue from \cite{Intema2017} and, in the right-hand panel, from the rescaled TGSS-ADR catalogue from \cite{Hurley-Walker2017-TGSS}. This shows the differences for the lower (red) and upper (blue) parts of the field, defined as split in declination at -4.5$^{\circ}$. The black dotted line indicates the median ratio of LOFAR to TGSS-ADR fluxes across the whole field. Errors in distances from the pointing centre show the bin width, whereas those in flux ratios indicate the 16th and 84th percentiles in the flux ratios in that bin. }}
\label{fig:fluxoffsetsposition}
\end{figure*}

{To further investigate whether these large offsets between LOFAR to TGSS-ADR fluxes occur as a result of declination, we investigate this ratio as a function of distance} from the centre of the field, comparing the lower and upper half of our observations.  The results of this comparison can be seen in Figure~\ref{fig:fluxoffsetsposition} and suggests that over the majority of the field there appears to be no difference in the flux ratios between LOFAR and TGSS-ADR in the upper or lower half of the field. There is however a difference between the upper and lower halves of the field at large angular distances ($> 3^{\circ}$) from the pointing centre, when compared with TGSS-ADR. At these large angular distances from the centre, the lower half of the field has values of {$S_{\textrm{LOFAR}}/S_{\textrm{TGSS}}$} more similar to 1, and more comparable with previous observations. In the upper half of the field and $> 3^{\circ}$, however, we observe higher {$S_{\textrm{LOFAR}}/S_{\textrm{TGSS}}$} ratios than the median value. This may suggest that the LOFAR primary beam model at the large extents of the field in the north-south directions overcompensates in the upper parts of the field compared to the lower field. However as this is only an effect at the largest angles, this is a small issue for most of the field and should not affect much of the region in which we have multiwavelength {data. If we instead look at this with the rescaled TGSS-ADR catalogue from \cite{Hurley-Walker2017-TGSS}, there is a much larger dichotomy in the flux ratios in the upper and lower halves of the field. 
}

{From these flux comparisons, there is conflict as to what flux scale correction should be applied. As such, we do not apply any correction factor in the fluxes given in Table \ref{tab:catalogue}. In all future analyses we will discuss the impact if our fluxes from LOFAR are in fact too large.} 

\begin{figure}
\begin{center}
\centering
\begin{minipage}[b]{0.5\textwidth}
\includegraphics[width=9cm]{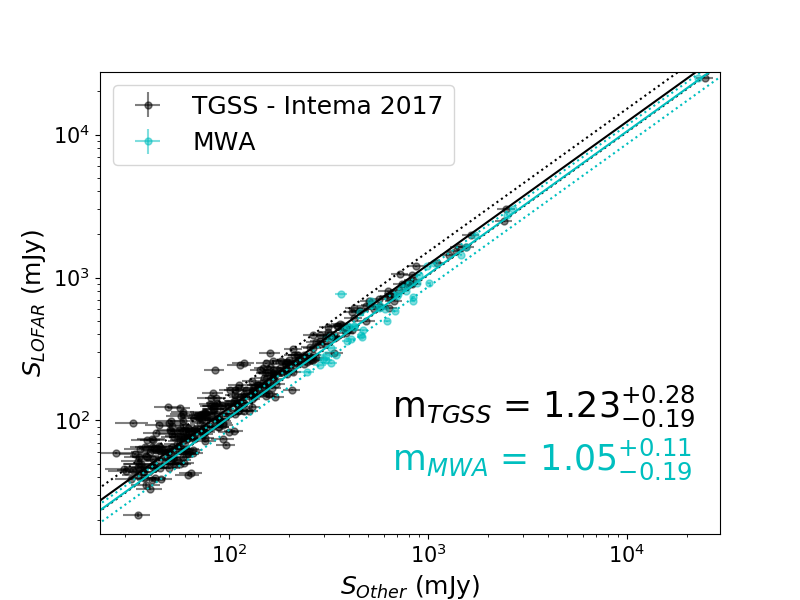}
\subcaption{}
\end{minipage}
\newline
\begin{minipage}[b]{0.5\textwidth}
\includegraphics[width=9cm]{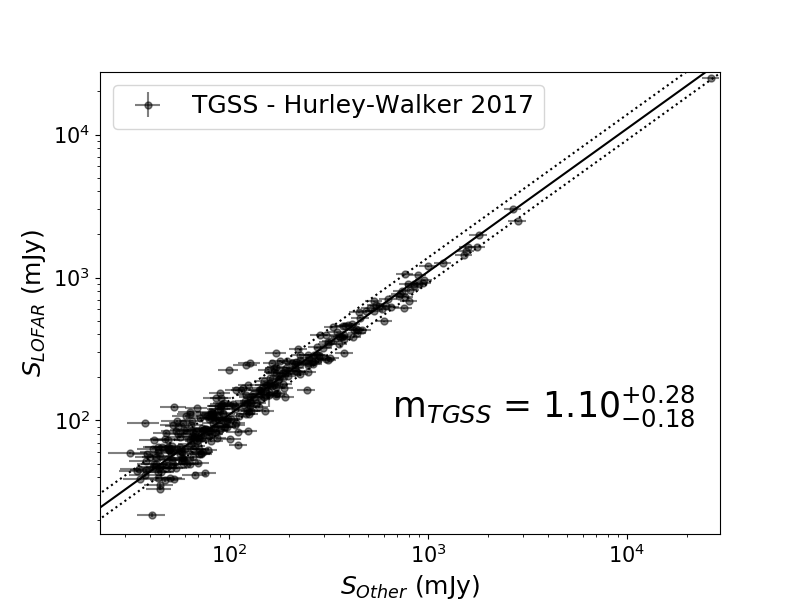}
\subcaption{}
\end{minipage}
\end{center}
\caption{{Comparisons of the integrated flux recorded by LOFAR to that recorded by TGSS-ADR (black) and MWA (cyan) in Figure (a) and from the rescaled TGSS-ADR catalogue from \protect \cite{Hurley-Walker2017-TGSS} in Figure(b). The black/cyan dotted lines show the fit to the data of the ratio between the two fluxes when the median offset of the ratio between LOFAR to TGSS-ADR/MWA is used. The dotted lines indicate the errors associated with these fits (generated using the 16th and 84th percentiles). We quote the value of the median $S_{\textrm{LOFAR}}/S_{\textrm{TGSS}}$ and $S_{\textrm{LOFAR}}/S_{\textrm{MWA}}$ values and their errors in the lower right corner. }}
\label{fig:fluxoffsets1}
\end{figure}

\subsection{Positional offsets}
\label{sec:positions}

\subsubsection{Positional offset to other radio surveys}
Next we investigate the differences in apparent positions of sources of our observations compared to the surveys in Table \ref{tab:matches}. Differences in the apparent positions are necessary to be understood for optical identification and subsequent spectroscopic study, such as WEAVE-LOFAR \citep{Smith2016}. To do this we compared our positional offsets to the TGSS-ADR and other multi-frequency radio data over the XMM-LSS field. Data sets at higher radio frequencies are especially useful to compare the positional accuracy {because they typically have higher resolution and the ionosphere, which can move the apparent position of these sources, has a larger effect at lower frequencies. }

\begin{figure}
\begin{center}
\centering
\includegraphics[width=10cm]{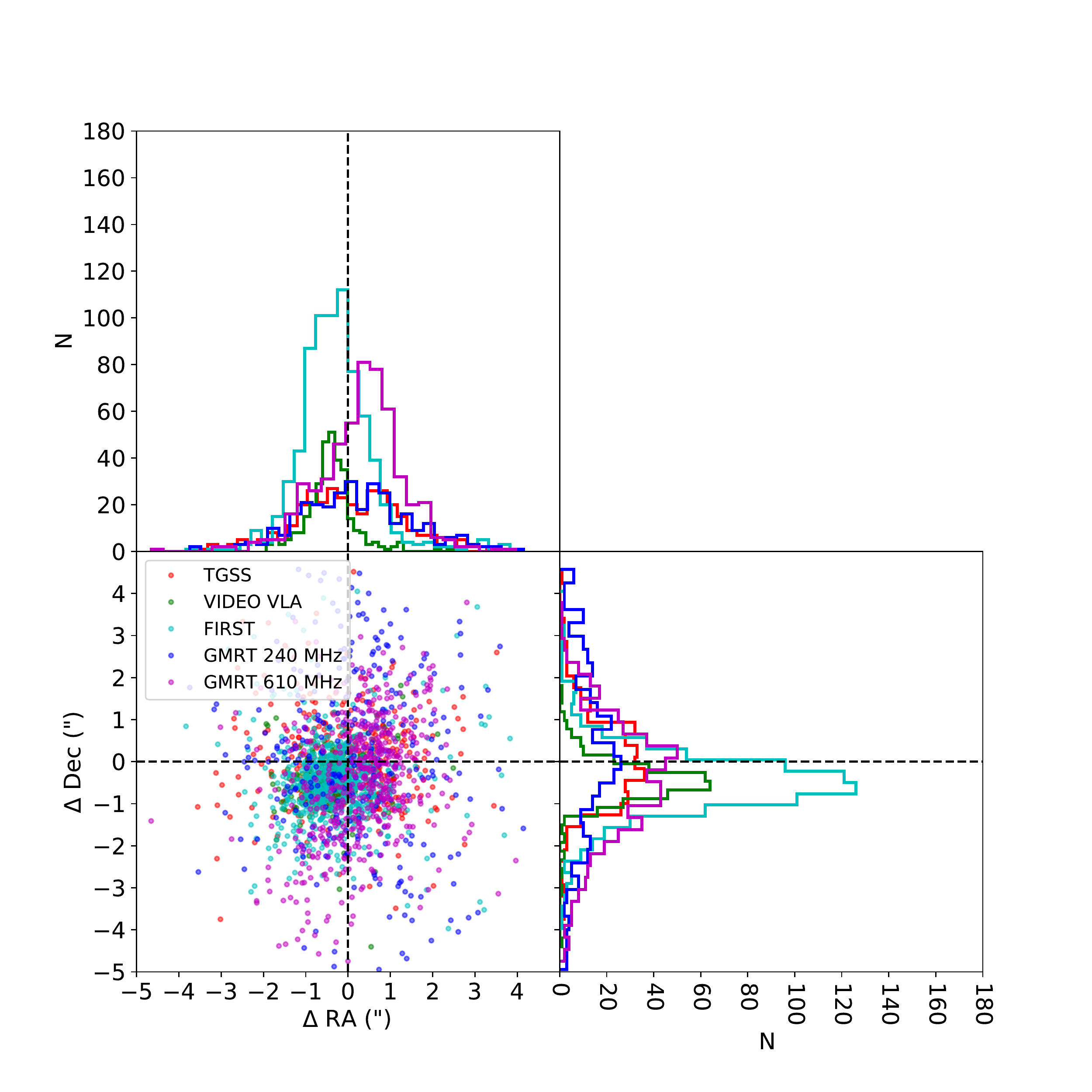}
\end{center}
\caption{ {Positional RA and Dec offsets of the LOFAR sources compared to TGSS-ADR \protect \citep[][red]{Intema2017}, VIDEO VLA observations \protect \cite[][green]{HeywoodVLA}, FIRST \protect \cite[][cyan]{Becker1995, Helfand2015}, and GMRT observations \protect \citep[][]{Tasse2007} at 240 MHz (blue) and 610 MHz (magenta). The histograms are both modelled as a Gaussian and the median values are used to quantify the average offsets. The values of these are given in Table \ref{tab:pos_offsets}. For the histogram, the errors are calculated by the 16th and 84th percentiles. }}
\label{fig:posoffset}
\end{figure}

Again, we continue to use the small, isolated, high signal-to-noise source criteria used in Section~\ref{sec:fluxes}{, using the respective beam sizes for the poorest resolution data}. In all cases, we define the offset as $\Delta$RA= RA$_{\textrm{LOFAR}}$ - RA$_2$, where RA$_2$ is the RA of the {matched catalogue. The parameter $\Delta$ Dec is defined in the same way}. {The offsets and number of sources used to calculate these values are shown in~Table \ref{tab:pos_offsets} for each of the external catalogues.} The offsets are also plotted in Figure~\ref{fig:posoffset}. We fit the distribution of positional offsets by a {Gaussian} and compute its mean and standard deviation. We also record the median offset and calculated associated errors using the 16th and 84th percentiles. 

The {TGSS-ADR and GMRT 240 MHz observations have the smallest offsets compared to our LOFAR observations, having both RA and Dec offsets constrained to $\lesssim 0.15$". However, these two catalogues suggest offsets in opposite directions to one another. For the highest frequency observations (GMRT 610 MHz, FIRST, and VIDEO VLA), the measured offsets suggest that the LOFAR declinations are offset with respect to those measured with other instruments, by approximately 5\% of the synthesised beam size. Both FIRST and the VLA observations find an $\sim 0.4$" in RA and $\sim 0.5$" offset in Dec. Whilst the GMRT 610 MHz sources have similar magnitude offsets, this is in the opposite direction for the RA offsets to that of VLA and FIRST. The Dec offsets from GMRT however act in the same direction and are the same order of magnitude.  }

{We include a correction to the LOFAR positions using a constant offset from the FIRST histogram (i.e. $\Delta$RA$=-$0.34" and $\Delta$Dec$=-$0.51") to our final catalogue, which can be seen in the final two columns of Table \ref{tab:catalogue}.}

\subsubsection{Positional offset to multiwavelength data}
{Finally, we compare the positions of our sources to those derived from multiwavelength data.} To do this, we used the VIDEO \citep[][]{Jarvis2013} survey and compared the positional offsets of sources in that field. For point source, the host galaxy in VIDEO that is associated with the LOFAR galaxy should be nearly co-incident with the source. For more extended sources and for those with complicated jet morphology, this may not be the case. As true matching to multiwavelength counterparts involves processes such as visual identification and likelihood ratio matching \citep[see e.g. ][]{McAlpine2012, Williams2018, Williams2018a, Prescott2018}, we consider the offsets assuming that the closest VIDEO source is the counterpart for our LOFAR galaxy as well as considering the distribution of positional offsets for all sources within a 10" radius. 

\begin{figure*}
\begin{center}
\begin{minipage}[b]{0.5\textwidth}
\includegraphics[width=10cm]{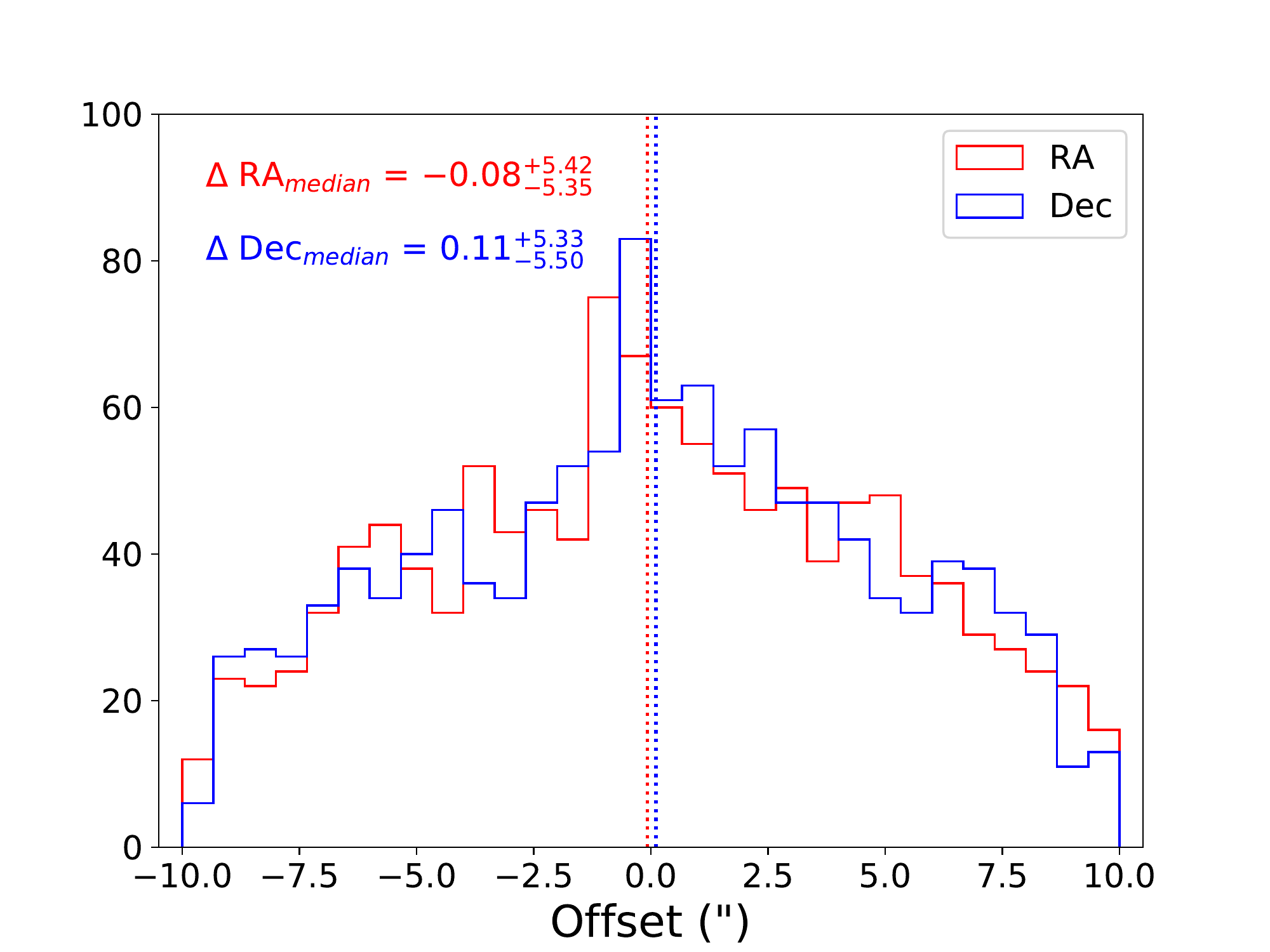}
\subcaption{}
\end{minipage}% 
\begin{minipage}[b]{0.5\textwidth}
\includegraphics[width=10cm]{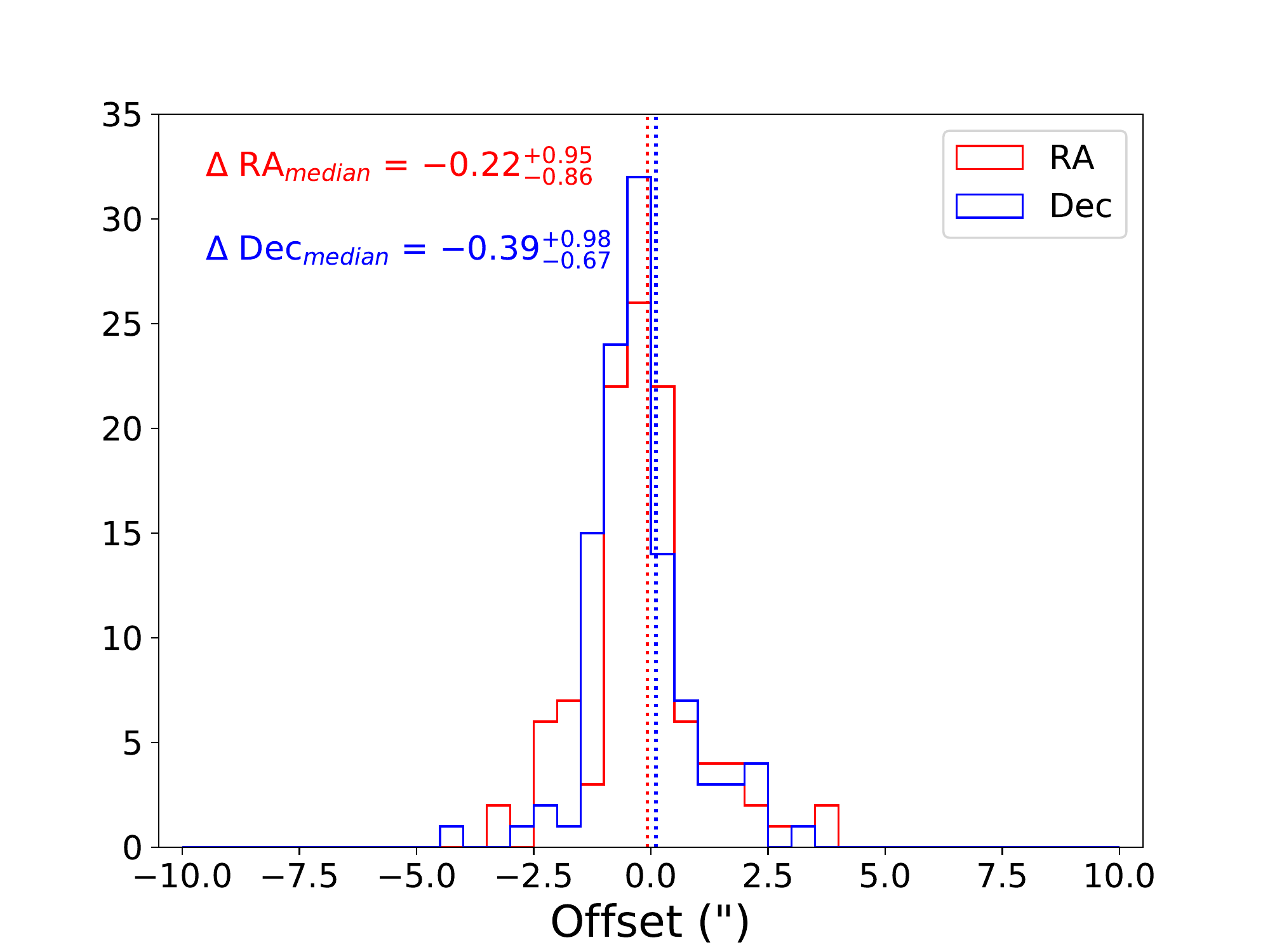}
\subcaption{}
\end{minipage}% 
\end{center}
\caption{{Positional RA (red) and Dec (blue) offsets of the LOFAR sources compared to VIDEO sources \protect \citep{Jarvis2013}. Shown in (a) is the histogram for all matches within 10" and in (b) is a histogram of offsets for the best-matched VIDEO source for each LOFAR source that had a VIDEO source within 10". The upper left shows the median offsets and the errors associated with these from the 16th and 84th percentiles. The dashed line indicates the median offset value for either the RA offsets (red) or Dec offsets (blue). Offsets are defined as RA$_{\textrm{LOFAR}}$ - R$_{\textrm{VIDEO}}$ and similar for Dec.}}
\label{fig:video_offset}
\end{figure*}

For this investigation, we do not make any cuts on our LOFAR sources;{ i.e. we do not only use those that are compact,} isolated, and have high-signal-to-noise. We do however constrain our sources to be within the {region $ 33.85 ^{\circ}\geq$ RA $\leq 37.15 ^{\circ}$ and  $-5.35^{\circ} \geq$ Dec $\leq -4.20^{\circ}$ to overlap with the VIDEO region\footnote{{Although there may be some holes in the VIDEO sources due to artefacts from stars, etc.}}. }The results of this can be seen in Figure \ref{fig:video_offset}. The left-hand panel (Figure \ref{fig:video_offset}a) shows a histogram of the offsets for all VIDEO sources that are located within 10" of a LOFAR source. The right-hand panel (Figure \ref{fig:video_offset}b) shows only the offset for the closest VIDEO source to each LOFAR source. Again, these positional offsets have an average positional offset constrained to $\lesssim$0.5" ($-0.21^{+0.95}_{-0.86}$ for RA and $-0.39^{+0.98}_{-0.67}$ for Dec; these are quoted as the median value and the errors from the 16th and 84th percentiles). {This suggests a marginally larger declination offset compared to the right ascension offset, but both of the values are smaller in magnitude than we find when we compared to FIRST, VLA, and GMRT 610 MHz observations.} Again, we emphasise that the nearest VIDEO source may not be the correct host galaxy for our LOFAR sources and a true understanding of the positional offsets with respect to the multiwavelength host galaxy will be performed once these galaxies have been properly cross-matched. Nevertheless, these results give confidence that the positions are accurate to within 0.4''.

\begin{table*}
\begin{center}
\begin{tabular}{c c c c c c c }
\centering
Catalogue &Frequency  & Matches & RA offset (") & RA offset (") & Dec offset (") & Dec offset (") \\  
&(MHz)& within 5" & \textit{Histogram} & \textit{Gaussian} & \textit{Histogram} & \textit{Gaussian} \\ \hline \hline
TGSS-ADR & 150 &  319 & $-0.06_{-1.10}^{+1.16}$& $\mu=-0.00$, $\sigma=1.19$& $-0.03_{-0.93}^{+1.06}$& $\mu=-0.04$, $\sigma=0.97$\\
GMRT  &240 &   310 & $+0.10_{-1.20}^{+1.38}$& $\mu=+0.11$, $\sigma=1.26$& $+0.14_{-2.02}^{+2.09}$& $\mu=+0.22$, $\sigma=1.99$\\  
GMRT  &610 &  532  & $+0.36_{-1.06}^{+0.76}$& $\mu=+0.41$, $\sigma=0.81$& $-0.41_{-1.43}^{+1.29}$& $\mu=-0.37$, $\sigma=1.36$\\
FIRST &1400 &   738 & $-0.34_{-0.64}^{+0.72}$ &$\mu=-0.35$, $\sigma=0.66$  & $-0.51_{-0.60}^{+0.65}$& $\mu=-0.51$, $\sigma=0.58$ \\
VLA  &1500 &   317 & $-0.40_{-0.47}^{+0.37}$& $\mu=-0.41$, $\sigma=0.35$& $-0.47_{-0.42}^{+0.43}$& $\mu=-0.50$, $\sigma=0.39$\\
\end{tabular}
\caption{{Positional RA and Dec offsets of LOFAR compared to other radio surveys as presented in Figure \ref{fig:posoffset} for compact, isolated, high signal-to-noise sources matched within 5". The offsets are described in two ways, firstly by modelling them as Gaussians and secondly using the median, 16th, and 84th percentiles of the offsets.}}
\label{tab:pos_offsets}
\end{center}
\end{table*}

\subsection{Spectral Indices}
\label{sec:specindex}

\begin{figure}
\begin{center}
\centering
\includegraphics[width=10cm]{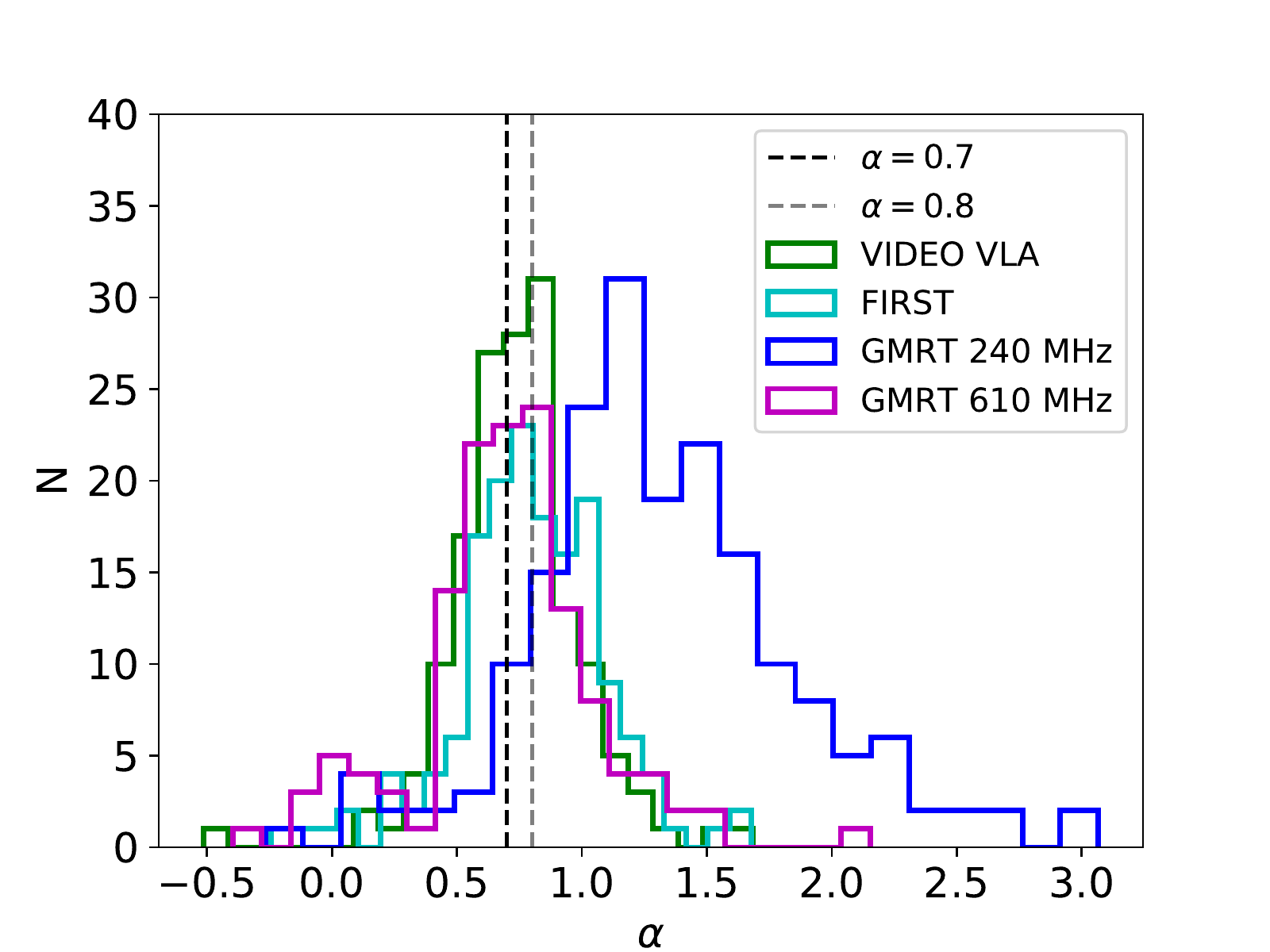}
\end{center}
\caption{{Histograms showing the measured spectral index, $\alpha$, of sources in this field through matching to VIDEO VLA observations \protect \cite[][green]{HeywoodVLA}, {FIRST \protect \cite[][cyan]{Becker1995,Helfand2015},} and GMRT observations \protect \citep[][]{Tasse2007} at 240 MHz (blue) and 610 MHz (magenta) within a 5" radius.}}
\label{fig:spectralindex}
\end{figure}

{Finally, we use the multi-frequency radio observations to investigate the spectral properties of the LOFAR sources. For this we assume a typical synchrotron single power-law distribution, where other surveys typically find an average $\alpha \sim 0.7-0.8$ \citep[see e.g.][]{Intema2017,CalistroRivera2017}. We calculate the} spectral index for each matched source and the histograms of $\alpha$ for the LOFAR sources matched to the other four external catalogues are shown in Figure~\ref{fig:spectralindex}. We assume a single power-law index for each source. Again we use the same source selection as in Sections \ref{sec:fluxes} and \ref{sec:positions}, that is compact, isolated, and high signal to noise. {As studying spectral indices is inherently biased because of the differing flux limits between surveys, we only include sources that could be detected in both catalogues used to calculate the spectral index. To do this we impose a flux cut on the LOFAR sources being compared so that any sources that could have a value of $\alpha =2$ would have been detected in both surveys using 5$\sigma$ sensitivities from Table \ref{tab:matches}. }The properties of the observed spectral indices shown in Figure~\ref{fig:spectralindex} are also given in Table~\ref{tab:spec_index}. Again we model these as both a {Gaussian and record the median and errors from the quantiles} of the data as in Section~\ref{sec:positions}.

{From this it can be seen that the values of $\alpha$ obtained from GMRT at 610 MHz, FIRST, and VLA are peaked at $\sim 0.7-0.8$. On the other hand the $\alpha$ values derived from GMRT measurements at 240 MHz are peaked at a much higher value, $1.26_{-0.36}^{+0.52}$. This suggests that the flux densities in the GMRT 240 MHz survey have flux scalings that may be offset to the flux densities observed in this work. For the three catalogues from which we obtain spectral indices $\sim 0.7-0.8$ we record $\alpha$ values of $0.70_{-0.24}^{+0.27}$ (GMRT 610 MHz), $0.79_{-0.22}^{+0.26}$ (FIRST), and $0.76_{-0.24}^{+0.18}$ (VLA). These are similar to what we expect from previous continuum radio surveys. }

{If we had assumed a constant factor for the relationship between the LOFAR and TGSS-ADR fluxes, these spectral indices would have been different. If it is the case that the flux should be corrected to match TGSS-ADR (or the rescaled catalogue) then the spectral indices would have been smaller. For the GMRT 240 MHz data, with this flux correction, the spectral index would be lower and likely to be more consistent with typical values. For spectral indices calculated at frequencies close to 144 MHz, the influence of flux corrections would be particularly important. If the largest correction, 1.23 from TGSS-ADR comparisons, is used, then the spectral index values we obtain are $\alpha_{240}^{144} \sim 0.85$, $\alpha_{610}^{144} \sim 0.56$, $\alpha_{1400}^{144} \sim 0.70,$  and finally $\alpha_{1500}^{144} \sim 0.67$. Therefore for our measurements with VLA and FIRST, the changes in $\alpha$ are less and the values remain consistent within the typical $\alpha \sim 0.7-0.8$ values. For GMRT measurements, the effect is larger, however, because of the closer frequency values. }

\begin{table*}
\begin{center}
\begin{tabular}{c c c c c c}
\centering
Catalogue  &Frequency  & Matches & $\alpha$ & $\alpha$   \\ 
&(MHz)& within 5" & \textit{Histogram} & \textit{Gaussian} \\ \hline \hline
GMRT &240 &  192 & $1.26_{-0.36}^{+0.52}$ & $\mu=1.25$, $\sigma=0.41$ \\ 
GMRT  &610 & 141 & $0.70_{-0.24}^{+0.27}$ & $\mu=0.72$, $\sigma=0.22$ \\
FIRST &1400 &  157 & $0.79_{-0.22}^{+0.26}$& $\mu=0.81$, $\sigma=0.24$ \\
VLA &1500 &  155 & $0.76_{-0.24}^{+0.18}$ & $\mu=0.73$, $\sigma=0.20$\\
\end{tabular}
\caption{{Spectral indices, $\alpha$ generated between the LOFAR fluxes and other radio surveys, as presented in Figure \ref{fig:spectralindex}. The median, 16th and 84th percentiles of the offsets have been used to describe the average spectral indices observed. These are from comparisons with GMRT \protect \citep[][]{Tasse2007}, {FIRST \protect \citep[][]{Becker1995,Helfand2015}}, and VLA \protect \citep[][]{HeywoodVLA}.}}
\label{tab:spec_index}
\end{center}
\end{table*}

\section{Source counts}
\label{sec:sourcecounts}

Next, we investigate the source counts of radio sources in the field. Source counts are important for understanding the population of radio galaxies. This is especially true at faint {flux densities}, where SFGs and the radio quiet quasar population become dominant \citep[see e.g.][]{Wilman2008, White2015}. The distribution of sources at such faint {flux densities} is important for predictions for future telescopes, e.g. the Square Kilometre Array (SKA), where source confusion noise may be an issue.

{The directly measured source counts do not reflect the true source counts expected for extragalactic radio sources as the chance of detection of sources of different sizes/{flux densities} is a function of location within the map as well as their signal to noise. Hence, corrections need to be applied to the direct source counts from the \textsc{PyBDSF} derived catalogue. For these investigations, we calculated two corrections to be applied to the source counts. Firstly we applied a correction to account for false detection of sources over the field. Secondly we made simulations of sources in the image to account for completeness, i.e. how well can we detect sources in the image, taking into account resolution and visibility area effects. These are described in further detail below. }

\subsection{False detection}
\label{sec:fdr}
{The false detection rate (FDR) accounts for noise spikes and artefacts in the image} that are extracted by the source detection package, \textsc{PyBDSF}, as sources. Symmetry means that positive noise spikes have counterpart negative spikes that are detectable as sources in the inverse (negative) image. We exploit this property to correct for the FDR by using the inverse image. Using the inverted image, \textsc{PyBDSF} was re-run using the same parameters as used to obtain the source catalogue (see Table~\ref{tab:PyBDSF}). However this  overestimates the FDR because {\textsc{PyBDSF} takes regions of higher rms around bright sources into account}. In the inverse image there are no such bright sources and so it may be more likely that negative artefacts are included as sources, when the positive counterparts may not have been. {This leads to higher correction factors than would be appropriate in bins with high flux densities.}

Indeed, when we investigated the FDR, it became apparent that some of the bright {sources in the field had strong artefacts around them} that were giving large FDRs at high {flux densities}. This is not expected as the FDR should typically only be an issue at faint {flux densities}, where noise peaks may have been confused with sources. {Therefore we exclude regions} around bright sources from this analysis. {To do this we masked out circular regions of radius $200"$ around each source in the original, raw \textsc{PyBDSF} catalogue that had a {flux density} greater than 0.1 Jy. This ensures} that we are measuring source counts only in regions where we are not {influenced} by artefacts, {which} could be biasing our results. 

{To calculate the FDR, the number of sources observed in the inverted image for each {flux density} bin in this catalogue is calculated. This can be compared to the number of sources detected in the real image to calculate the fraction of real sources. This fraction of real sources in the i$^{\textrm{th}}$ {flux density} bin of the source counts is given by}

\begin{equation}  f_{\textrm{real, i}} = \frac{N_{\textrm{catalogue, i}} - N_{\textrm{inv., i}}}{N_{\textrm{catalogue, i}}} \end{equation}
where $N_{\rm inv.}$ denotes the detected sources from the inverted image. This fraction is applied to the source counts in the extracted catalogue by multiplying the source counts in each bin by the real fraction corresponding to that bin. The errors associated with the false detection are calculated using the Poissonian errors in the individual number counts. The resulting false detection correction factors are shown in Figure \ref{fig:sc_corrections}.

\subsection{Completeness simulations}
\label{sec:completeness}

{We also carried out} simulations to investigate how well sources can be detected in the image, i.e. the completeness of our catalogue. {These simulations need} to take into account the fact that extended sources are affected by resolution bias. This relates to the fact that for sources larger than the beam size, as the size increases, the ratio of the peak to total flux decreases. This means that for larger sources with the same total {flux density} compared to smaller {sources,} the peak {flux density (units of Jy beam$^{-1}$)} may decrease sufficiently that the source may be unable to be detected over the noise. This affects {our ability to detect extended} sources. 

In \cite{Williams2016} completeness corrections were calculated through injecting sources into the residual image; resolution bias was calculated using similar methods to those of \cite{Prandoni2001} and \cite{Mahony2016} who make use of relations from \cite{Windhorst1990}. For this study however, we decided to make use of the SKA Design Study Simulated Skies \citep[S$^3$;][]{Wilman2008}. The S$^3$ simulations provide mock catalogues of sources and their associated {flux densities} at five different radio frequencies. It also contains information on the expected {sizes} of the sources. These provide us with a {realistic catalogue to inject sources of varying sizes} into our simulations. 

We created 100 simulations of the field where each simulation had 1,500 sources added to the image at random positions. {These are elliptical components that have major/minor axes} from S$^3$ and random position angles were assigned. The simulated sources were convolved with the restoring beam before being injected into the image. The {flux density} for each source was also taken from S$^3$ and was calculated by scaling the 1.4~GHz {flux densities} to 144 MHz using $\alpha=0.7$. The 1.4 GHz {flux densities} were used as these have been more robustly compared to and are in good agreement with previous data, especially at faint {flux densities} \citep[see e.g.][]{Wilman2008,Smolcic2015}, whereas the 151 MHz {flux densities} from S$^3$ have not been as well tested at fainter {flux densities} and may suffer from too steep spectral curvature.

{We only injected S$^3$ sources with total flux densities above 1 mJy at 144~MHz into the image. We however used a 2 mJy flux density limit in our source counts, as we believe our source counts are reliable down to this flux density level. Injecting sources into the final image \citep[not the residual image used by ][]{Williams2016} also allowed us to account for confusion of sources. Using these simulations we also inherently took into account the visibility area of sources at different flux density limits. This relates to the fact that because of the varying noise across the field, {in particular} the primary beam attenuation, faint sources will not be detected over the full radio image.}

{For each simulated image, the sources were re-extracted using \textsc{PyBDSF} with the same detection criteria as used to create the real source catalogue. The correction factor was then calculated by comparing the number of injected sources to the number of recovered sources in each flux density bin. This can be carried out with two methods, using the general principle that the correction factor is given by}

\begin{equation}
 \textrm{Completeness \ correction}_{\textrm{i}}  =  \frac{N_{\textrm{injected,i}}}{N_{\textrm{recovered\_from\_injected,i}}}.
\label{eq:completeim} \end{equation}
Here, $N_{\textrm{recovered\_from\_injected,i}}$ is the number of sources extracted from the \textsc{PyBDSF} catalogue of the simulation once the number of sources from the original \textsc{PyBDSF} catalogue of the field {has} been subtracted. This accounts for the fact that there are already sources that existed in the image pre-simulation.

{In Method 1, the correction was calculated by directly comparing the source counts of the extracted \textsc{PyBDSF} catalogue to the source counts from the injected simulated sources. This was calculated once the source counts of objects that were originally in the image were subtracted. By doing this, the effect of factors such as Eddington bias (where faint sources on noise spikes may appear at higher fluxes) and the merging of sources were already taken into account. This also considered the effect of the recovery of sources from \textsc{PyBDSF} and any differences in the flux density \textsc{PyBDSF} recovers to what was simulated. This can therefore produce correction factors $<$1 when the flux densities of sources move in and out of the flux densities bins used to determine the source counts. }

{In Method 2, we directly compared whether a simulated source is recovered, irrespective of the flux density with which the source is measured with. To do this, we first cross-matched the \textsc{PyBDSF} catalogue generated for each simulation to the original image \textsc{PyBDSF} catalogue, within a 5" radius, before matching to the input catalogue of simulated sources. The first positional match attempts to remove any sources that were originally present in the image. For the second match, we used a 15" positional radius ($\sim 2 \times$ the beam size) to match the simulated sources to those we recover with \textsc{PyBDSF}. This 15" positional radius ensures that in the cases where simulated sources may have merged with another neighbouring source, they are still included. We can then see what fraction of our original simulated input catalogue is recovered for a given flux density bin. This however has limitations such as if sources have merged together with those objects already in the image. For the sources that are determined to have not been in the original image before we simulated the sources, typically $\sim$98\% of these are matched to the catalogue of input simulated sources\footnote{{This might be slightly lower when we use the larger sized source in Section \ref{sec:scmodel}}}. This lack of 100\% match may bias our results slightly higher, but is a small correction. }

\begin{figure}
\begin{center}
\centering
\includegraphics[height=6.5cm]{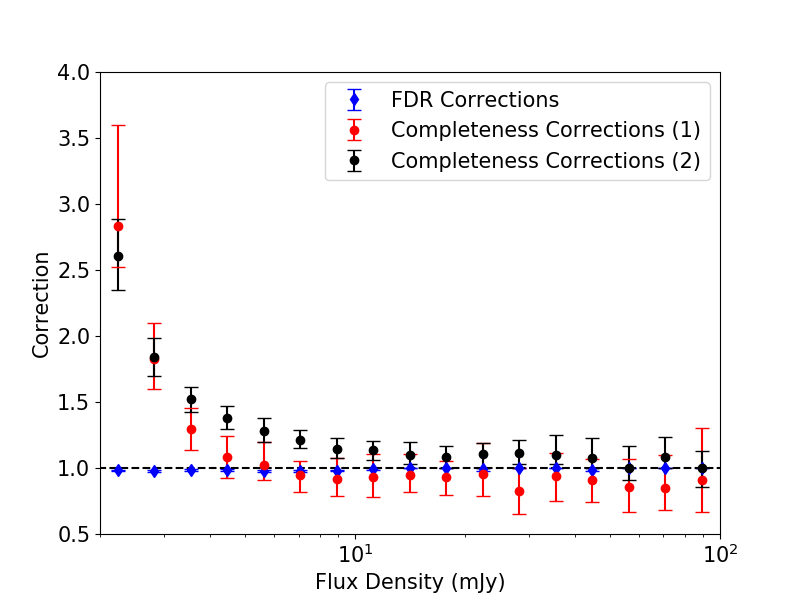}
\caption{{Correction factors calculated and then applied to the source counts from the catalogue using the methods described in Sections \ref{sec:fdr} and \ref{sec:completeness}: FDR corrections (blue) and completeness/resolution correction using Method 1 (red) and Method 2 (black). }}
\label{fig:sc_corrections}
\end{center}
\end{figure}

{In both cases, to obtain} an overall estimate for our completeness correction and its associated errors, the completeness corrections that were estimated from all simulations were combined. The completeness correction factor is given as the median value in each flux density bin, from all simulations, and the associated {errors were} calculated from the 16th and 84th percentiles. These correction factors can be seen in Figure \ref{fig:sc_corrections}.

\begin{figure*}
\begin{center}
\centering
\begin{minipage}{\textwidth}
\centering
\includegraphics[height=10.5cm]{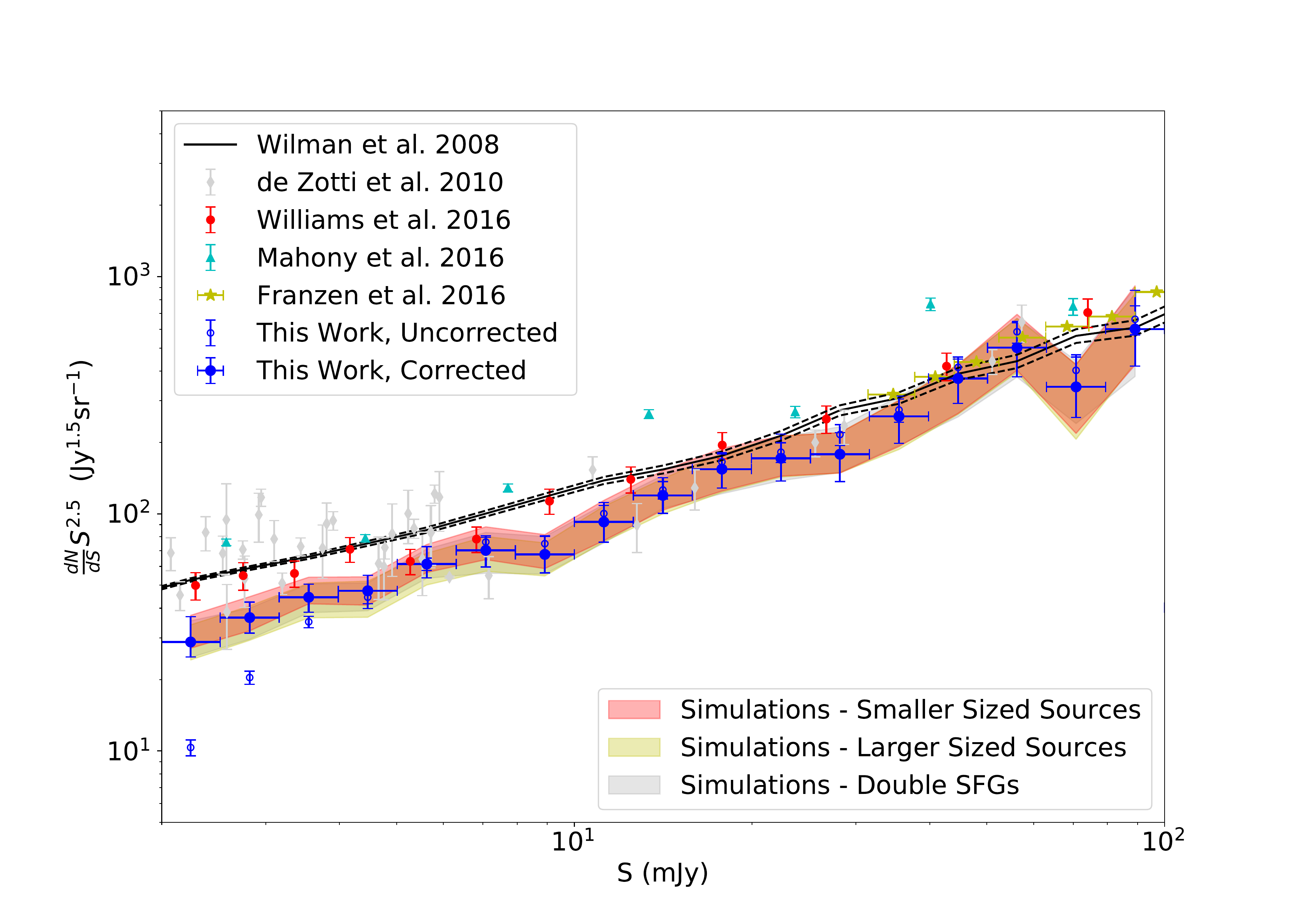}
\subcaption{Corrected using FDR correction and the completeness correction Method 1.}
\end{minipage}%
\newline
\begin{minipage}{\textwidth}
\centering
\includegraphics[height=10.5cm]{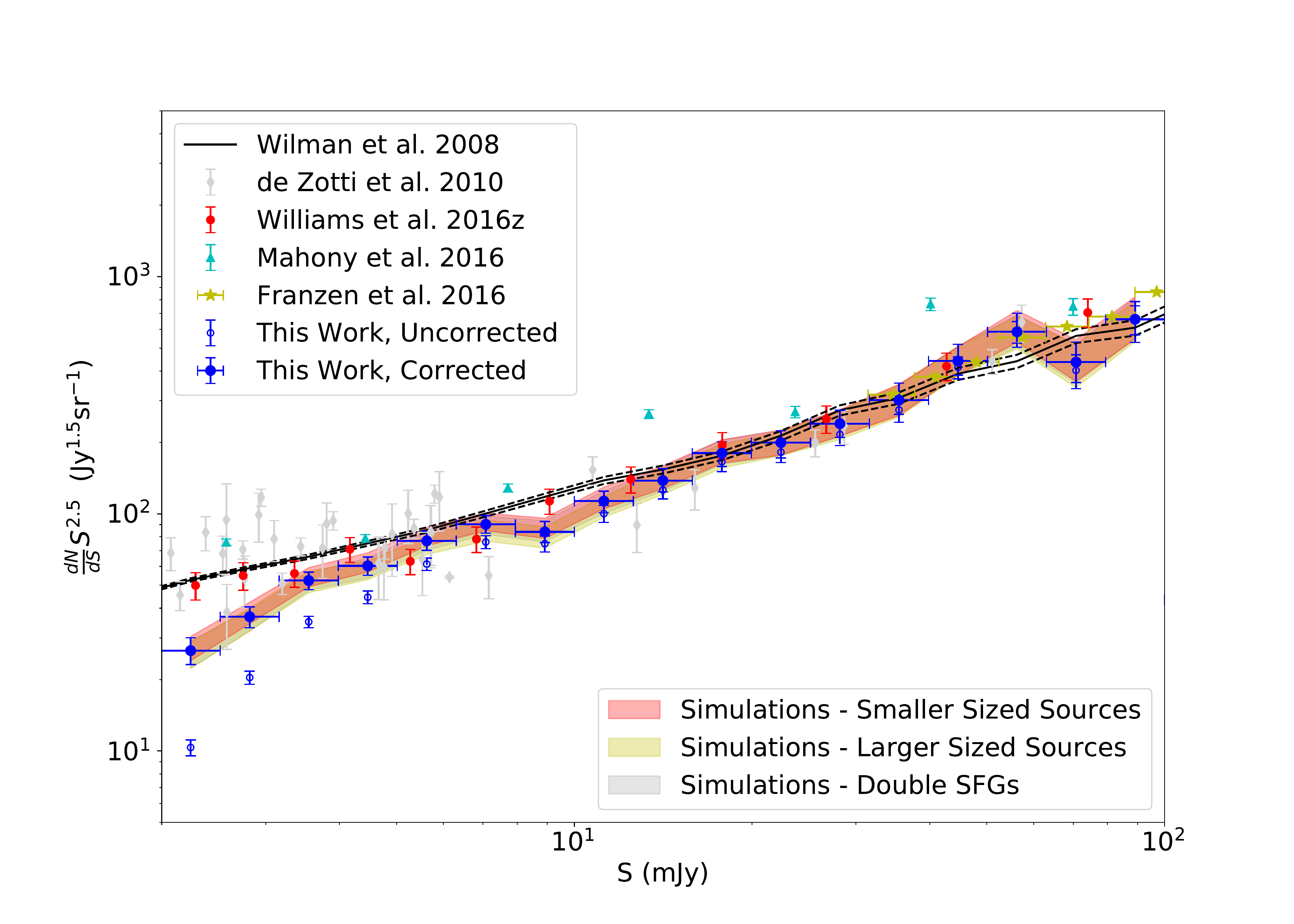}
\subcaption{Corrected using FDR correction and the completeness correction Method 2.}
\end{minipage}%
\caption{{Measured and corrected source counts of the catalogue and comparisons to previous studies. Shown are the measured source counts (open blue circles); corrected source counts using the FDR and completeness simulations (filled blue circles); source counts from \protect \cite{Franzen2016} (yellow stars); source counts from \protect \cite{Williams2016} (red circles); source counts from \protect \cite{deZotti2010} scaled to $\sim$150 MHz assuming $\alpha=0.7$ (grey diamonds); source counts from \protect \cite{Mahony2016} (cyan diamonds), and finally simulated source counts from the S$^3$ \protect \cite{Wilman2008} simulations (red line). The shaded red and yellow regions reflect the source count corrections using smaller/larger sources from Section \ref{sec:scmodel}. The results using the Method 1 completeness correction are shown in panel (a) and those of Method 2 are shown in panel (b).}}
\label{fig:sourcecounts}
\end{center}
\end{figure*}

\subsection{Final source counts}

The FDR and completeness correction factors were combined together with our value for the source counts multiplicatively. The errors were calculated {by} combining the errors of the source counts, FDR, and completeness corrections in quadrature. We used the catalogue without removing noise/artefacts as these may be removed by the masking described in Section \ref{sec:fdr} or should be accounted for by the FDR. The associated source counts and correction factors are given in Table~\ref{tab:scounts_method1} for Method 1 and Table~\ref{tab:scounts_method2} for Method 2. We did not include the corrected source counts at {flux densities below 2 mJy as these flux densities }had much larger correction factors that may be less reliable.

The source counts (in the range $\sim$2-100 mJy) can be seen in Figure \ref{fig:sourcecounts}, which includes the derived source counts with no correction factors applied (blue open circles) and with the corrections from Sections \ref{sec:fdr} and \ref{sec:completeness} applied (blue filled circles). Previous work shown for comparison includes that of \cite{Williams2016} whose observations of the Bo\"otes field, at $\sim 120 \mu$Jy beam$^{-1}$ rms at the centre, are comparably deep LOFAR observations of another blank field at $\sim$150~MHz (red circles). In addition, LOFAR direction-independent observations of the Lockman Hole field at 150 MHz are also plotted (cyan triangles\footnote{using the W90 data points from \cite{Mahony2016}}). {We also use source counts from recent observations} with the Murchison Widefield Array {\citep[MWA;][shown as yellow stars]{MWA}}. These observations \citep[][]{Franzen2016} cover 540 deg$^2$ using the MWA's wide field of view in only 12 hours of observations. Their source counts reach a depth of 30 mJy. Finally we also show the \cite{deZotti2010} catalogues of radio source counts from a collection of surveys at different frequencies from 150 MHz to 8.4 GHz (grey diamonds). We scale these to 144 MHz using $\alpha =0.7$. Finally we also make comparisons to the simulated catalogues of S$^3$ from \cite{Wilman2008}, using the 1.4 GHz {flux densities} and scaled to 144 MHz again using a spectral index of 0.7 (black line). 

    {The dominant effect in correcting our source counts arises from the completeness simulations. As can be seen in Figure \ref{fig:sc_corrections}, the completeness corrections are $\sim 1$ for {flux densities} $\gtrsim$4 mJy in Method 1 and for  $\gtrsim$10 mJy in Method 2. Below these {flux densities} we have large completeness corrections that increase to factors of $\sim 3$ in our faintest flux density bin ($\sim 2$ mJy). The FDR corrections, on the other hand, are approximately 1 even at low flux densities. At high flux densities we do not expect much of a completeness correction, however we note that there are still small completeness correction factors of $<$1 for Method 1. We attribute this to natural scatter in \textsc{PyBDSF} accurately recovering the source flux, especially when the source model is not inherently Gaussian. For compact galaxies (typically the faint, star-forming sources at low flux densities) we expect that these, when convolved with the beam, will be well modelled by Gaussian sources. The larger sources, typically the large ellipse components for FRI and FRII sources, may not be well approximated as a Gaussian source when the source is convolved with the beam. It is therefore important to note that the model chosen to inject sources into our simulation may affect our estimates of corrected source counts. For Method 2, the correction factors are typically larger than in Method 1, except at the faintest flux density bin. By construction, these correction factors must all be greater than 1.}

 {As can be seen in Figure \ref{fig:sourcecounts}, source counts from the LOFAR XMM-LSS field are consistent with previous observations at flux densities above $\sim 10$~mJy. These include the source counts from \cite{Williams2016}, \cite{Franzen2016}, \cite{deZotti2010}, and the simulations of \cite{Wilman2008}. At flux densities lower than 10 mJy the source counts recovered using Method 1 (Figure \ref{fig:sourcecounts}a) are typically lower than those from \cite{Williams2016} and \cite{Wilman2008}. They are however consistent with some of the scaled source counts from \cite{deZotti2010}. These points were from observations by \cite{Seymour2008} of the XMM 13 h field to 7.5 $\mu$Jy with VLA A and B configuration observations and with MERLIN observations. A and B configurations are the most extended configurations of the VLA, providing the best angular resolution, but may lose flux from extended emission. The faint source counts in \cite{Seymour2008} may therefore be a result of not detecting the total flux density from an extended source. The fact that our source counts are lower may suggest that there is an underlying correction that has not been fully accounted for. The source counts from \cite{Mahony2016} \citep[those which use the size relation from][]{Windhorst1990} however are typically larger than those measured in this work. }
 
{With our second method to account for completeness corrections (Figure \ref{fig:sourcecounts}b) our measured source counts are similar to previous measurements from \cite{Williams2016} down to $\sim$3mJy. We therefore note that our measured source
counts are affected by how completeness is defined and whether we take into account the bias we have in recovering sources due to the effects from the noise (such as Eddington bias) and source extraction software. We also mention that if the correction factors of flux from TGSS-ADR or the rescaled TGSS-ADR had been used, these source counts would have moved further from previous measurements. The fact that our source counts are similar to those from previous works when the flux densities have not been rescaled suggests that the flux densities measured by LOFAR may not need such large corrections. }

\subsection{Simulating the effects of the source count model}
\label{sec:scmodel}

We also investigate how our source model from S$^3$ may be affecting our source count corrections. To do this we ran three more sets of 100 simulations as in Section \ref{sec:completeness}. The three simulations are the same as in Section \ref{sec:completeness} but make different assumptions about the S$^3$ model that could be affecting the source counts we recover. These different models are
\begin{enumerate}
\item {We assume S$^3$ has too few SFGs and therefore include a larger number of SFGs in the input S$^3$ catalogue than in S$^3$ }.
\item {We assume S$^3$ was overestimating the sizes of sources and therefore use smaller sizes than suggested in S$^3$ }.
\item {We assume S$^3$ was underestimating the sizes of sources and therefore use larger sizes than suggested in S$^3$ }.
\end{enumerate}

{The first simulation tests the effect of having underestimated the number of SFGs in the S$^3$ simulations. We did this because evidence from recent radio surveys \citep[][]{Smolcic2017} suggests that S$^3$ }may be underpredicting the expected source counts; this may come from a lack of SFGs or RQQs in the models. We therefore tested for the effect that our chosen source-count model has on our corrected source counts. To do this, we modified our input S$^3$ catalogues by doubling the number of SFGs. For the second two simulations we modelled the effect of the sizes in S$^3$ being either under- or overestimated. To quantify this, we ran two more sets of simulations in which we first increased and then decreased the sizes from S$^3$ by {20\%}. Choosing 20\% is not a physically motivated value, however it is used to show how significant {changes in the intrinsic source size distribution} may affect the sources we recover. By making the sources larger, they are less likely to be detected as the peak {flux density per beam} decreases and thus may fall below the 5$\sigma$ threshold limit. These therefore produce larger corrections. For the smaller sources, these have brighter peak {flux densities per beam} and so smaller correction factors are expected.

Performing these three further tests allows us to understand how any uncertainty in our assumed model may propagate through to our estimates of the corrections needed to the source counts. The completeness corrections found using these models can be found in Tables \ref{tab:scounts_method1} and \ref{tab:scounts_method2} and are shown in Figure \ref{fig:sourcecounts} as the shaded regions. Although there are small differences in how the simulations affect the source counts, these are typically small and the values are consistent. This therefore suggests that our source count model is having little effect on our recovered source counts and we can be confident in the correction factors we have derived. However again we note that this is assuming a certain profile for the elliptical components in S$^3$ of constant surface brightness. Using different profiles, such as Gaussians, however may introduce differences in the source count correction factors suggested.

\section{Clustering in the XMM-LSS Field}
\label{sec:clustering}
{The depth and area of the LOFAR XMM-LSS field makes it ideal for investigating the large-scale clustering of radio sources. Therefore we investigated the clustering of radio sources in the XMM-LSS field and compared} this to the clustering from other radio observations. To do this, we followed the same method as in \cite{Hale2018} and calculated the angular two-point correlation function, $\omega(\theta)$. This quantifies the excess probability of the clustering of galaxies at certain angular scales compared to a random distribution of galaxies in the same field \citep[see e.g.][]{Peebles1980,Blake2002,Lindsay2014}. To construct the random catalogue of galaxies, we considered the varying rms across the image as in \cite{Hale2018}. We used the final data catalogue in which artefacts were removed and multiple components of sources had been merged.

We used an estimator \citep[][]{Landy1993} in which the number of pairs of galaxies within a certain angular separation in the data catalogue ($DD(\theta)$) are compared to the number of galaxy pairs in a random catalogue ($RR(\theta)$) as well as the number of galaxy pairs between the two catalogues ($DR(\theta)$). The equation for the angular two-point correlation function, $\omega(\theta)$ is given in Equation \ref{eq:tpcf} and is calculated using the package \textsc{TreeCorr} \citep[][]{Jarvis2015}.

\begin{equation}
\omega(\theta) = \frac{DD(\theta) -2DR(\theta) + RR(\theta)}{RR(\theta)}
\label{eq:tpcf}
\end{equation} 
We also construct bootstrapped errors before fitting a power-law model with an integral constraint {of $\omega(\theta) = A\theta^{-\gamma}$ \citep[see e.g.][]{RocheEales, Hale2018}.}

\begin{figure}
\begin{center}
\centering
\includegraphics[height=6.5cm]{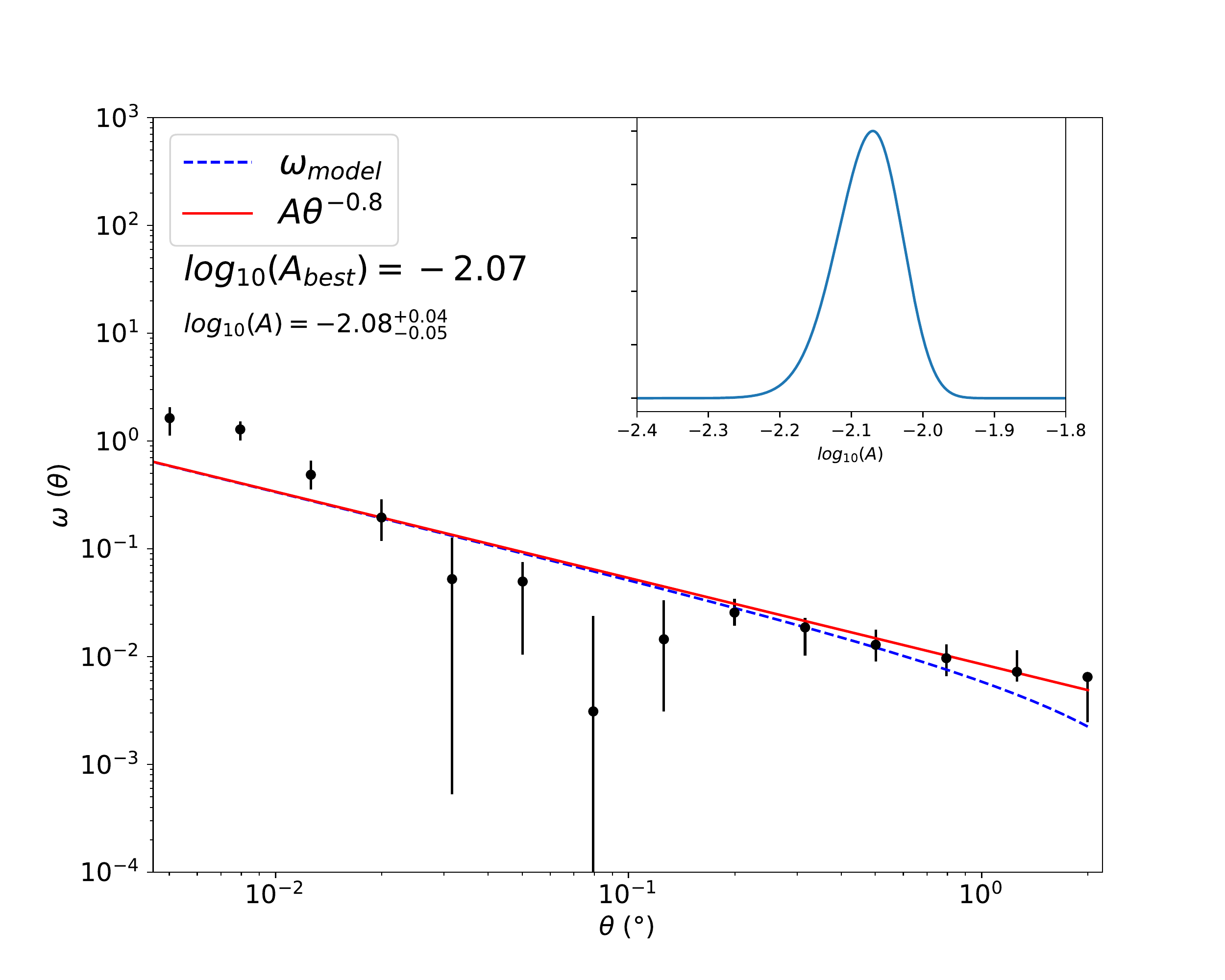}
\caption{Clustering in the XMM-LSS field for the merged source galaxy catalogue for a fixed slope of $\gamma$=0.8. The insets in the upper right-hand corner quantifies the probability distribution derived from the $\chi^2$ values  when fitting of the amplitude of the correlation function, $A$.}
\label{fig:tpcf_fit}
\end{center}
\end{figure}

Our clustering results can be seen in Figure \ref{fig:tpcf_fit} along with the fit to the correlation function as a one-parameter model where we fix $\gamma$=0.8. {This is a value predicted by theory \citep[][]{Peebles1974} and observed and used in many clustering observations \citep[see e.g.][]{Norberg2002,NVSSClustering,Hale2018}}. The fixed slope results appear to fit the large angular scales but appears to underestimate the clustering at small angular separations. This may suggest that we have an excess of clustering at small angular scales or that we have not fully joined all possible two-component sources. In the first case, this excess clustering could be related to the clustering of sources within the same dark matter halo \citep[the `1-halo' term;][]{CooraySheth,Yang,Zheng2007, Hatfield2016}. In the second case, although we attempted to match multiple components of the same source together, this excess clustering suggests that there may be components that should have been merged together {\citep[see e.g.][]{Overzier2003}}. This may {preferentially occur,} especially in regions where we did not have the multiwavelength coverage from \cite{HeywoodVLA} and \cite{Jarvis2013}, for example. {The ancillary data over those regions were useful in assessing whether components needed to be combined.}

\begin{figure}
\begin{center}
\centering
\includegraphics[width=8cm]{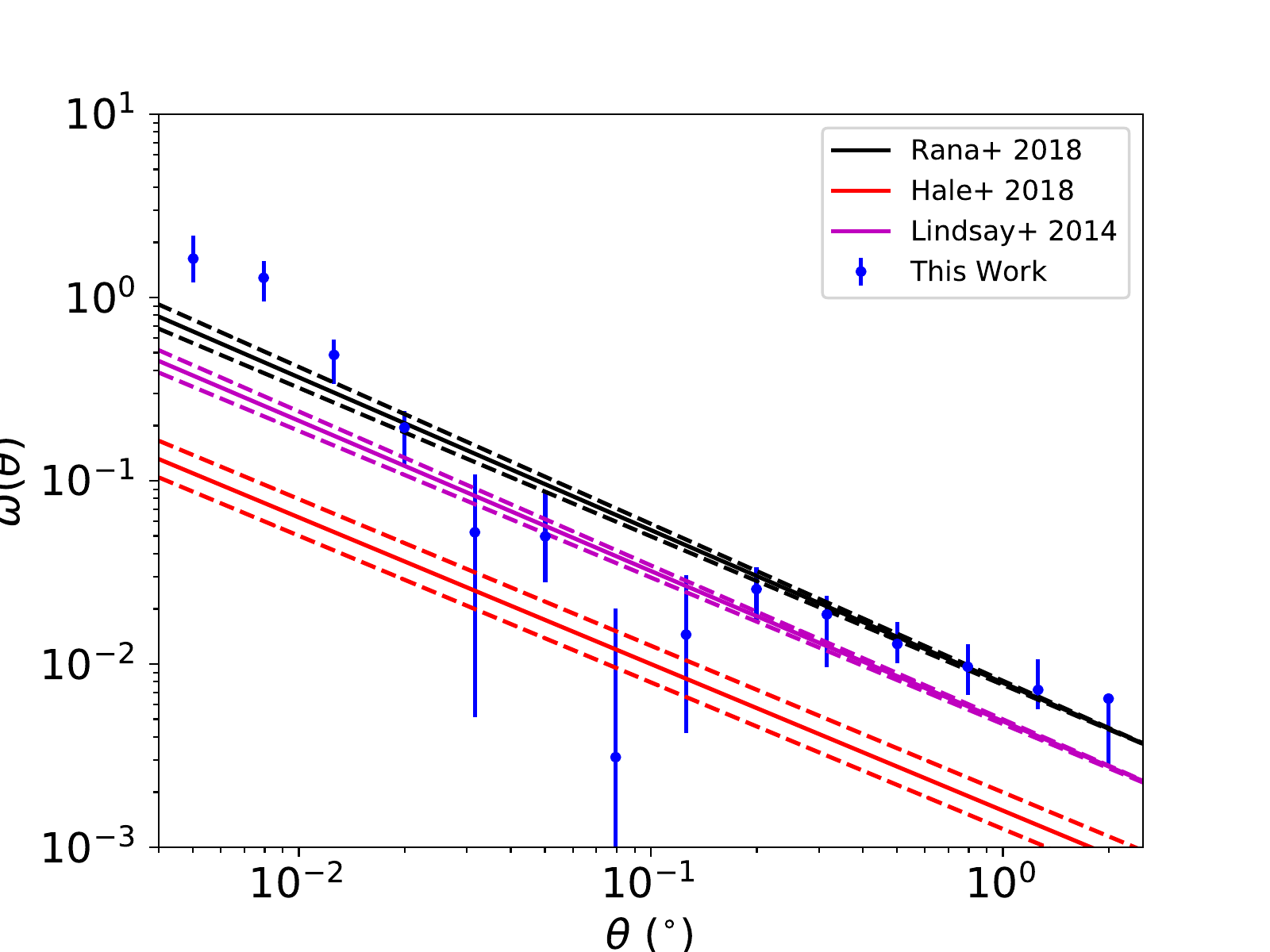}
\end{center}
\caption{Comparing the clustering of radio galaxies in this catalogue to other radio observations from \protect \cite{Hale2018}, \protect \cite{Lindsay2014}, and \protect \cite{Rana2018}.}
\label{fig:tpcf_comparison}
\end{figure}

In Figure \ref{fig:tpcf_comparison} we compare the clustering found in this work to a number of other studies: \cite{Hale2018}, who investigated the clustering of radio sources in the COSMOS field using the recently released 3 GHz Survey \citep[][]{Smolcic2017};  \cite{Rana2018}, who investigated the clustering of the TGSS-ADR survey \citep[][]{Intema2017}; and \cite{Lindsay2014}, who investigated the clustering of {FIRST \cite{Becker1995,Helfand2015}} sources in the GAMA \citep[][]{Driver2011} field, but also included a measurement for all sources that is shown in Figure \ref{fig:tpcf_comparison}. This allows us to compare to {surveys at both} differing frequencies and sensitivities. Surveys with better sensitivities typically observe smaller clustering amplitudes \citep[$A$; see e.g. ][ for relations between clustering amplitude and {flux density}]{Wilman2003}\footnote{Although this is not necessarily true as the redshift distribution of sources also needs to be taken into account; see e.g. \cite{Lindsay2014, Magliocchetti2017, Hale2018}}. This {is because} deeper observations are able to observe fainter objects, which typically have lower stellar masses and reside in less massive (and hence less-clustered) haloes. 

{ \cite{Rana2018} measured the clustering of radio sources at similar frequencies to that observed in this work, but at slightly higher {flux density} limits. The lowest {flux density} limit was at 50 mJy (and is shown in Figure \ref{fig:tpcf_comparison})}. In Figure \ref{fig:tpcf_comparison}, {we see that over a large range of angles} our clustering agrees with the analysis of \cite{Rana2018}. We, however, observe a small excess of clustering at angles $\lesssim 0.01 ^{\circ}$, which may be due to multi-component sources as mentioned earlier. At some intermediate angles (0.03 - 0.2$^{\circ}$) {there appears to be a lack of clustering on these scales}. This is currently unexplained but may be related to the rms properties of the field because if there is an effect that we have not taken into account (e.g. around bright sources) then this may be affecting our results. This is especially true on the largest scales, although our clustering measurements appear to have good agreement (within the errors) with \cite{Rana2018}. 

Comparing our results to those at different frequencies, our clustering measurements have a much {larger amplitude to those from \cite{Hale2018}}. {This is likely to reflect the higher flux density limits in this work compared to the JVLA 3 GHz COSMOS Survey \citep[][]{Smolcic2017}; this survey has an rms $\sim 2.3 \ \mu$Jy beam$^{-1}$ at 3 GHz, which corresponds to a limit of $\sim 13 \ \mu$Jy beam$^{-1}$ in \cite{Hale2018} as a 5.5$\sigma$ limit was used ($\sim$0.1 mJy beam$^{-1}$ at 144 MHz assuming $\alpha=0.7$)}. Therefore we are likely tracing higher mass haloes with higher clustering amplitudes. The work by \cite{Lindsay2014} is more similar to ours over the majority of angular scales, again with our two-point correlation function exhibiting an excess at small angles. For their work, \cite{Lindsay2014} used sources from FIRST {\citep{Becker1995,Helfand2015}}, which have a {flux density} limit of 1 mJy at 1.4 GHz ($\sim$ 5 mJy at 144 MHz) and so we expect our results to be similar to this.

However we are unable to just compare clustering amplitudes if we want to truly understand the typical haloes traced by these galaxies because we also need {to account for differences} in the redshifts of the sources being observed. Including redshift information is necessary as we are only investigating the projected clustering {and need} to investigate spatial clustering or use Limber inversion \citep[see e.g. ][]{Limber1953, Lindsay2014}  to relate the angular to spatial clustering to get a handle on the large-scale structure. We defer investigations into how these clustering measurements relate to halo masses and biases to {future work. We present} the clustering to show that we obtain clustering measurements similar to previous results and with similar slopes, $\gamma$. {This suggests that we have adequately understood} the noise properties in order to generate the random catalogue. 

\section{Conclusions}
\label{sec:conclusions}
In this paper we {have presented observations} with LOFAR of the XMM-LSS field at 144 MHz over an area of $\sim$ 27 deg$^2$. This field is at particularly low elevation for the LOFAR stations (centred at Dec = -4.5$^{\circ}$) {making it a challenging field to observe and produce images of a similar fidelity to those at higher declinations}. However, these observations will be useful on account of the wealth of ancillary data available over the area of the observation. Once cross-matched, this will be important for studies into the evolution of SFGs and AGN across cosmic time. We present images and catalogues from three four-hour pointings with LOFAR, producing a final image and catalogue with 3,044 sources. This reaches an rms depth of $\sim 0.28 $ mJy beam$^{-1}$ in the centre and an average rms of $\sim 0.4 $ mJy beam$^{-1}$. 

{Our observations show that the flux scale is marginally consistent with previous observations of this field at 150 MHz and has a relation of log($S_{LOFAR}$)/log($S_{TGSS}$) of $\sim$1.23; however, when compared to the MWA, a ratio of log($S_{LOFAR}$)/log($S_{MWA}$) of $\sim$1.05 is observed}. We also show that there are positional offsets typically constrained within $\sim 0.4$" when compared to, for example \cite{Becker1995} and \cite{Tasse2007}. This suggests good agreement with the positional information of known radio sources in the field, however offsets are important for optical identification and subsequent spectroscopic study with surveys such as WEAVE-LOFAR. This {accuracy should be achieved} through cross-matching of the data with ancillary near-IR and optical observations of the field using the wealth of ancillary data available. Finally, the spectral indices found by comparing the {flux densities observed with LOFAR to deep surveys at $\sim$1.5 GHz have $\alpha \sim 0.7$, consistent with typical radio continuum sources.} 

{We also presented the source counts over this field, making comparisons with previous observations at both 150 MHz and higher frequencies. Once corrected for false detection and completeness (with two differing methods) our source counts are, in general, in good agreement with previous measurements at high flux densities. However at some low flux densities, below $S\sim4$ mJy we observe slightly lower (a factor of $\sim 1-2$) source counts than previously observed at 150 MHz when we use a completeness correction method that considers bias from the noise and source extraction software. Using a completeness method that instead is similar to previous source count measurements at 150 MHz \citep{Williams2016}, we recover source counts similar to their work. We also investigated how the source model used may be affecting measurements of the source counts, looking at effects such as the assumed source sizes and the numbers of SFGs we assume in our models. {These assumptions make little difference to the source counts from our LOFAR image}. However, we note that differences in the assumed surface brightness profiles of sources have not been considered here and it may be important to understand those in order to see whether they have a large effect on the source counts we expect to obtain. }

{Finally we compare the clustering in this field to recent observations at 150 MHz \citep[][]{Rana2018} and to observations of clustering at other frequencies from \cite{Lindsay2014, Hale2018}.} Although differences in the clustering of these surveys exist as a result of the different sensitivities and redshift distributions, we observe that using a clustering slope of 0.8 \citep[as typically used in the literature, see e.g.][]{Peebles1974, Peebles1980, Blake2002, Hale2018} appears to be appropriate for the clustering measured in this work over a large variety of angles. We note that we observe a small excess of clustering at the smallest angular scales, which may suggest that some of our sources still may need to be combined into one source with other components or that we are distinguishing between the 1- and 2-halo terms. Combining the LOFAR sources in this paper with the deep multiwavelength data available to obtain redshifts for these sources will allow future work with LOFAR into the bias and halo mass of radio galaxy populations observed over a wide range of redshifts.

The XMM-LSS field is a key field to investigate, with a vast wealth of ancillary data that allows deep studies of galaxy properties and investigations into galaxy evolution. Deeper radio observations in this field \citep[e.g.][]{Jarvis2017} as well as multiwavelength observations such as those with WEAVE-LOFAR \citep[][]{Smith2016} will enhance our knowledge of these radio galaxies and our understanding of galaxy evolution in the
future. Our investigations have shown that keeping in mind differences in observation times and known declination effects, we are able to reach sensitivities with LOFAR at low declinations, which are expected compared to previous works \citep[][]{Williams2016, Hardcastle2016}. 

\begin{acknowledgements}
This paper is based on data obtained with the International LOFAR Telescope (ILT) under LCO\_015, LT5\_007 and LC7\_024. LOFAR \citep[][]{LOFAR} is the LOw Frequency ARray designed and constructed by ASTRON. It has observing, data processing, and data storage facilities in several countries, which are owned by various parties (each with their own funding sources) and are collectively operated by the ILT foundation under a joint scientific policy. The ILT resources have benefitted from the following recent major funding sources: CNRS-INSU, Observatoire de Paris, and Universit\'e d'Orl\'eans, France; BMBF, MIWF-NRW, MPG, Germany; Science Foundation Ireland (SFI), Department of Business, Enterprise and Innovation (DBEI), Ireland; NWO, The Netherlands; The Science and Technology Facilities Council, UK. CLH would like to acknowledge the support given from the Science and Technology Facilities Council (STFC) for their support to the first author through an STFC studentship. The data used in work was in part processed on the Dutch national e-infrastructure with the support of SURF Cooperative through grant e-infra 160022 \& 160152. This research has made use of the University of Hertfordshire high-performance computing facility (\url{http://uhhpc.herts.ac.uk/}) and the LOFAR-UK computing facility located at the University of Hertfordshire and supported by STFC [ST/P000096/1]. MJH acknowledges support from the UK Science and Technology Facilities Council [ST/M001008/1]. LKM acknowledges support from Oxford Hintze Centre for Astrophysical Surveys, which is funded through generous support from the Hintze Family Charitable Foundation. This publication arises from research partly funded by the John Fell Oxford University Press (OUP) Research Fund. IP acknowledges support from INAF through the PRIN SKA/CTA project `FORECaST'. PNB and JS are grateful for support from the UK STFC via grant ST/M001229/1. The Leiden LOFAR team acknowledges support from the ERC Advanced Investigator programme NewClusters 321271. RJvW acknowledges support from the VIDI research programme with project number 639.042.729, which is financed by the Netherlands Organisation for Scientific Research (NWO). This work made use of code from the package \textsc{astropy} \cite{astropy}.
\end{acknowledgements}

\bibliography{xmmlss_lofar}
\bibliographystyle{aa}

\begin{landscape}
\begin{table}
\centering
\begin{tabular}{cccccccccc}
\hline
$S$  & $S_c$  & N & $\frac{dN}{dS}S^{2.5}$ & FDR&Completeness  & Corrected $\frac{dN}{dS}S^{2.5}$ &Completeness& Completeness & Completeness \\[6pt]
(mJy) &  (mJy) &  & (Jy$^{1.5}$) & & Correction & (Jy$^{1.5}$)& SFG &  Small &  Large \\[6pt] \hline \hline \\
2.00 - 2.51 & 2.24 & 169$_{-13}^{+13}$ & $10.34_{-0.80}^{+0.80}$ & $0.98 \pm 0.01$ & 2.84$_{-0.32}^{+0.76}$ & $28.81_{-3.91}^{+8.03}$& 2.90$_{-0.40}^{+0.54}$& 3.11$_{-0.42}^{+0.57}$& 2.84$_{-0.40}^{+0.47}$\\[6pt] 
 2.51 - 3.16 & 2.82 & 236$_{-15}^{+15}$ & $20.40_{-1.30}^{+1.30}$ & $0.98 \pm 0.01$ & 1.83$_{-0.23}^{+0.27}$ & $36.51_{-5.11}^{+5.94}$& 1.73$_{-0.22}^{+0.25}$& 1.91$_{-0.29}^{+0.33}$& 1.69$_{-0.18}^{+0.31}$\\[6pt] 
 3.16 - 3.98 & 3.55 & 287$_{-16}^{+16}$ & $35.04_{-1.95}^{+1.95}$ & $0.98 \pm 0.01$ & 1.29$_{-0.16}^{+0.16}$ & $44.48_{-6.01}^{+6.03}$& 1.31$_{-0.19}^{+0.14}$& 1.39$_{-0.18}^{+0.17}$& 1.25$_{-0.17}^{+0.22}$\\[6pt] 
 3.98 - 5.01 & 4.47 & 258$_{-16}^{+16}$ & $44.50_{-2.76}^{+2.76}$ & $0.98 \pm 0.01$ & 1.08$_{-0.16}^{+0.16}$ & $47.35_{-7.53}^{+7.69}$& 1.02$_{-0.11}^{+0.14}$& 1.08$_{-0.12}^{+0.14}$& 1.02$_{-0.16}^{+0.15}$\\[6pt] 
 5.01 - 6.31 & 5.62 & 252$_{-15}^{+15}$ & $61.39_{-3.65}^{+3.65}$ & $0.98 \pm 0.01$ & 1.02$_{-0.11}^{+0.18}$ & $61.41_{-7.72}^{+11.35}$& 1.02$_{-0.11}^{+0.15}$& 1.07$_{-0.11}^{+0.16}$& 0.97$_{-0.12}^{+0.15}$\\[6pt] 
 6.31 - 7.94 & 7.08 & 221$_{-14}^{+14}$ & $76.05_{-4.82}^{+4.82}$ & $0.98 \pm 0.01$ & 0.94$_{-0.13}^{+0.11}$ & $70.15_{-10.51}^{+9.48}$& 0.93$_{-0.17}^{+0.18}$& 1.02$_{-0.16}^{+0.16}$& 0.89$_{-0.10}^{+0.17}$\\[6pt] 
 7.94 - 10.00 & 8.91 & 154$_{-12}^{+12}$ & $74.85_{-5.83}^{+5.83}$ & $0.98 \pm 0.01$ & 0.92$_{-0.13}^{+0.16}$ & $67.43_{-11.07}^{+12.95}$& 0.92$_{-0.14}^{+0.16}$& 0.93$_{-0.11}^{+0.17}$& 0.92$_{-0.16}^{+0.08}$\\[6pt] 
 10.00 - 12.59 & 11.22 & 146$_{-12}^{+12}$ & $100.24_{-8.24}^{+8.24}$ & $0.99 \pm 0.01$ & 0.93$_{-0.15}^{+0.18}$ & $92.43_{-16.55}^{+19.22}$& 0.92$_{-0.15}^{+0.16}$& 0.93$_{-0.14}^{+0.20}$& 0.91$_{-0.12}^{+0.15}$\\[6pt] 
 12.59 - 15.85 & 14.13 & 130$_{-11}^{+11}$ & $126.08_{-10.67}^{+10.67}$ &1.00 & 0.95$_{-0.13}^{+0.16}$ & $119.61_{-19.23}^{+22.45}$& 0.97$_{-0.11}^{+0.16}$& 1.00$_{-0.17}^{+0.20}$& 0.97$_{-0.16}^{+0.13}$\\[6pt] 
 15.85 - 19.95 & 17.78 & 121$_{-11}^{+11}$ & $165.76_{-15.07}^{+15.07}$ &1.00 & 0.93$_{-0.13}^{+0.12}$ & $154.32_{-26.00}^{+24.35}$& 0.88$_{-0.12}^{+0.17}$& 0.93$_{-0.16}^{+0.19}$& 0.91$_{-0.14}^{+0.15}$\\[6pt] 
 19.95 - 25.12 & 22.39 & 94$_{-9}^{+9}$ & $181.90_{-17.42}^{+17.42}$ & $0.99 \pm 0.01$ & 0.95$_{-0.17}^{+0.24}$ & $171.41_{-34.03}^{+45.51}$& 0.96$_{-0.17}^{+0.18}$& 1.00$_{-0.19}^{+0.17}$& 0.98$_{-0.17}^{+0.18}$\\[6pt] 
 25.12 - 31.62 & 28.18 & 79$_{-8}^{+8}$ & $215.94_{-21.87}^{+21.87}$ &1.00 & 0.83$_{-0.17}^{+0.17}$ & $178.31_{-41.62}^{+41.73}$& 0.88$_{-0.16}^{+0.18}$& 0.86$_{-0.16}^{+0.14}$& 0.84$_{-0.12}^{+0.16}$\\[6pt] 
 31.62 - 39.81 & 35.48 & 71$_{-8}^{+8}$ & $274.13_{-30.89}^{+30.89}$ &1.00 & 0.94$_{-0.19}^{+0.18}$ & $257.50_{-59.46}^{+56.14}$& 0.89$_{-0.16}^{+0.22}$& 0.90$_{-0.17}^{+0.15}$& 0.89$_{-0.17}^{+0.17}$\\[6pt] 
 39.81 - 50.12 & 44.67 & 76$_{-8}^{+8}$ & $414.49_{-43.63}^{+43.63}$ & $0.99 \pm 0.01$ & 0.91$_{-0.17}^{+0.16}$ & $371.98_{-80.09}^{+77.26}$& 0.84$_{-0.19}^{+0.17}$& 0.85$_{-0.16}^{+0.15}$& 0.81$_{-0.12}^{+0.19}$\\[6pt] 
 50.12 - 63.10 & 56.23 & 76$_{-8}^{+8}$ & $585.48_{-61.63}^{+61.63}$ &1.00 & 0.86$_{-0.19}^{+0.21}$ & $501.84_{-123.40}^{+136.13}$& 0.85$_{-0.18}^{+0.28}$& 0.89$_{-0.19}^{+0.29}$& 0.88$_{-0.18}^{+0.25}$\\[6pt] 
 63.10 - 79.43 & 70.79 & 37$_{-6}^{+6}$ & $402.63_{-65.29}^{+65.29}$ &1.00 & 0.85$_{-0.17}^{+0.25}$ & $342.90_{-88.26}^{+115.04}$& 0.81$_{-0.17}^{+0.26}$& 0.79$_{-0.19}^{+0.21}$& 0.76$_{-0.19}^{+0.24}$\\[6pt] 
 79.43 - 100.00 & 89.13 & 43$_{-6}^{+6}$ & $660.95_{-92.23}^{+92.23}$ &1.00 & 0.91$_{-0.24}^{+0.40}$ & $600.86_{-180.84}^{+274.99}$& 0.88$_{-0.28}^{+0.46}$& 0.93$_{-0.26}^{+0.45}$& 0.91$_{-0.22}^{+0.35}$\\[6pt] 
 \end{tabular}
\caption{{Calculated source counts and associated errors for our observations of the XMM-LSS field. This includes information on the flux densities of the bin, raw counts (and associated error) in the given flux density bin, source counts, and correction factors (Method 1) as described in Sections \ref{sec:fdr} and \ref{sec:completeness}. Finally the corrected source counts using these corrections are presented, as well as estimates for the completeness corrections using the models in Section \ref{sec:scmodel}.}}
\label{tab:scounts_method1}
\end{table}
\end{landscape}

\pagebreak

\begin{landscape}
\begin{table}
\centering
\begin{tabular}{cccccccccc}
\hline
$S$  & $S_c$  & N & $\frac{dN}{dS}S^{2.5}$ & FDR&Completeness  & Corrected $\frac{dN}{dS}S^{2.5}$ &Completeness& Completeness & Completeness \\[6pt]
(mJy) &  (mJy) &  & (Jy$^{1.5}$) & & Correction & (Jy$^{1.5}$)& SFG &  Small &  Large \\[6pt] \hline \hline \\
2.00 - 2.51 & 2.24 & 169$_{-13}^{+13}$ & $10.34_{-0.80}^{+0.80}$ & $0.98 \pm 0.01$ & 2.61$_{-0.26}^{+0.28}$ & $26.49_{-3.35}^{+3.50}$& 2.49$_{-0.23}^{+0.31}$& 2.67$_{-0.23}^{+0.25}$& 2.51$_{-0.24}^{+0.25}$\\[6pt]
 2.51 - 3.16 & 2.82 & 236$_{-15}^{+15}$ & $20.40_{-1.30}^{+1.30}$ & $0.98 \pm 0.01$ & 1.84$_{-0.15}^{+0.14}$ & $36.81_{-3.74}^{+3.71}$& 1.77$_{-0.12}^{+0.16}$& 1.92$_{-0.16}^{+0.17}$& 1.78$_{-0.13}^{+0.19}$\\[6pt]
 3.16 - 3.98 & 3.55 & 287$_{-16}^{+16}$ & $35.04_{-1.95}^{+1.95}$ & $0.98 \pm 0.01$ & 1.52$_{-0.10}^{+0.10}$ & $52.32_{-4.49}^{+4.43}$& 1.51$_{-0.10}^{+0.09}$& 1.56$_{-0.11}^{+0.14}$& 1.49$_{-0.11}^{+0.15}$\\[6pt]
 3.98 - 5.01 & 4.47 & 258$_{-16}^{+16}$ & $44.50_{-2.76}^{+2.76}$ & $0.98 \pm 0.01$ & 1.38$_{-0.08}^{+0.09}$ & $60.33_{-5.27}^{+5.45}$& 1.36$_{-0.11}^{+0.09}$& 1.42$_{-0.10}^{+0.11}$& 1.33$_{-0.09}^{+0.07}$\\[6pt]
 5.01 - 6.31 & 5.62 & 252$_{-15}^{+15}$ & $61.39_{-3.65}^{+3.65}$ & $0.98 \pm 0.01$ & 1.28$_{-0.08}^{+0.10}$ & $76.92_{-6.75}^{+7.60}$& 1.27$_{-0.07}^{+0.10}$& 1.32$_{-0.07}^{+0.11}$& 1.23$_{-0.07}^{+0.07}$\\[6pt]
 6.31 - 7.94 & 7.08 & 221$_{-14}^{+14}$ & $76.05_{-4.82}^{+4.82}$ & $0.98 \pm 0.01$ & 1.21$_{-0.06}^{+0.08}$ & $90.24_{-7.41}^{+8.09}$& 1.20$_{-0.06}^{+0.09}$& 1.24$_{-0.06}^{+0.09}$& 1.15$_{-0.08}^{+0.09}$\\[6pt]
 7.94 - 10.00 & 8.91 & 154$_{-12}^{+12}$ & $74.85_{-5.83}^{+5.83}$ & $0.98 \pm 0.01$ & 1.14$_{-0.07}^{+0.08}$ & $83.92_{-8.34}^{+8.94}$& 1.15$_{-0.07}^{+0.09}$& 1.18$_{-0.07}^{+0.09}$& 1.09$_{-0.07}^{+0.06}$\\[6pt]
 10.00 - 12.59 & 11.22 & 146$_{-12}^{+12}$ & $100.24_{-8.24}^{+8.24}$ & $0.99 \pm 0.01$ & 1.14$_{-0.08}^{+0.06}$ & $113.14_{-12.11}^{+11.35}$& 1.12$_{-0.07}^{+0.09}$& 1.18$_{-0.08}^{+0.08}$& 1.08$_{-0.04}^{+0.06}$\\[6pt] 
 12.59 - 15.85 & 14.13 & 130$_{-11}^{+11}$ & $126.08_{-10.67}^{+10.67}$ &1.00 & 1.10$_{-0.06}^{+0.10}$ & $138.09_{-14.15}^{+17.05}$& 1.11$_{-0.08}^{+0.09}$& 1.14$_{-0.09}^{+0.09}$& 1.08$_{-0.08}^{+0.11}$\\[6pt]
 15.85 - 19.95 & 17.78 & 121$_{-11}^{+11}$ & $165.76_{-15.07}^{+15.07}$ &1.00 & 1.09$_{-0.09}^{+0.08}$ & $180.18_{-21.82}^{+21.04}$& 1.09$_{-0.06}^{+0.11}$& 1.11$_{-0.07}^{+0.09}$& 1.05$_{-0.05}^{+0.08}$\\[6pt]
 19.95 - 25.12 & 22.39 & 94$_{-9}^{+9}$ & $181.90_{-17.42}^{+17.42}$ & $0.99 \pm 0.01$ & 1.11$_{-0.11}^{+0.08}$ & $199.62_{-27.49}^{+24.37}$& 1.09$_{-0.06}^{+0.12}$& 1.11$_{-0.07}^{+0.09}$& 1.10$_{-0.06}^{+0.07}$\\[6pt]
 25.12 - 31.62 & 28.18 & 79$_{-8}^{+8}$ & $215.94_{-21.87}^{+21.87}$ &1.00 & 1.11$_{-0.08}^{+0.10}$ & $239.93_{-29.95}^{+33.15}$& 1.11$_{-0.07}^{+0.11}$& 1.12$_{-0.08}^{+0.10}$& 1.11$_{-0.11}^{+0.10}$\\[6pt]
 31.62 - 39.81 & 35.48 & 71$_{-8}^{+8}$ & $274.13_{-30.89}^{+30.89}$ &1.00 & 1.10$_{-0.07}^{+0.15}$ & $301.54_{-38.87}^{+53.91}$& 1.10$_{-0.06}^{+0.10}$& 1.11$_{-0.11}^{+0.12}$& 1.08$_{-0.08}^{+0.15}$\\[6pt]
 39.81 - 50.12 & 44.67 & 76$_{-8}^{+8}$ & $414.49_{-43.63}^{+43.63}$ & $0.99 \pm 0.01$ & 1.08$_{-0.08}^{+0.15}$ & $440.65_{-56.35}^{+78.40}$& 1.08$_{-0.08}^{+0.13}$& 1.08$_{-0.08}^{+0.11}$& 1.06$_{-0.06}^{+0.13}$\\[6pt]
 50.12 - 63.10 & 56.23 & 76$_{-8}^{+8}$ & $585.48_{-61.63}^{+61.63}$ &1.00 & 1.00$_{-0.09}^{+0.17}$ & $585.48_{-81.99}^{+115.41}$& 1.00$_{-0.09}^{+0.14}$& 1.06$_{-0.13}^{+0.14}$& 1.00$_{-0.11}^{+0.13}$\\[6pt]
 63.10 - 79.43 & 70.79 & 37$_{-6}^{+6}$ & $402.63_{-65.29}^{+65.29}$ &1.00 & 1.08$_{-0.08}^{+0.15}$ & $436.18_{-78.29}^{+94.14}$& 1.08$_{-0.08}^{+0.12}$& 1.10$_{-0.10}^{+0.17}$& 1.07$_{-0.14}^{+0.13}$\\[6pt]
 79.43 - 100.00 & 89.13 & 43$_{-6}^{+6}$ & $660.95_{-92.23}^{+92.23}$ &1.00 & 1.00$_{-0.14}^{+0.13}$ & $660.95_{-132.82}^{+124.41}$& 1.00$_{-0.11}^{+0.18}$& 1.00$_{-0.10}^{+0.20}$& 1.00$_{-0.13}^{+0.13}$\\[6pt]
 \end{tabular}
\caption{{Calculated source counts and associated errors for our observations of the XMM-LSS field. This includes information on the flux densities of the bin, raw counts (and associated error) in the given flux density bin, source counts, and correction factors (Method 2) as described in Sections \ref{sec:fdr} and \ref{sec:completeness}. Finally the corrected source counts using these corrections are presented, as well as estimates for the completeness corrections using the models in Section \ref{sec:scmodel}.}}
\label{tab:scounts_method2}
\end{table}

\end{landscape}

\begin{appendix}
\onecolumn
\section{Large sources in the field}
\label{sec:large_sources}

Images of some of the largest sources in the field are presented. For plotting purposes, these have all been presented with a flux scale of -3$\sigma_{local}$ and 10$\sigma_{local}$, where $\sigma_{local}$ is the rms at the location of the source.

\noindent \begin{minipage}[b]{0.2\textwidth}
\centering
\includegraphics[height=3.8cm]{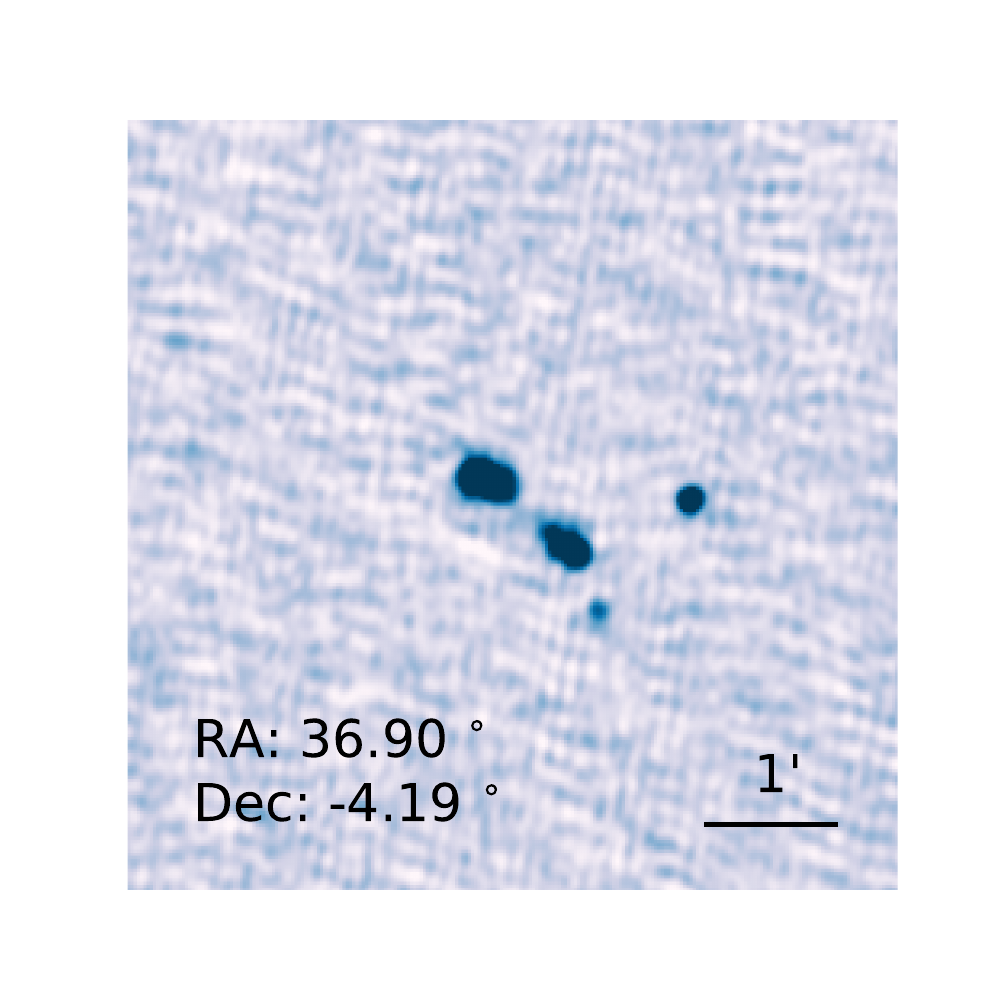}
\end{minipage}% 
\begin{minipage}[b]{0.2\textwidth}
\centering
\includegraphics[height=3.8cm]{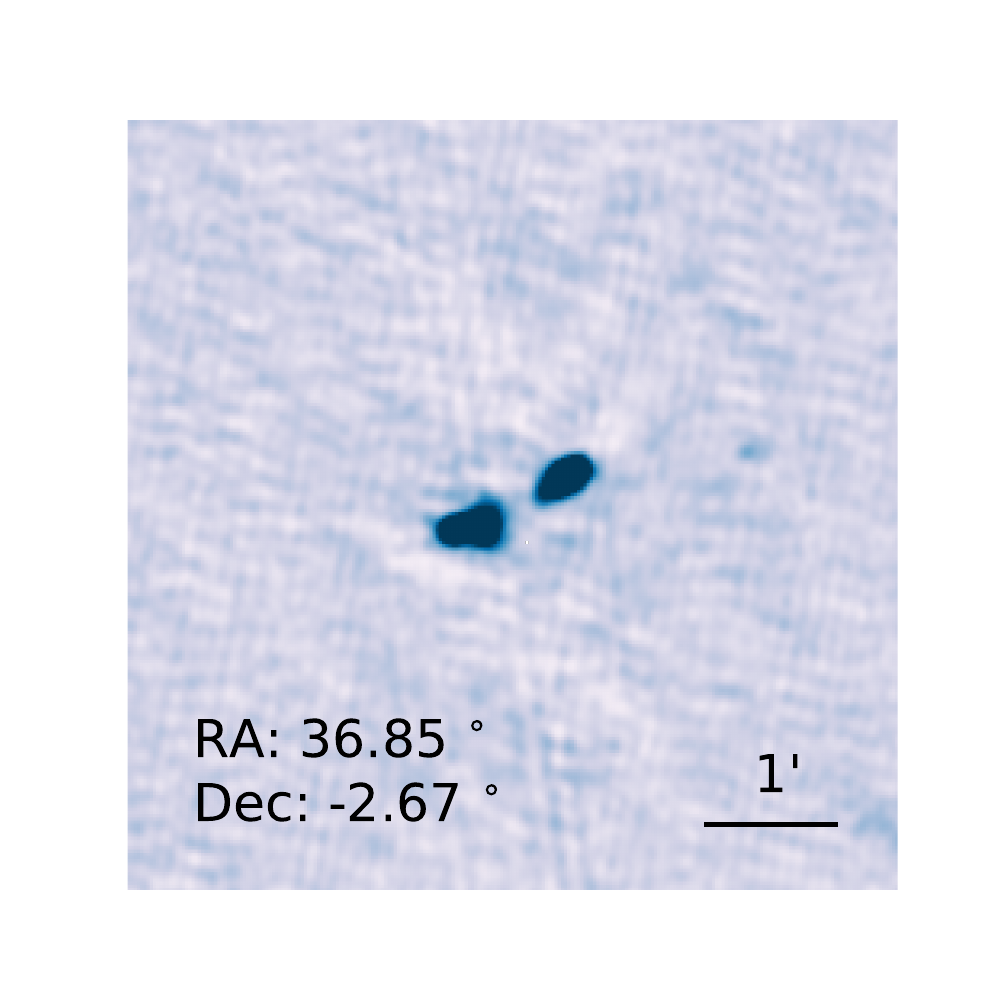}
\end{minipage}% 
\begin{minipage}[b]{0.2\textwidth}
\centering
\includegraphics[height=3.8cm]{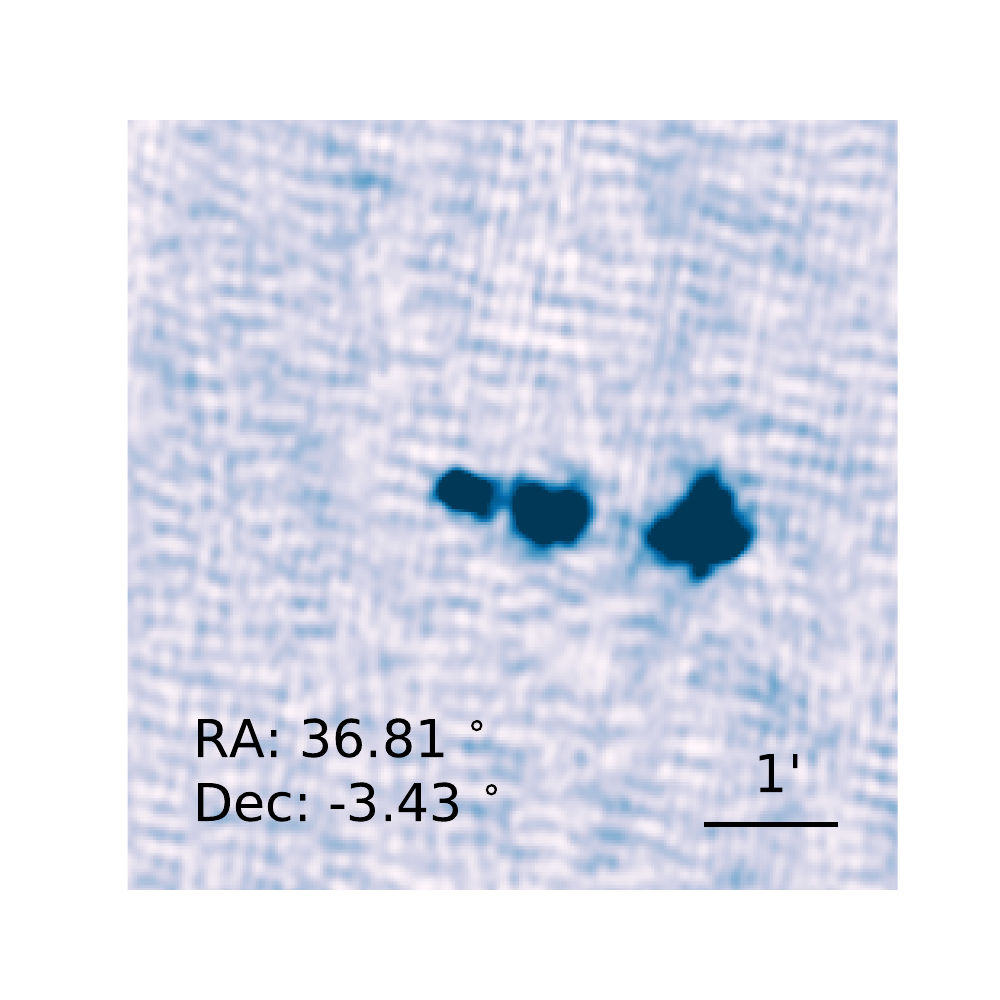}
\end{minipage}% 
\begin{minipage}[b]{0.2\textwidth}
\centering
\includegraphics[height=3.8cm]{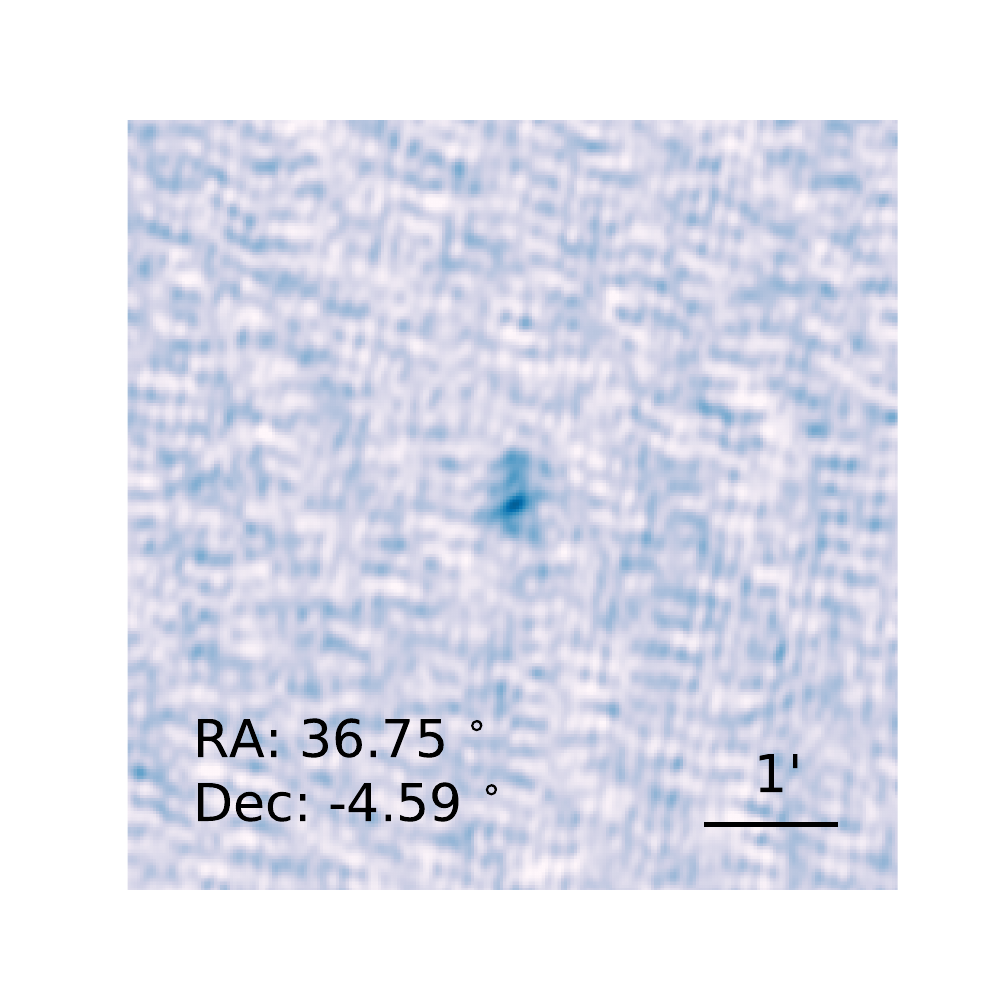}
\end{minipage}% 
\begin{minipage}[b]{0.2\textwidth}
\centering
\includegraphics[height=3.8cm]{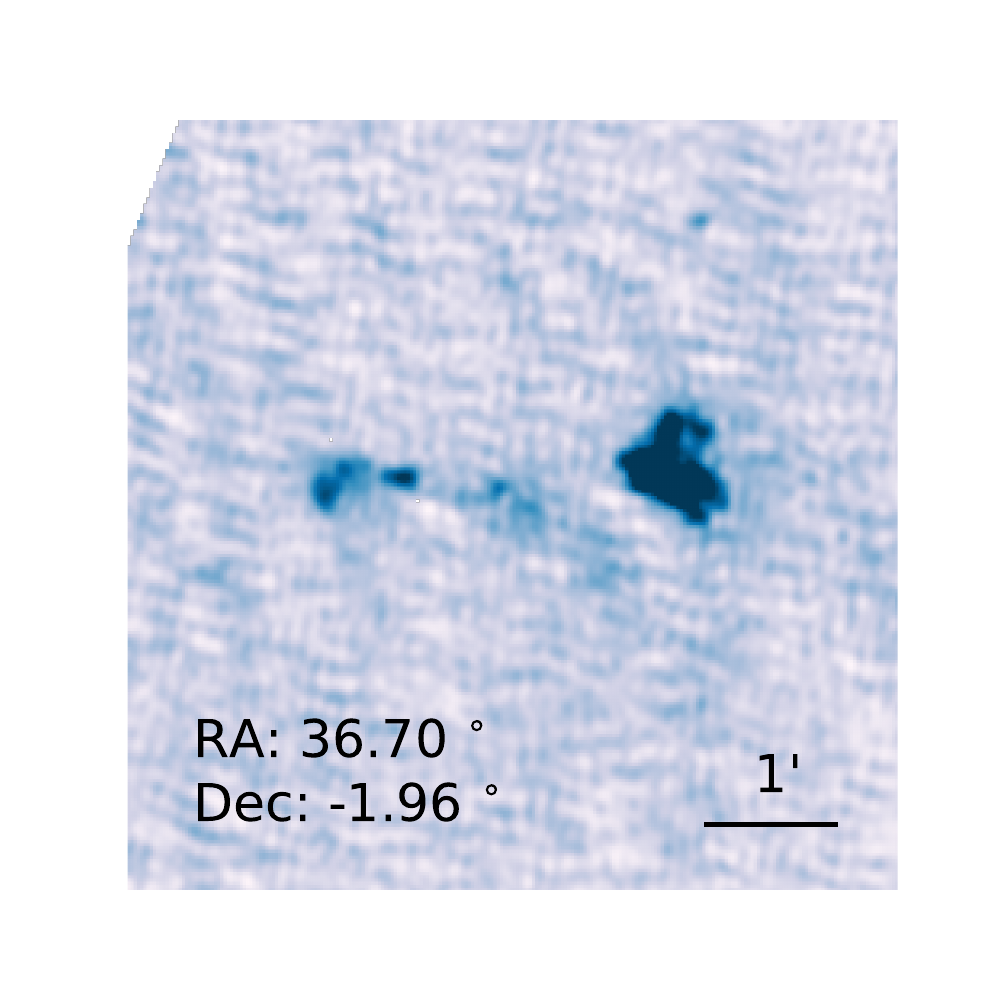}
\end{minipage}% 
\newline
\begin{minipage}[b]{0.2\textwidth}
\centering
\includegraphics[height=3.8cm]{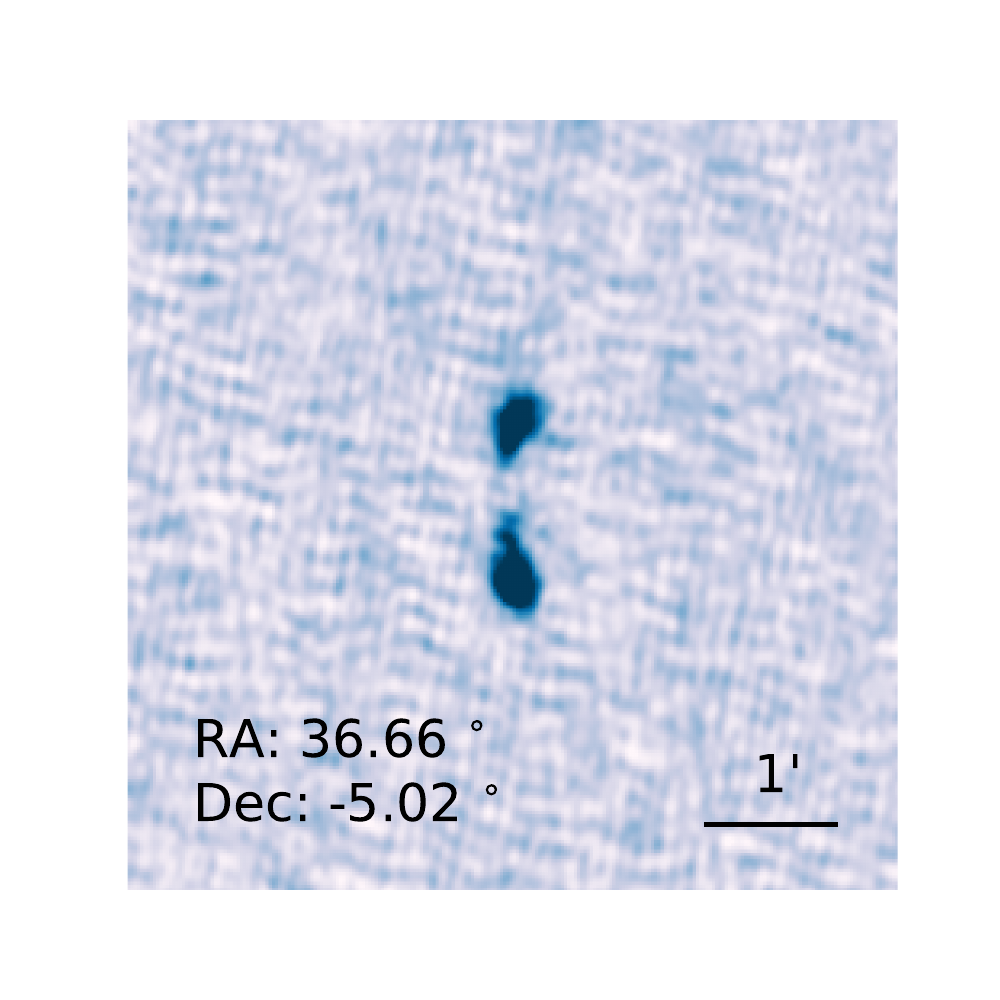}
\end{minipage}% 
\begin{minipage}[b]{0.2\textwidth}
\centering
\includegraphics[height=3.8cm]{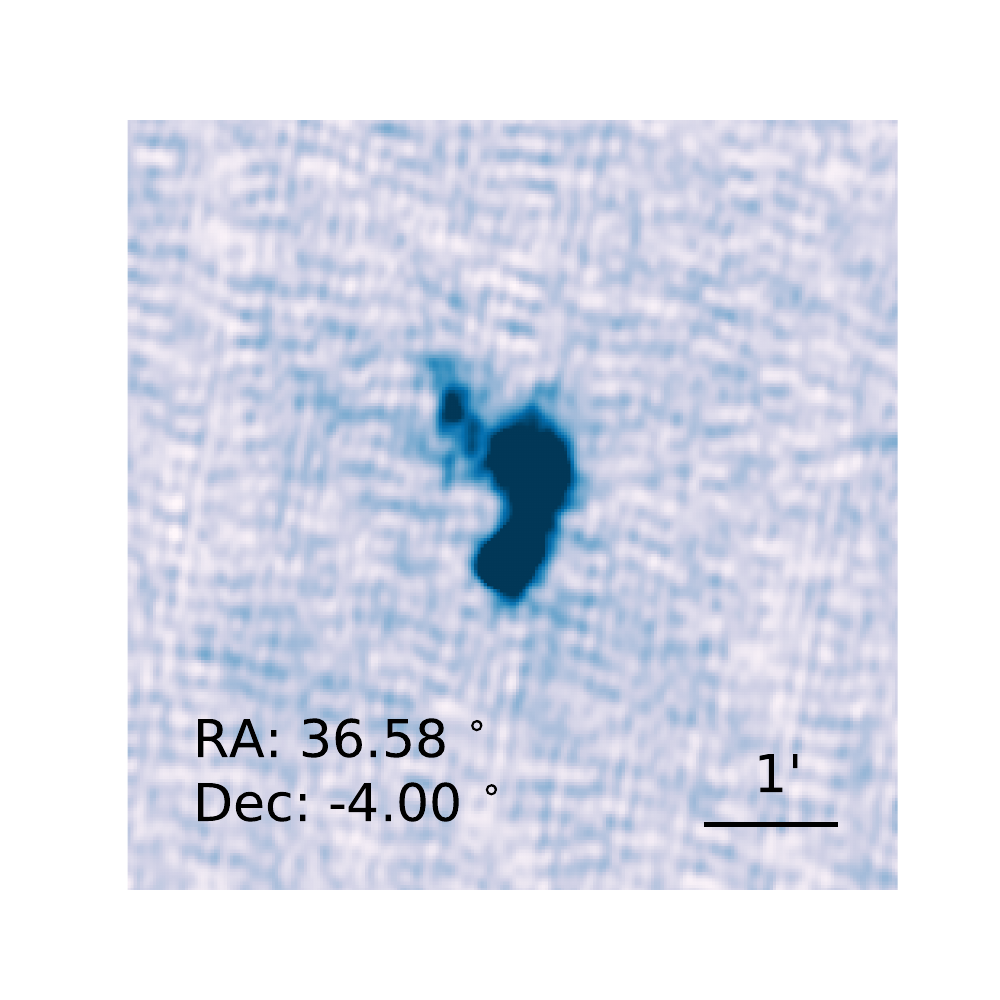}
\end{minipage}% 
\begin{minipage}[b]{0.2\textwidth}
\centering
\includegraphics[height=3.8cm]{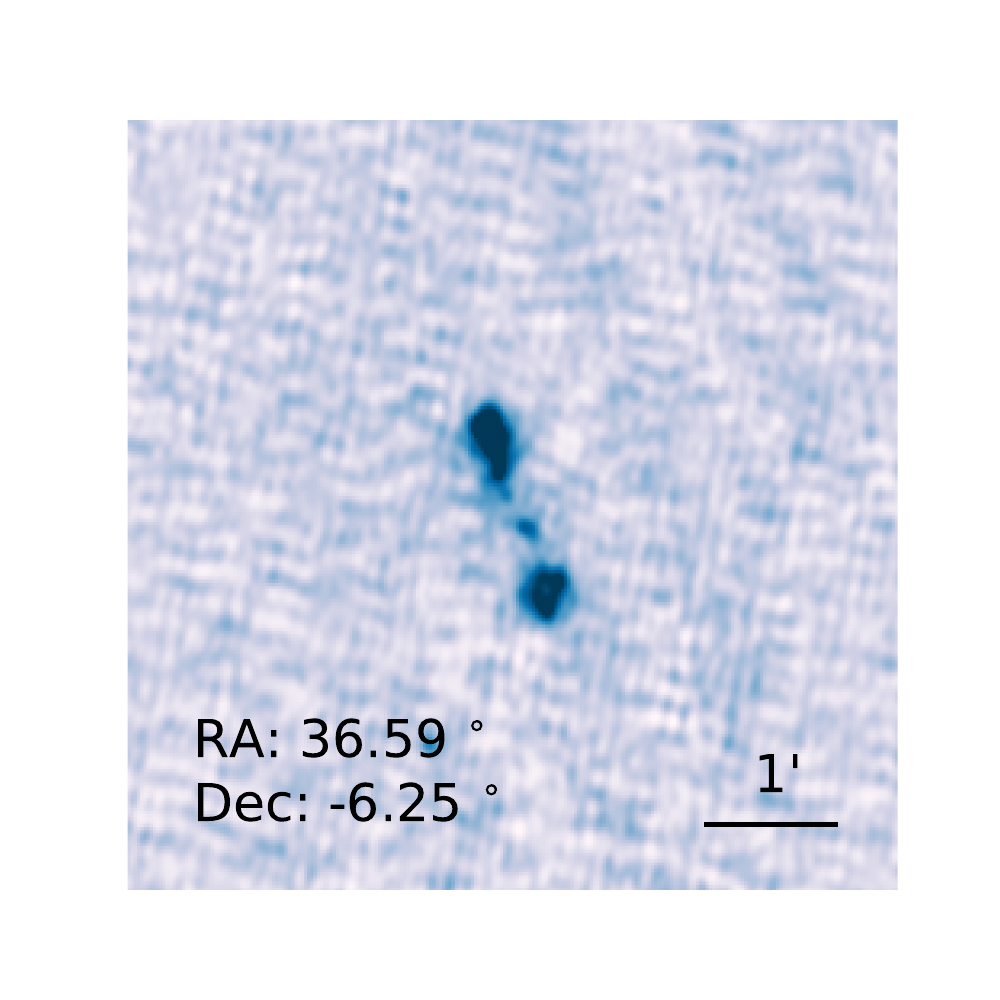}
\end{minipage}% 
\begin{minipage}[b]{0.2\textwidth}
\centering
\includegraphics[height=3.8cm]{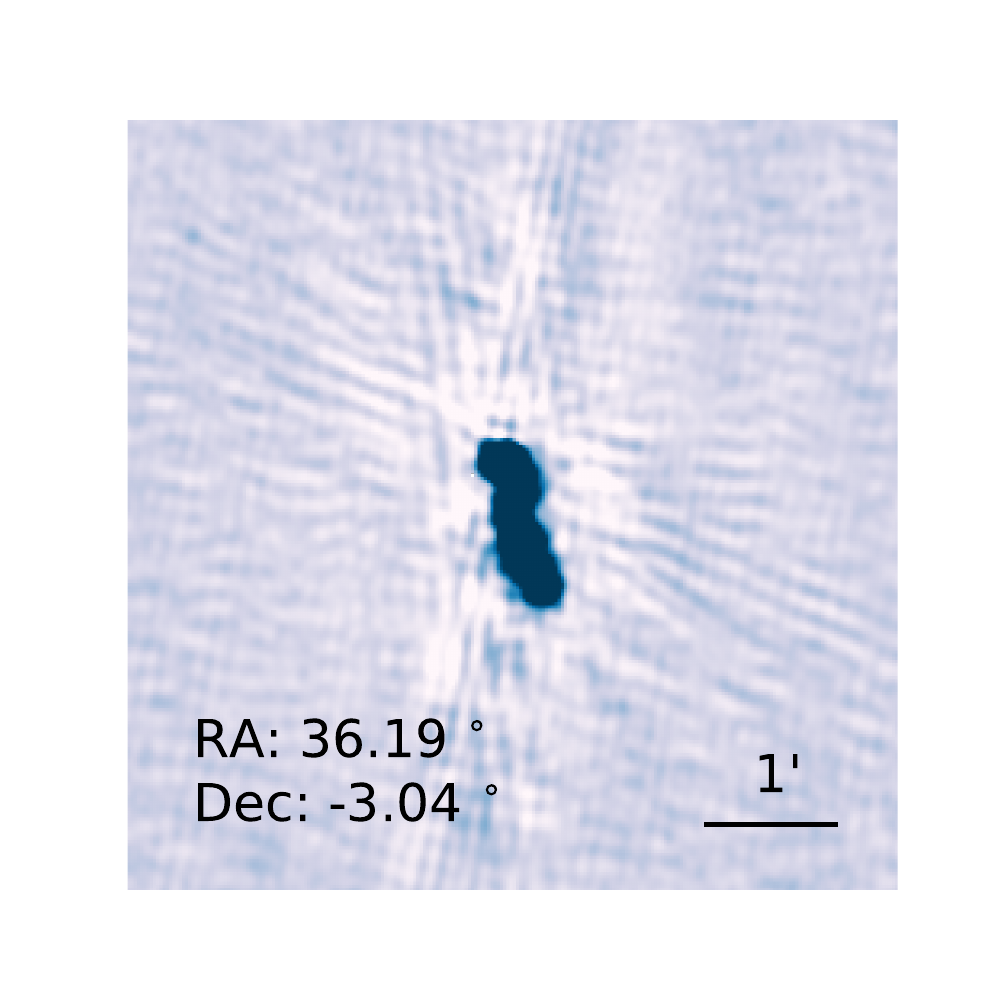}
\end{minipage}% 
\begin{minipage}[b]{0.2\textwidth}
\centering
\includegraphics[height=3.8cm]{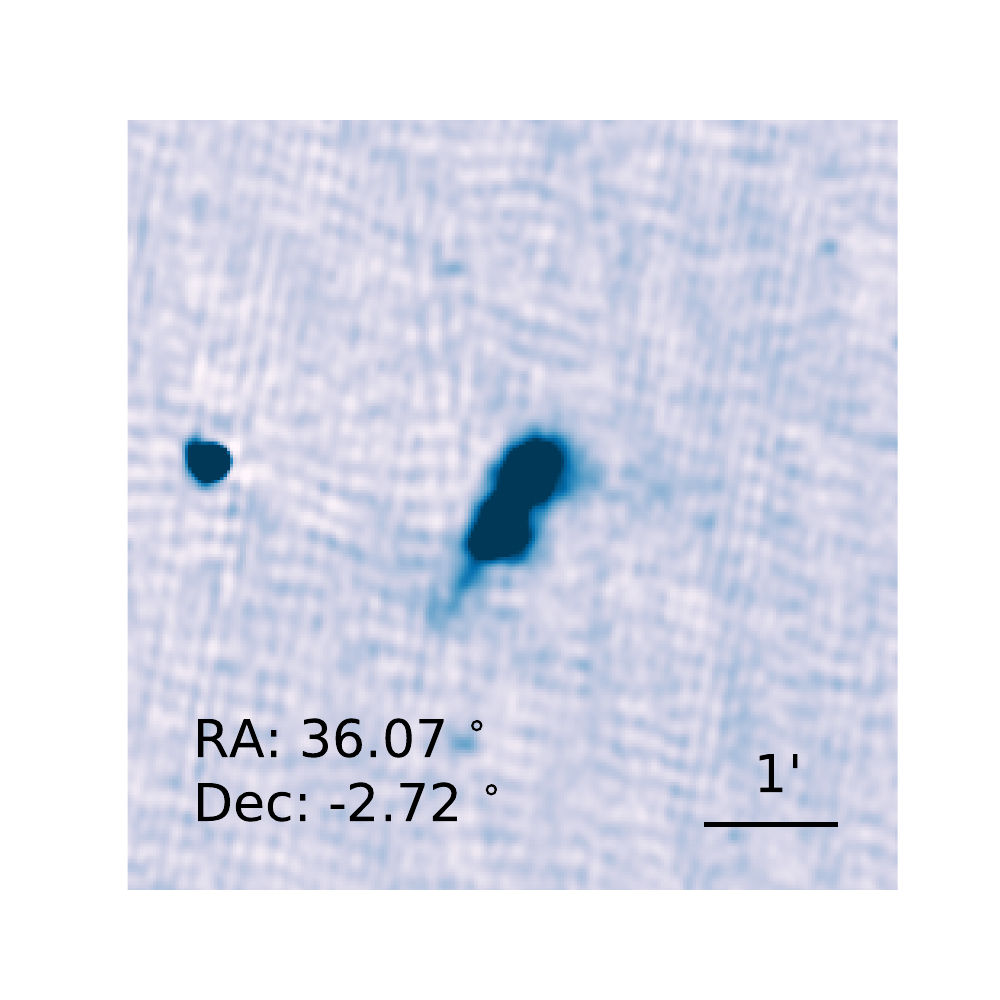}
\end{minipage}% 
\newline
\begin{minipage}[b]{0.2\textwidth}
\centering
\includegraphics[height=3.8cm]{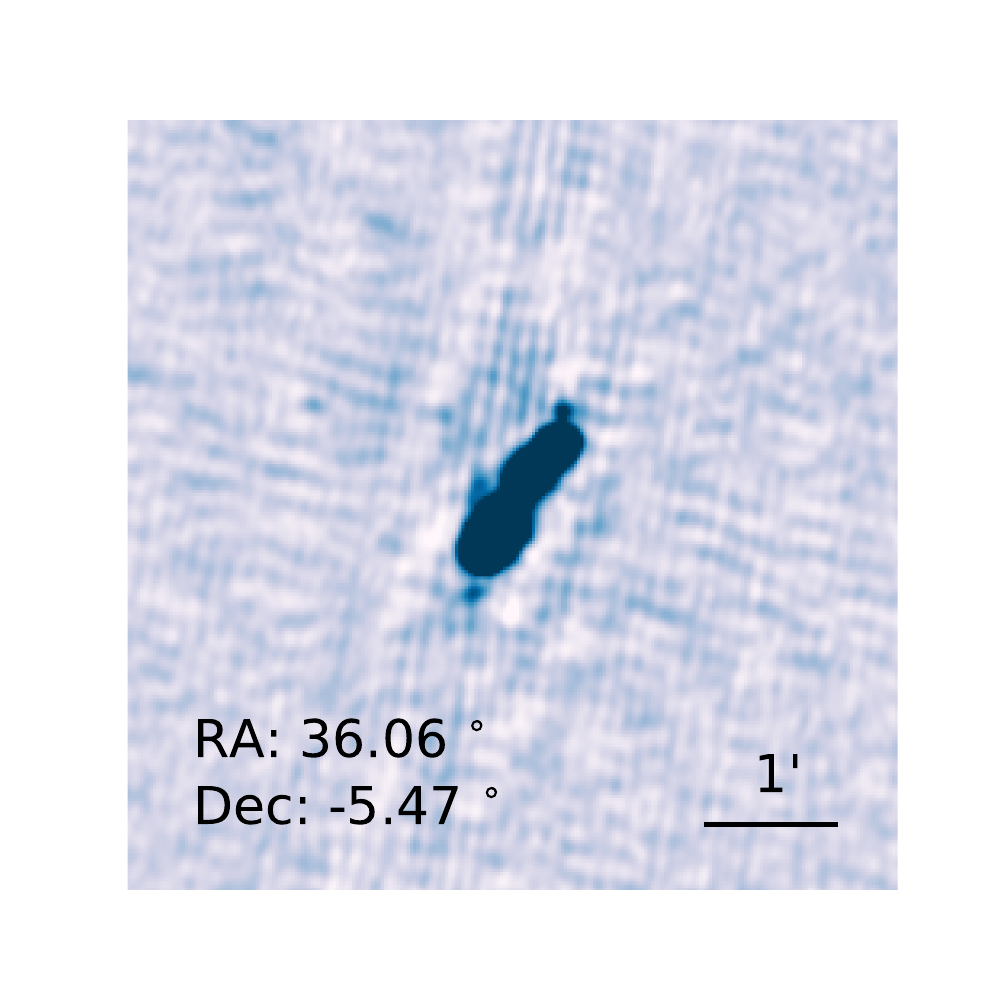}
\end{minipage}% 
\begin{minipage}[b]{0.2\textwidth}
\centering
\includegraphics[height=3.8cm]{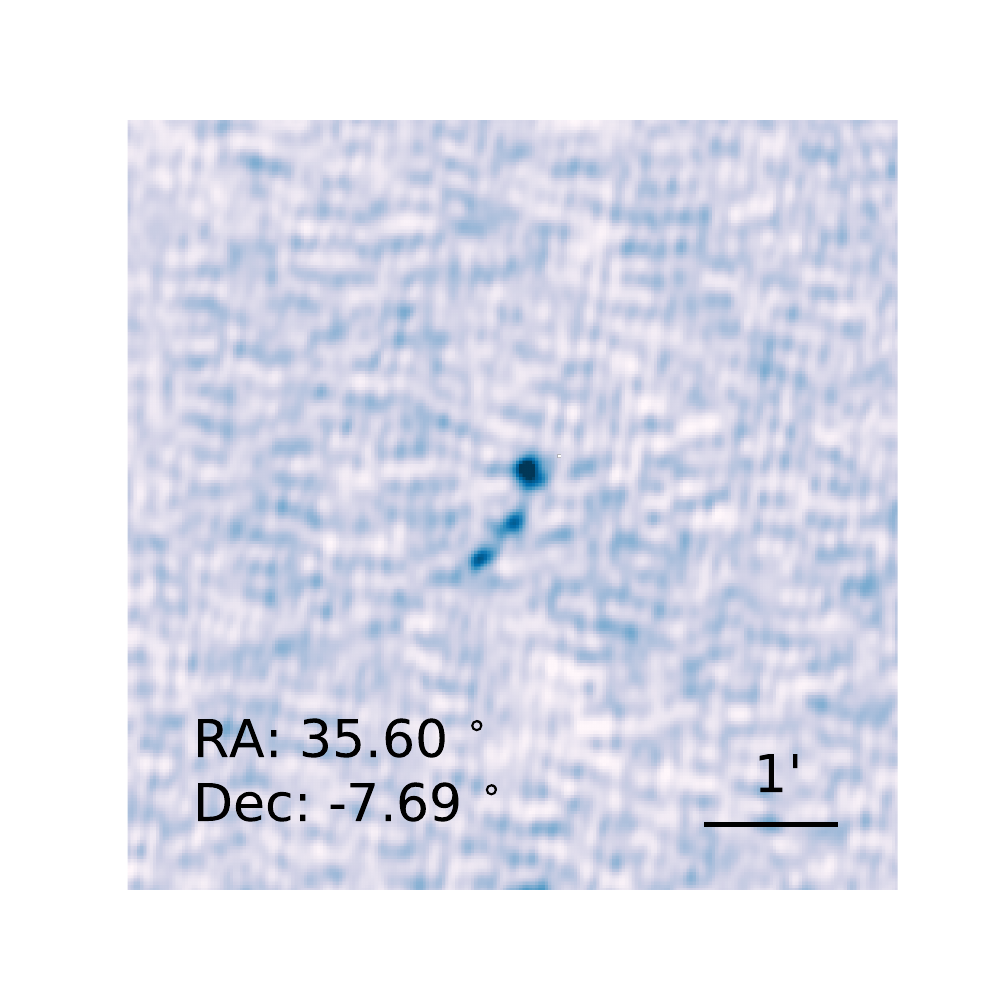}
\end{minipage}% 
\begin{minipage}[b]{0.2\textwidth}
\centering
\includegraphics[height=3.8cm]{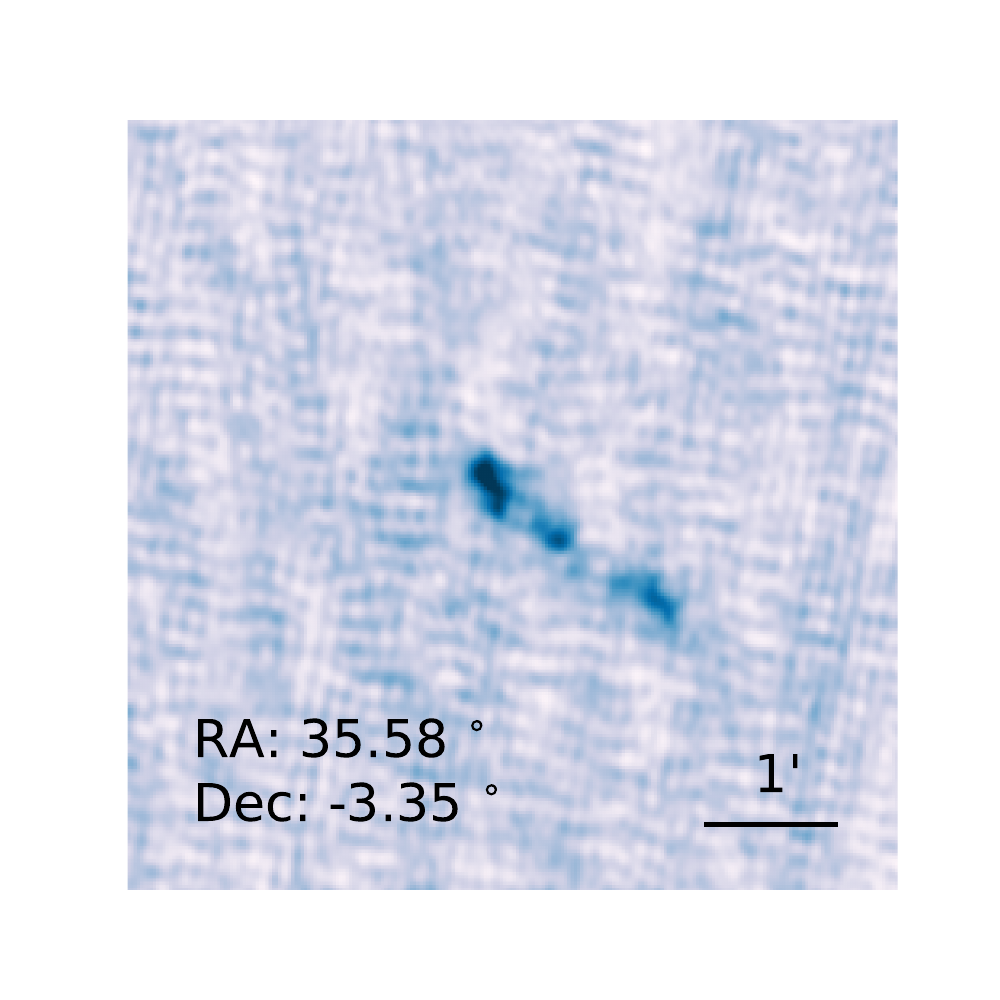}
\end{minipage}% 
\begin{minipage}[b]{0.2\textwidth}
\centering
\includegraphics[height=3.8cm]{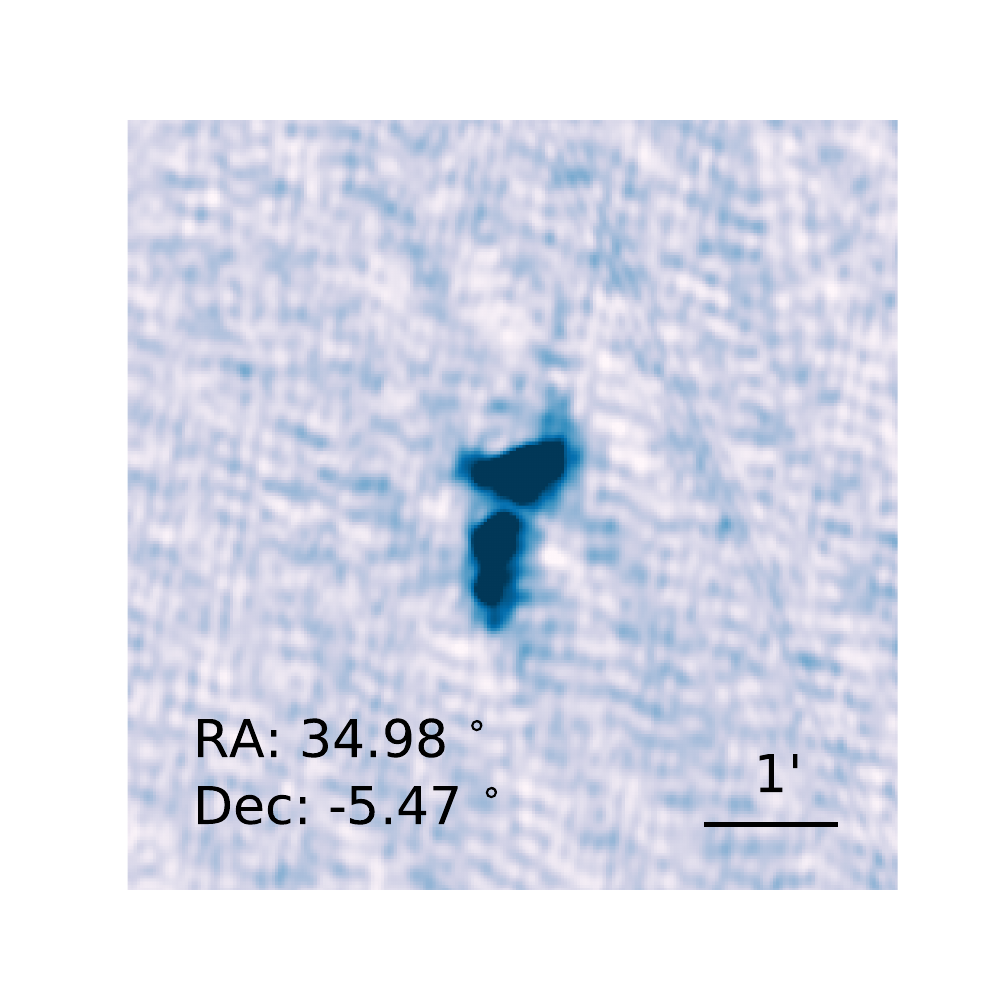}
\end{minipage}%
\begin{minipage}[b]{0.2\textwidth}
\centering
\includegraphics[height=3.8cm]{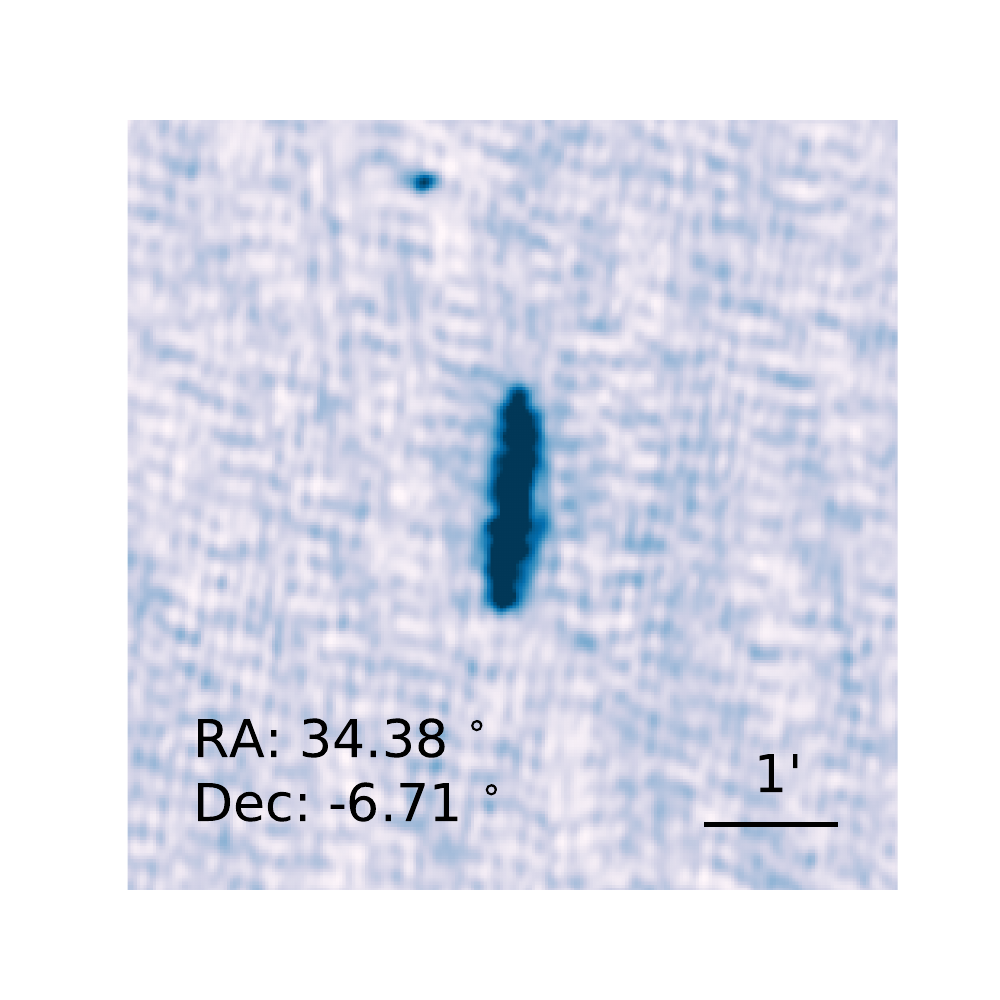}
\end{minipage}% 
\newline
\begin{minipage}[b]{0.2\textwidth}
\centering
\includegraphics[height=3.8cm]{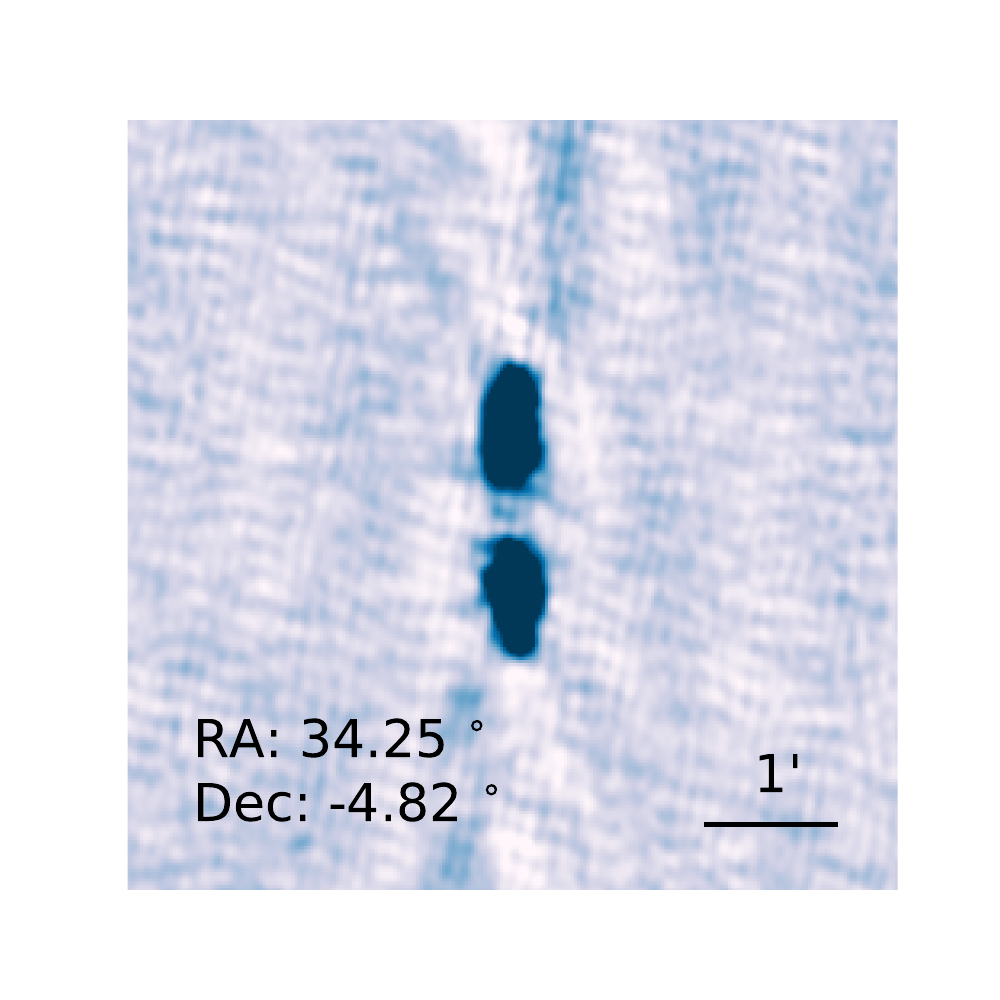}
\end{minipage}% 
\begin{minipage}[b]{0.2\textwidth}
\centering
\includegraphics[height=3.8cm]{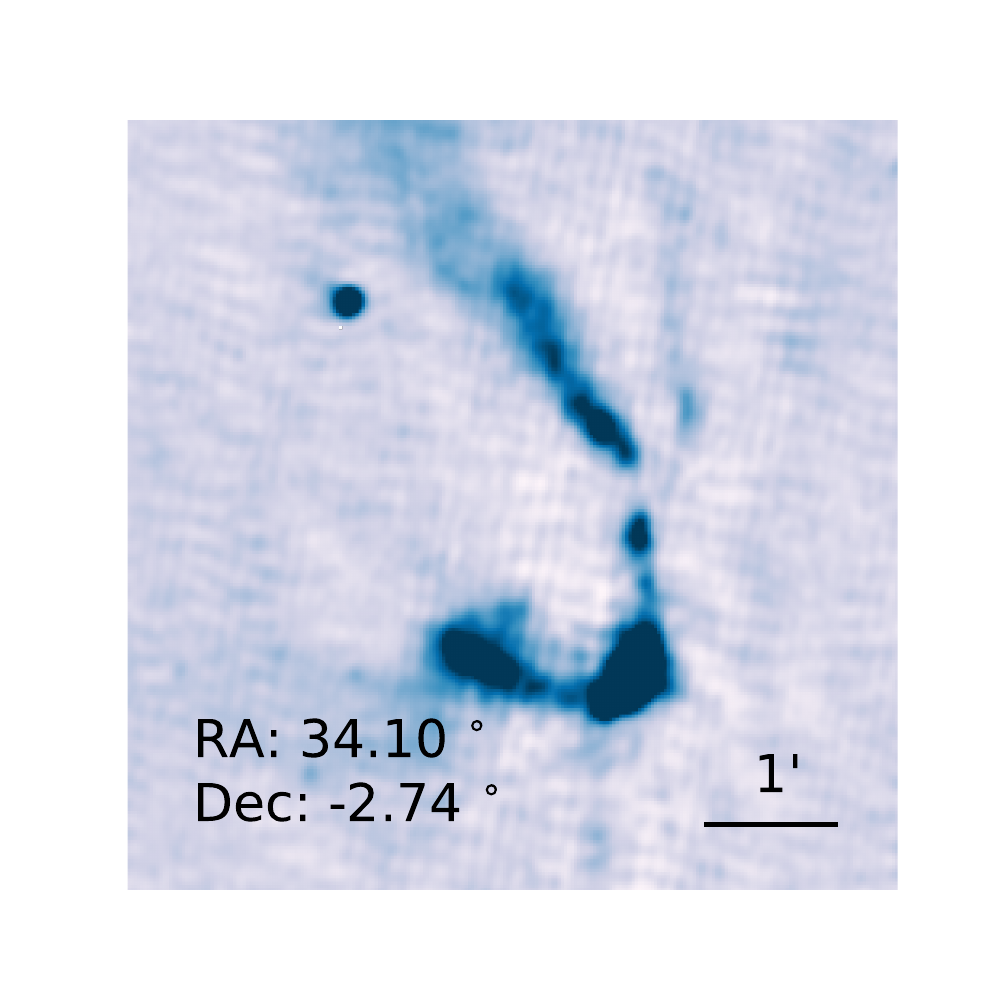}
\end{minipage}% 
\begin{minipage}[b]{0.2\textwidth}
\centering
\includegraphics[height=3.8cm]{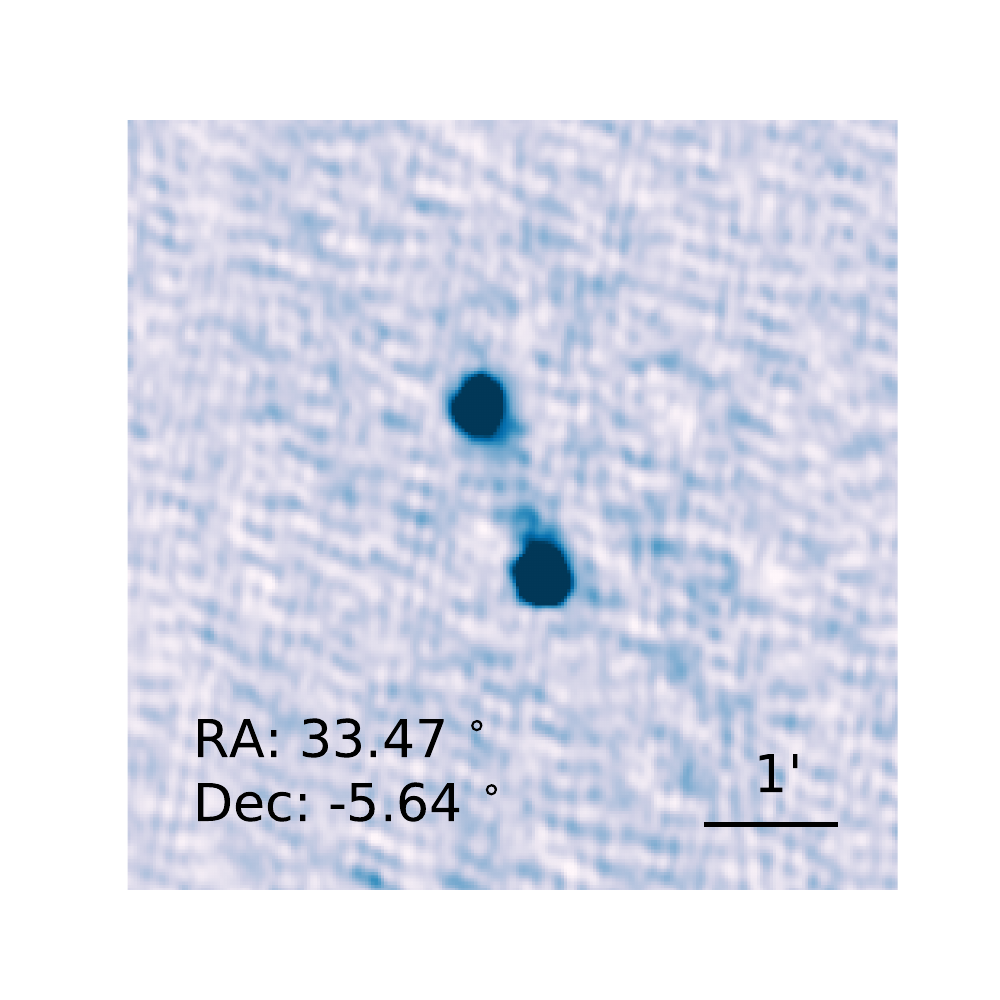}
\end{minipage}% 
\begin{minipage}[b]{0.2\textwidth}
\centering
\includegraphics[height=3.8cm]{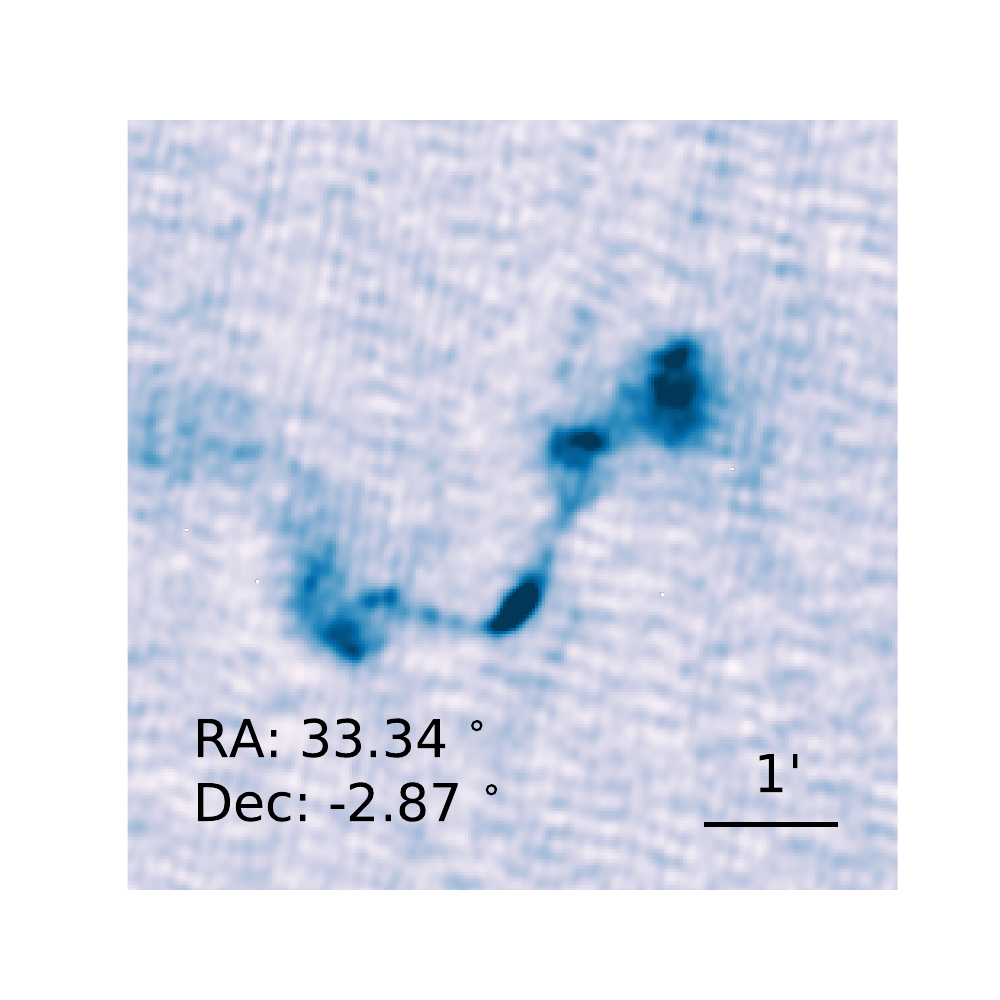}
\end{minipage}% 
\begin{minipage}[b]{0.2\textwidth}
\centering
\includegraphics[height=3.8cm]{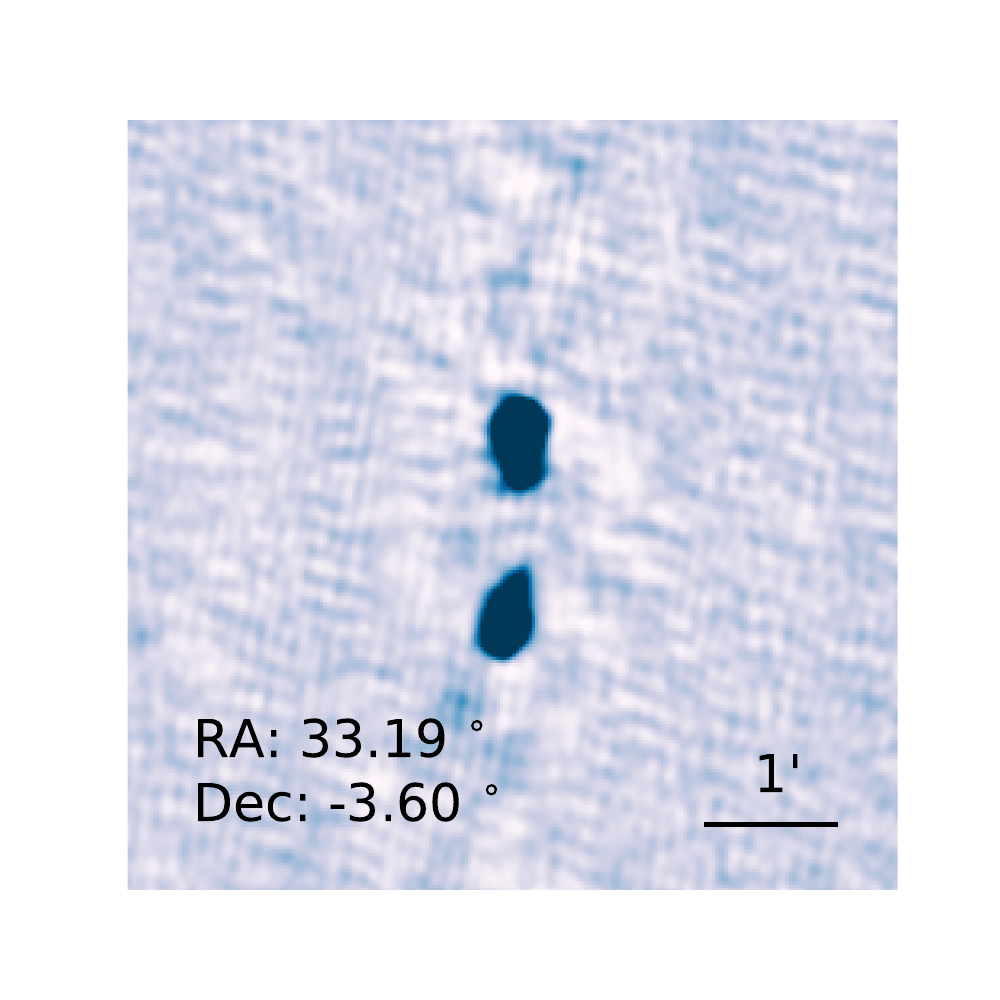}
\end{minipage}% 
\newline
\begin{minipage}[b]{0.2\textwidth}
\centering
\includegraphics[height=3.8cm]{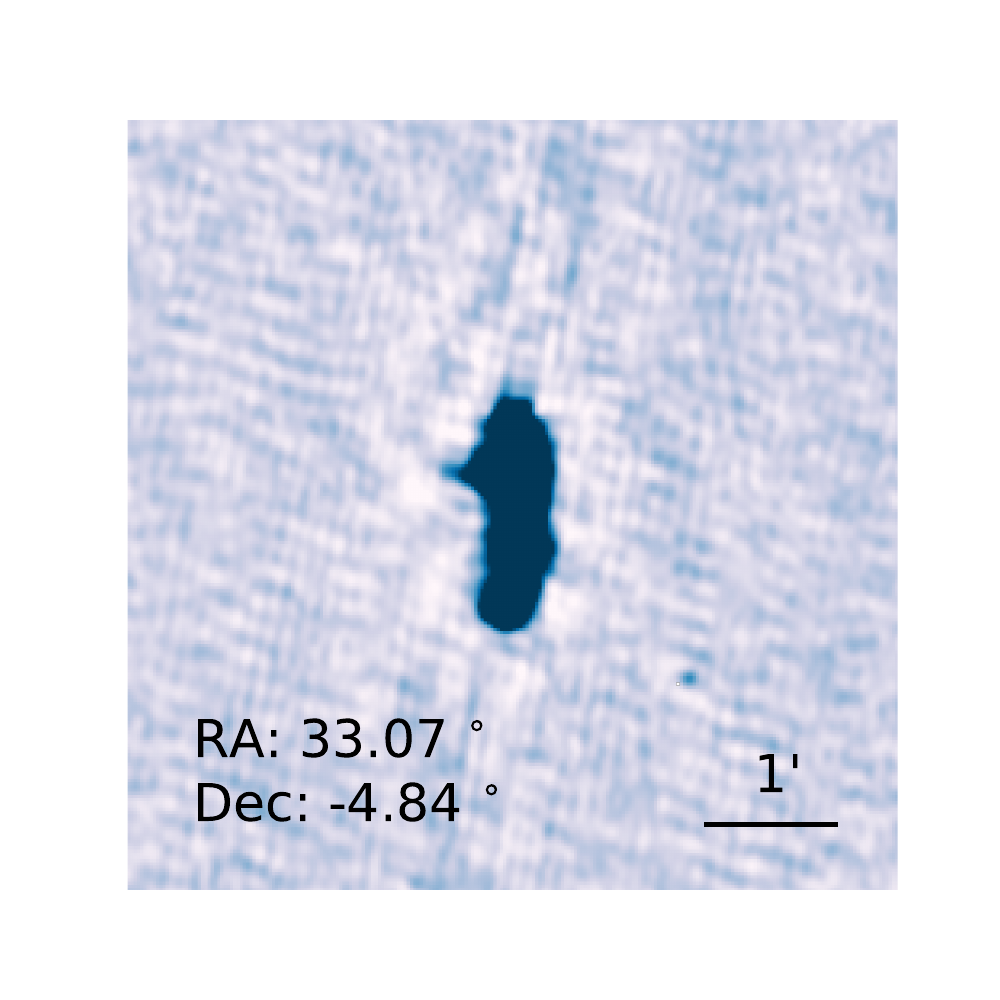}
\end{minipage}% 
\begin{minipage}[b]{0.2\textwidth}
\centering
\includegraphics[height=3.8cm]{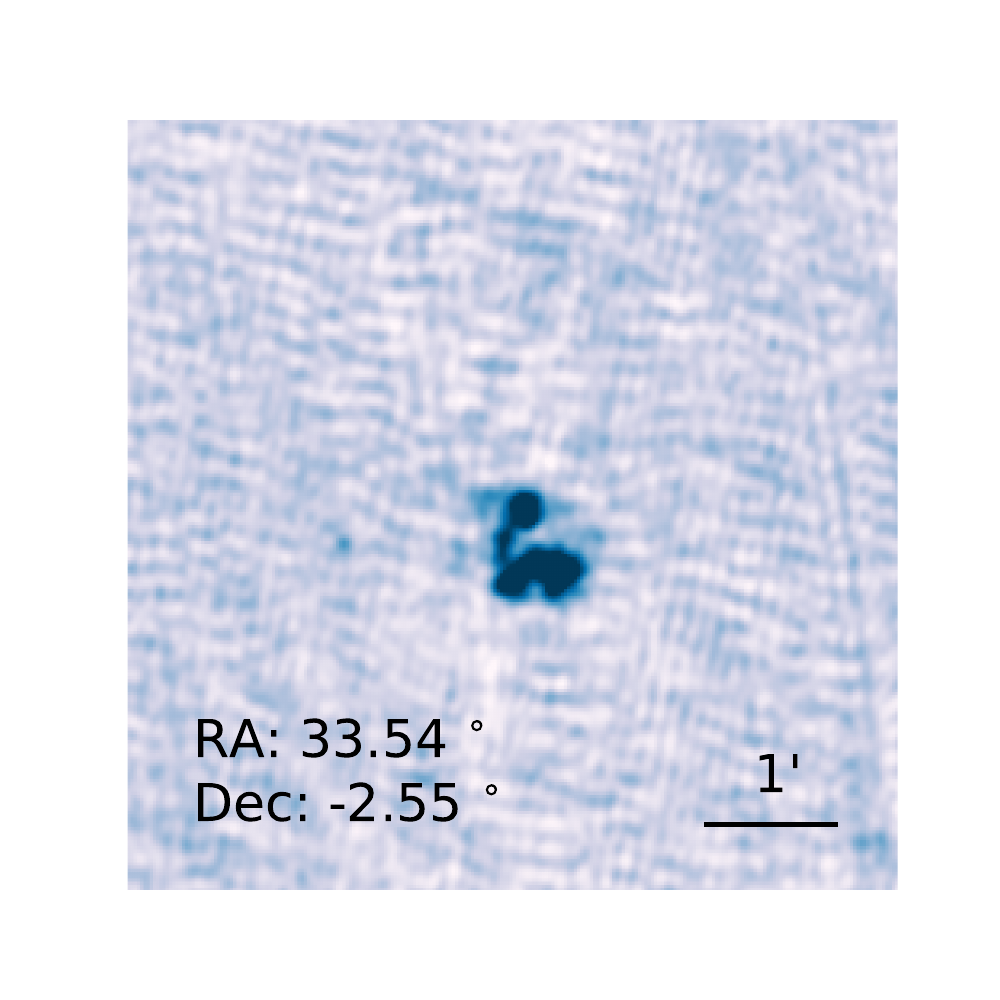}
\end{minipage}%
\FloatBarrier

\end{appendix}

\end{document}